\documentclass[a4paper,fleqn,usenatbib]{mnras}
\usepackage{amsmath}
\usepackage{dsfont}
\usepackage{graphicx,amsmath,bm,natbib,color, multirow}
\usepackage{amssymb}
\usepackage{amsmath} 
\usepackage{appendix}
\usepackage{xspace}
\usepackage{multirow}
\usepackage{mathptmx}
\usepackage{url}
\usepackage{xcolor}
\usepackage{footnote}
\usepackage{hyperref}  
\usepackage{graphicx}
\usepackage{wrapfig}
\usepackage{epstopdf}
\usepackage{acronym}
\usepackage{tablefootnote}
\usepackage{lineno}
\usepackage{eso-pic}
\usepackage{threeparttable}

\usepackage{caption}
\def\reff@jnl#1{{\rm#1\/}}
\def\aj{\reff@jnl{AJ}}         
\def\araa{\reff@jnl{ARA\&A}}      
\def\apj{\reff@jnl{ApJ}}        
\def\apjl{\reff@jnl{ApJ}}        
\def\apjs{\reff@jnl{ApJS}}       
\def\aap{\reff@jnl{A\&A}}        
\def\aapr{\reff@jnl{A\&A~Rev.}}     
\def\aaps{\reff@jnl{A\&AS}}       
\def\mnras{\reff@jnl{MNRAS}}      
\def\physrep{\reff@jnl{Physics Reports}}
\def\prd{\reff@jnl{Phys.Rev.D}}     
\def\prl{\reff@jnl{Phys.Rev.Lett}}   
\def\pasp{\reff@jnl{PASP}}       
\def\pasj{\reff@jnl{PASJ}}       
\def\nat{\reff@jnl{Nature}}       
\def\jcap{\reff@jnl{JCAP}}   
\def\memsai{\reff@jnl{MemSAI}} 
\def\na{\reff@jnl{New Astronomy}}       
\def\procspie{\reff@jnl{SPIE}}       
\def\pasa{\reff@jnl{PASA}}       

\def\Sref#1{Sec.~\ref{#1}\xspace}

\def\Fref#1{Fig.~\ref{#1}\xspace}

\def\Tref#1{Table~\ref{#1}\xspace}

\def\Eref#1{Eq.~(\ref{#1})\xspace}

\def\Aref#1{Appendix~\ref{#1}\xspace}
\def\Cref#1{Chapter~\ref{#1}\xspace}

\begin{document}

\title[A Unified Analysis of Four Cosmic Shear Surveys]{A Unified Analysis of Four Cosmic Shear Surveys}
\author[C.~Chang et al.]{
Chihway Chang,$^{1,*}$ 
Michael Wang,$^{2}$ 
Scott Dodelson,$^{3}$
Tim Eifler,$^{4,5}$ 
Catherine Heymans,$^{6}$ 
\newauthor
Michael Jarvis,$^{7}$ 
M.~James Jee,$^{8,9}$ 
Shahab Joudaki,$^{10}$ 
Elisabeth Krause,$^{4,5}$ 
Alex Malz,$^{11,12}$ 
\newauthor
Rachel Mandelbaum,$^{3}$ 
Irshad Mohammed,$^{2}$ 
Michael Schneider,$^{13}$ 
Melanie Simet,$^{4,14}$ 
\newauthor
Michael Troxel$^{15,16}$ and 
Joe Zuntz$^{6}$ 
\newauthor \large{(The LSST Dark Energy Science Collaboration)} \\ \\
$^{1}$Kavli Institute for Cosmological Physics, University of Chicago, Chicago, IL 60637 \\
$^{2}$Fermi National Accelerator Laboratory, P. O. Box 500, Batavia, IL 60510, USA \\
$^{3}$McWilliams Center for Cosmology, Department of Physics, Carnegie Mellon University, Pittsburgh, PA 15213, USA \\
$^{4}$Jet Propulsion Laboratory, California Institute of Technology, 4800 Oak Grove Dr., Pasadena, CA 91109, USA \\
$^{5}$Department of Physics, California Institute of Technology, Pasadena, CA 91125, USA \\
$^{6}$Institute for Astronomy, University of Edinburgh, Edinburgh EH9 3HJ, UK \\
$^{7}$Department of Physics and Astronomy, University of Pennsylvania, Philadelphia, PA 19104, USA \\
$^{8}$Department of Astronomy, Yonsei University, Yonsei-ro 50, Seoul, Korea \\
$^{9}$Department of Physics, University of California, Davis, California, USA \\
$^{10}$Department of Physics, University of Oxford, Denys Wilkinson Building, Keble Road, Oxford OX1 3RH, UK \\
$^{11}$Center for Cosmology and Particle Physics, New York University, 726 Broadway, New York, 10003, USA \\
$^{12}$Department of Physics, New York University, 726 Broadway, New York, 10003, USA \\
$^{13}$Lawrence Livermore National Laboratory, Livermore, CA 94551, USA \\
$^{14}$University of California, Riverside, 900 University Avenue, Riverside, CA 92521, USA \\
$^{15}$Center for Cosmology and Astro-Particle Physics, The Ohio State University, Columbus, OH 43210, USA \\
$^{16}$Department of Physics, The Ohio State University, Columbus, OH 43210, USA \\
$^{*}$e-mail address: chihway@kicp.uchicago.edu 
}

\date{Accepted XXX. Received YYY; in original form ZZZ}
\maketitle

\begin{abstract}
In the past few years, several independent collaborations have presented cosmological constraints from tomographic cosmic shear analyses. These analyses differ in many aspects: the datasets, the shear and photometric redshift estimation algorithms, the theory model assumptions, and the inference pipelines. To assess the robustness of the existing cosmic shear results, we present in this paper a unified analysis of four of the recent cosmic shear surveys: the Deep Lens Survey (DLS), the Canada-France-Hawaii Telescope Lensing Survey (CFHTLenS), the Science Verification data from the Dark Energy Survey (DES-SV), and the 450 deg$^{2}$ release of the Kilo-Degree Survey (KiDS-450). By using a unified pipeline, we show how the cosmological constraints are sensitive to the various details of the pipeline. We identify several analysis choices that can shift the cosmological constraints by a significant fraction of the uncertainties. For our fiducial analysis choice, considering a Gaussian covariance, conservative scale cuts, assuming no baryonic feedback contamination, identical cosmological parameter priors and intrinsic alignment treatments, we find the constraints (mean, 16\% and 84\% confidence intervals) on the parameter $S_{8}\equiv \sigma_{8}(\Omega_{\rm m}/0.3)^{0.5}$ to be $S_{8}=0.94_{-0.045}^{+0.046}$ (DLS), $0.66_{-0.071}^{+0.070}$ (CFHTLenS), $0.84_{-0.061}^{+0.062}$ (DES-SV) and $0.76_{-0.049}^{+0.048}$ (KiDS-450). From the goodness-of-fit and the Bayesian evidence ratio, we determine that amongst the four surveys, the two more recent surveys, DES-SV and KiDS-450, have acceptable goodness-of-fit and are consistent with each other. The combined constraints are $S_{8}=0.79^{+0.042}_{-0.041}$, which is in good agreement with the first year of DES cosmic shear results and recent CMB constraints from the \textit{Planck} satellite. \\
\end{abstract}

\section{Introduction}

The large-scale structure of the Universe bends the light rays emitted from distant galaxies according to General Relativity \citep{Einstein1936}. This effect, known as weak (gravitational) lensing, introduces coherent distortions in galaxy shapes, which carry information of the cosmic composition and history. 

One of the most common statistics used to extract this information is \textit{cosmic shear}, as inferred by the two-point correlation function of galaxy shapes $\xi_{\pm}(\theta)$ \citep{Bartelmann2001}. Assuming the flat-sky approximation, these two-point functions are connected to the lensing power spectrum $C(\ell)$ via
\begin{equation}
\xi^{ij}_{\pm}(\theta) = \frac{1}{2\pi}\int d\ell \, \ell J_{0/4}(\theta \ell) \, C^{ij}(\ell),
\label{eq:xipm}
\end{equation}
where $J_{0/4}$ is the 0th/4th-order Bessel functions of the first kind. The $i$ and $j$ indices specify the two samples of galaxies (or in the case of $i=j$, the galaxy sample) from which the correlation function is calculated. Usually these samples are defined by a certain redshift selection. Under the Limber approximation \citep{Limber1953,Loverde2008} and in a spatially flat universe\footnote{For a non-flat universe, one would replace $\chi$ by $f_{K}(\chi)$ in the following equations, where $K$ is the universe's curvature, $f_{K}(\chi)=K^{-1/2}\sin(K^{1/2}\chi)$ for $K>0$ and $f_{K}(\chi)=(-K)^{-1/2}\sinh((-K)^{1/2}\chi)$ for $K<0$.}, the lensing power spectrum encodes cosmological information through 
\begin{equation}
C^{ij}(\ell) = \int_{0}^{\chi_{H}} d\chi \frac{q^{i}(\chi)q^{j}(\chi)}{\chi^2} P_{NL}\left( \frac{\ell + 1/2}{\chi}, \chi \right),
\end{equation}
where $\chi$ is the radial comoving distance, $\chi_{H}$ is the distance to the horizon, $P_{NL}$ is the nonlinear matter power spectrum, and $q(\chi)$ is the lensing efficiency defined via
\begin{equation}
q^{i}(\chi) = \frac{3}{2} \Omega_{\rm m} \left( \frac{H_{0}}{c}\right)^{2} \frac{\chi}{a(\chi)} \int_{\chi}^{\chi_{H}}d\chi ' n^{i}(\chi') \frac{dz}{d\chi'} \frac{\chi' - \chi}{\chi'},
\label{eq:lensing_efficiency}
\end{equation}
where $\Omega_{\rm m}$ is the matter density today, $H_{0}$ is the Hubble parameter today, $a$ is the scale factor, and $n^{i}(\chi)$ is the redshift distribution of the galaxy sample $i$. 

Since the first detection of cosmic shear in \cite{Bacon2000,Kaiser2000,Wittman2000,Schneider2002}, the field has seen a rapid growth. In particular, a number of large surveys have delivered cosmic shear results with competitive cosmological constraints in the past few years \citep{Heymans2013, Becker2015,Jee2016,Joudaki2017,Troxel2017,Hildebrandt2017,DES2017}, while ongoing and future surveys will deliver data in much larger volumes and better quality [e.g.  
 the Dark Energy Survey \citep[DES,][]{Flaugher2005}, the Hyper SuprimeCam Survey \citep[HSC,][]{Aihara2017}, the Kilo-Degree Survey \citep[KiDS,][]{deJong2015} and the Large Synoptic Survey Telescope \citep[LSST,][]{Ivezic2008,Abell2009}]. 
 
 One of the surprises that has emerged in the past couple of years is that there seems to be a modest level of discordance between different cosmological probes \citep{MacCrann2015,Freedman2017,Raveri2018}. Even though in many of these cases, the level of tension between the different probes still needs to be quantified more rigorously, one consequence has been that the cosmology community has started to more carefully scrutinize how the datasets are analyzed. This is especially important as we expect the statistical power of the datasets to be orders of magnitude better in the near future. If there is indeed a tension between the different probes, it could point to an exciting new direction where the simple $\Lambda$CDM cosmology cannot explain all the observables and new physics is needed.

A variety of studies have been carried out to understand systematic effects in weak lensing measurements. This includes systematics from the instrument and the environment, from modeling the point-spread function (PSF) and measuring galaxy shapes, from estimating the redshift of each galaxy, from the theoretical modeling, and many more \citep[see][and references therein for a comprehensive list of studies]{Mandelbaum2017}. In this work, we focus on understanding the steps between the shear catalog and cosmological constraints: measuring the shear two-point correlation function [\Eref{eq:xipm}], estimating the covariance, modeling of the signal, and inferring cosmological parameters. We build a modular and robust pipeline using the \textsc{Pegasus} workflow engine \citep{Deelman2015} to analyze the datasets in a streamlined and transparent fashion -- this pipeline will serve as the first step towards building up cosmological analysis pipelines for the LSST Dark Energy Survey Collaboration (DESC).

In this paper we apply the pipeline to four publicly available datasets that are precursors to ongoing and future cosmic shear surveys: the Deep Lens Survey \citep[DLS,][]{Jee2016}, the Canada-France-Hawaii Telescope Lensing Survey \citep[CFHTLenS,][]{Joudaki2017}, the Science Verification data from the DES \citep[DES-SV,][]{Abbott2015}, and the 450 deg$^{2}$ release of the KiDS \citep[KiDS-450,][]{Hildebrandt2017}. All four surveys were carried out fairly recently and have comparable statistical power, so a uniform pipeline is a powerful way to identify any discrepancies and to understand their origin. A detailed look at the consistency between the four datasets can also inform us about potential systematic issues in the processing that produces the catalogs from which our pipeline begins. It is, however, not the scope of this paper to investigate these issues upstream to our pipeline, where a thorough pixel-level study for each survey may be required. 

The paper is organized as follows. In \Sref{sec:data} we describe the details of the four datasets used in this work. In \Sref{sec:pipeline} we describe the pipeline that is used to process the data. We then outline in \Sref{sec:comparison_framework} the framework in which we compare the datasets and the elements in the pipeline that are allowed to vary. Our results are shown and discussed in \Sref{sec:results} and we conclude in \Sref{sec:conclusion}.

\begin{figure*}
\includegraphics[width=1.6\columnwidth]{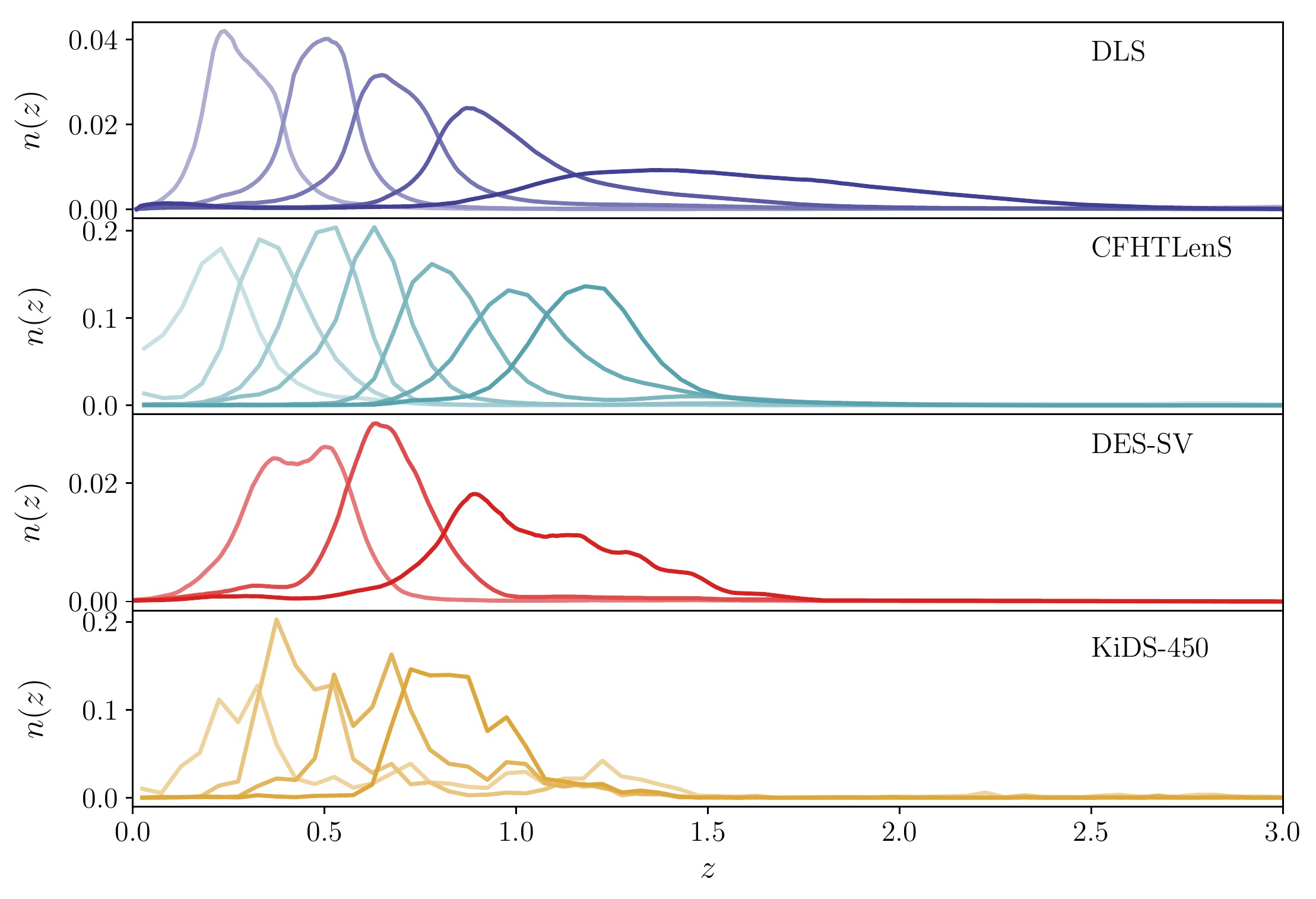}
\caption{Estimation of the the tomographic redshift distributions used in the four cosmic shear analyses. For DLS, CFHTLenS and DES-SV, stacked photometric redshift probability distribution functions (PDFs) were used; for KiDS-450, the redshift distribution of spectroscopic samples (weighted to match the source galaxies used for the cosmic shear analysis) were used. We see that DLS and CFHTLenS extend to higher redshift compared to DES-SV and KiDS-450.}
\label{fig:nofz}
\end{figure*}

\begin{figure*}
\includegraphics[width=1.6\columnwidth]{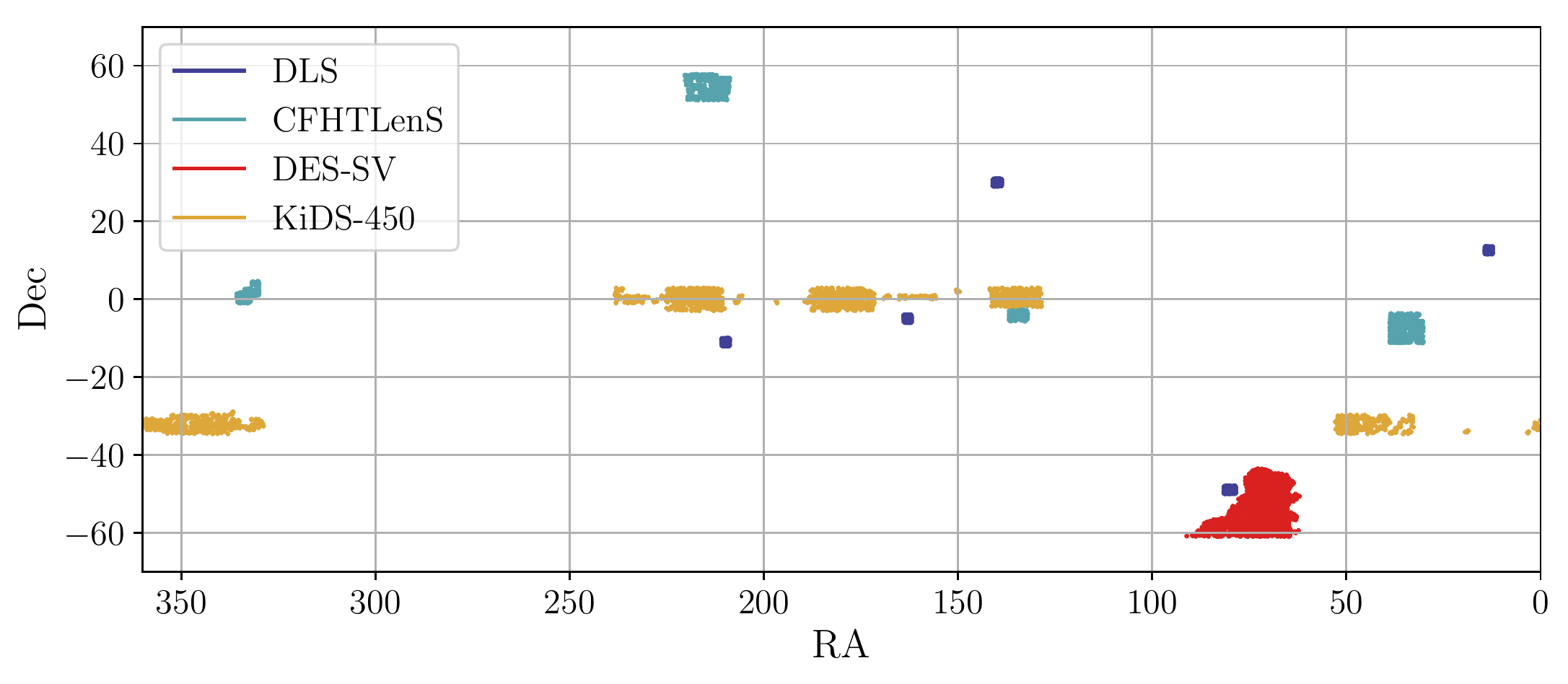}
\caption{The location and footprints of the four surveys analyzed in this paper. There is essentially no overlap between the footprints of the four surveys, except for a very small part of CFHTLenS and KiDS-450 at (RA, Dec) $\approx$ (130, -3) deg. We note that the projection in this plot does not reflect the relative area of the four surveys.} 
\label{fig:footprint}
\end{figure*}

\section{Precursor Surveys}
\label{sec:data}

We describe briefly the four datasets used in this work. In \Fref{fig:nofz} we show the estimated redshift distribution for each dataset. The number of tomographic bins in each case was chosen by the collaboration (and we keep that number fixed throughout), but the range does convey information about the depth of the surveys. For example, the DLS is much deeper and therefore is sensitive to shear at higher redshift. In \Fref{fig:footprint} we show the footprint of the four datasets on the sky. Since the footprints of these surveys are largely non-overlapping, they can be treated as independent. In \Tref{tab:survey_param} we list the main parameters used in each of the cosmic shear analyses.

\subsection{DLS: the Deep Lens Survey}

The DLS \citep{Wittman2000} consists of five $\sim 2 \times 2$ deg${^2}$ fields that add up to $\sim18$ deg$^{2}$. Two fields were observed by the Kitt Peak Mayall 4m telescope/Mosaic Prime-Focus Imager \citep{Muller1998}, and the other two by the Cerro Tololo Blanco 4m telescope/Mosaic Prime-Focus Imager. The total DLS dataset was taken over 140 nights of $B$, $V$, $R$ and $z$ imaging. The approximate limiting magnitudes for each band (at 5$\sigma$) are 26, 26, 27, 26 in $B$, $V$, $R$ and $z$, respectively. The average seeing is $\sim 0.9 \arcsec$ in $R$.

The cosmic shear cosmology analysis from DLS was first presented in \citet{Jee2012}, and later updated with \citet{Jee2016}, which is the analysis we focus on in this paper. The shear measurement method is described in \citet{Jee2012}, where an elliptical Gaussian galaxy model is used and image simulations \citep{Jee2011} were employed for calibration of the shear estimate. The photometric redshift (or, photo-z) estimation uses the \textsc{BPZ} code \citep{Benitez2000} and is validated against the PRIsm MUlti-object Survey \citep[PRIMUS,][]{Coil2011} in \citet{Jee2012}.

DLS is the deepest survey with the smallest area of all the four datasets used in this work. As we do not have access to the shape catalogs for DLS, we start from the pre-measured two-point correlation functions provided by the collaboration. 

\begin{table*}
\centering
\begin{threeparttable}
\caption{Characteristics of the four surveys and some of the modeling choices in each of the analyses.}
\label{tab:survey_param}
\begin{tabular}{lcccc}
\hline\hline
Parameters & DLS & CFHTLenS & DES-SV & KiDS-450 \\
\hline
Reference & \citet{Jee2016} & \citet{Joudaki2017} &\citet{Abbott2015} & \citet{Hildebrandt2017} \\
Area (deg$^{2}$)& 18 & 94 & 139 & 360 \\
Mean redshift &[0.45, 0.62, 0.75, & [0.35, 0.42, 0.50, 0.64 & [0.44, 0.67, 1.03] & [0.50, 0.49, 0.68, 0.85] \\
              & 1.04, 1.51]     & 0.88, 1.06, 1.20]  &  &  \\
$\sigma_{e}$ (per component)& [0.25, 0.25, 0.25, &  [0.28, 0.28, 0.28, 0.28, & [0.27, 0.28, 0.28] & [0.29, 0.28, 0.27, 0.28]\\
                                                 &  0.25, 0.25]       &   0.28, 0.28, 0.28]     &        &        \\
$n_{\rm eff}^{*}$ (arcmin$^{-2}$) &[3.15, 3.58, 3.02, & [1.28,1.10,1.11,1.59, & [1.97, 2.03, 2.13] & [2.31,1.83,1.80,1.45] \\
                                                    & 2.91, 3.75]      &  2.74,1.75,0.87]      &                &                      \\
$\xi_{+}$ $\theta_{\rm min}$ (arcmin)& 1.0& 1.0 & 4.6/2.0$^{\dagger}$ &0.5\\
$\xi_{+}$ $\theta_{\rm max}$ (arcmin)& 90.0& 60.0 & 60.0 &72.0\\
$\xi_{-}$ $\theta_{\rm min}$ (arcmin)& 1.0& 8.0 & 56.5/24.5$^{\dagger \dagger}$ & 4.2\\
$\xi_{-}$ $\theta_{\rm max}$ (arcmin)& 90.0& 120.0& 300.0 &300.0 \\
Data vector length & 240 & 280 & 36 & 130 \\
Covariance & 2048 simulations & 1988 simulations & 126 simulations & analytic \\
Cosmology inference & \textsc{CosmoPMC} & \textsc{CosmoMC} & \textsc{Cosmosis} & \textsc{CosmoMC} \\
\hline\hline
\end{tabular}
\begin{tablenotes}
\item[*] We adopt the definition used in \citet{Heymans2012}, where $n_{\rm eff} = A^{-1} (\Sigma_{i} w_{i} )^{2} / \Sigma_{i} w_{i}^{2}$. $A$ is the area of the survey while $w_{i}$ is the weight for source galaxy $i$. The summation runs over all source galaxies.
\item[$\dagger$] Only for $\xi_{+}^{23}$ and $\xi_{+}^{33}$, the small-scale cutoff is 2 arcmin.
\item[$\dagger \dagger$] Only for $\xi_{-}^{11}$ and $\xi_{-}^{12}$, the small-scale cutoff is 60 arcmin.
\end{tablenotes}
\end{threeparttable}
\end{table*}

\subsection{CFHTLenS: the Canada-France-Hawaii Telescope Lensing Survey}

The CFHTLenS data \citep{Erben2012, Heymans2012} spans four distinct contiguous fields of approximately 63.8, 22.6, 44.2 and 23.3 deg$^{2}$. Images are taken via the Canada-France-Hawaii 3.6m Telescope/MegaCam Imager in six filter bands: $u^{*}$, $g'$, $r'$, $i'$, $y'$, $z$. The limiting magnitudes for each band (at 5$\sigma$ in 2$\arcsec$ aperture) are 25.24, 25.58, 24.88, 24.54, 24.71, 23.46 in the six bands, respectively, while the average seeing is 
0.68$\arcsec$ in $i'$, where the shapes are measured.

The cosmic shear cosmology analysis from CFHTLenS was presented first in \citet{Fu2008} and later updated in \citet{Heymans2013,Kilbinger2012} and then \citet{Joudaki2017}, which is the focus of this paper. The shear measurement was based on the \textsc{LensFit} package \citep{Miller2007}, which is a likelihood-based model-fitting approach that allows for joint-fitting over multiple observations of the same galaxy. A two-component (disk plus bulge) model is used to fit the galaxy shape and to extract the galaxy ellipticity. The method marginalizes over nuisance parameters such as galaxy position, size, brightness and bulge fraction. \citet{Miller2013} describes the simulation-based calibrations that are applied to the shear catalog. The photo-z estimation was based on the \textsc{BPZ} code \citep{Benitez2000, Hildebrandt2012}. The catalogs are publicly available\footnote{\url{http://www.cfhtlens.org/astronomers/data-store}}.
As seen in \Fref{fig:nofz}, the CFHTLenS analysis uses the largest number of tomographic bins. 

In addition, in \citet{Joudaki2017} extensive explorations of the impact of different intrinsic alignment (IA) models, baryonic feedback models, and photo-z uncertainties were performed. When considered independently, only the IA amplitude was found to be substantially favored by the CFHTLenS data. However, with a 2$\sigma$ negative amplitude, this could be a sign of either simplistic modeling or unaccounted systematics. The CFHTLenS analysis further considered joint accounts of the systematic uncertainties, where the ``MIN'', ``MID'' and ``MAX'' cases included successively conservative treatments of the systematics modeling and scale cuts (along with a "fiducial" case that included no systematics). \citet{Joudaki2017} found that the $S_{8}$ constraints were sensitive to the specific treatment of the systematic uncertainties, where the level of concordance with \textit{Planck} ranged from decisive discordance (MIN) to substantial concordance (MAX). As a result, when quoting the nominal constraints from the collaboration, we show all three cases for CFHTLenS.

\subsection{DES-SV: the Dark Energy Survey Science Verification Data}

The DES-SV dataset was taken before the official DES run began and was designed to cover a smaller area ($\sim 250$ deg$^{2}$) to the full depth expected for DES. The area used in the cosmology analysis is a contiguous area of 139 deg$^{2}$. Images were taken with the Dark Energy Camera \citep{Flaugher2015} on the Cerro Tololo Blanco 4m telescope. Five filter bands: $g$, $r$, $i$, $z$, $Y$ were used to a median depth of $g \sim 24.0$, $r \sim 23.9$, $i \sim 23.0$ and $z \sim 22.3$, respectively. The average seeing is 1.11$\arcsec$ in $r$, 1.08$\arcsec$ in $i$ and 1.03$\arcsec$ in $z$ -- the DES-SV galaxy shapes used information from all three bands.

The cosmology analysis from weak lensing was presented in \citet{Abbott2015}, while the details and testing of the measurements were recorded in \citet{Becker2015}. Two independent shear catalogs were produced from the DES-SV data and have been extensively tested in \citet{Jarvis2015}. In this work we use the catalog produced by the shear measurement algorithm \textsc{ngmix} \citep{Sheldon2014}, which is a fast Bayesian fitting algorithm that models galaxies as a mixture of Gaussian profiles. The Gaussian profiles are chosen to approximate an exponential disk. Several photo-z algorithms were tested in \citet{Becker2015} and \citet{Bonnett2016} including \textsc{SkyNet} \citep{Bonnett2013} and \textsc{BPZ} \citep{Benitez2000}. In \citet{Abbott2015}, results from all shear and photo-z catalogs were presented and shown to be consistent. In this work we use only the \textsc{ngmix} catalog and the \textsc{SkyNet} photo-z, as these were recommended by DES as the fiducial catalogs with the best performance. All catalogs are publicly available\footnote{\url{https://des.ncsa.illinois.edu/releases/sva1/content}}.

The analysis pipeline used in \citet{Abbott2015} is based on \textsc{ComoSIS} \citep{Zuntz2014}, which is the same cosmology inference framework we use in this paper, so we expect very good agreement between our analysis and \citet{Abbott2015}.

\begin{figure*}
\includegraphics[width=1.6\columnwidth]{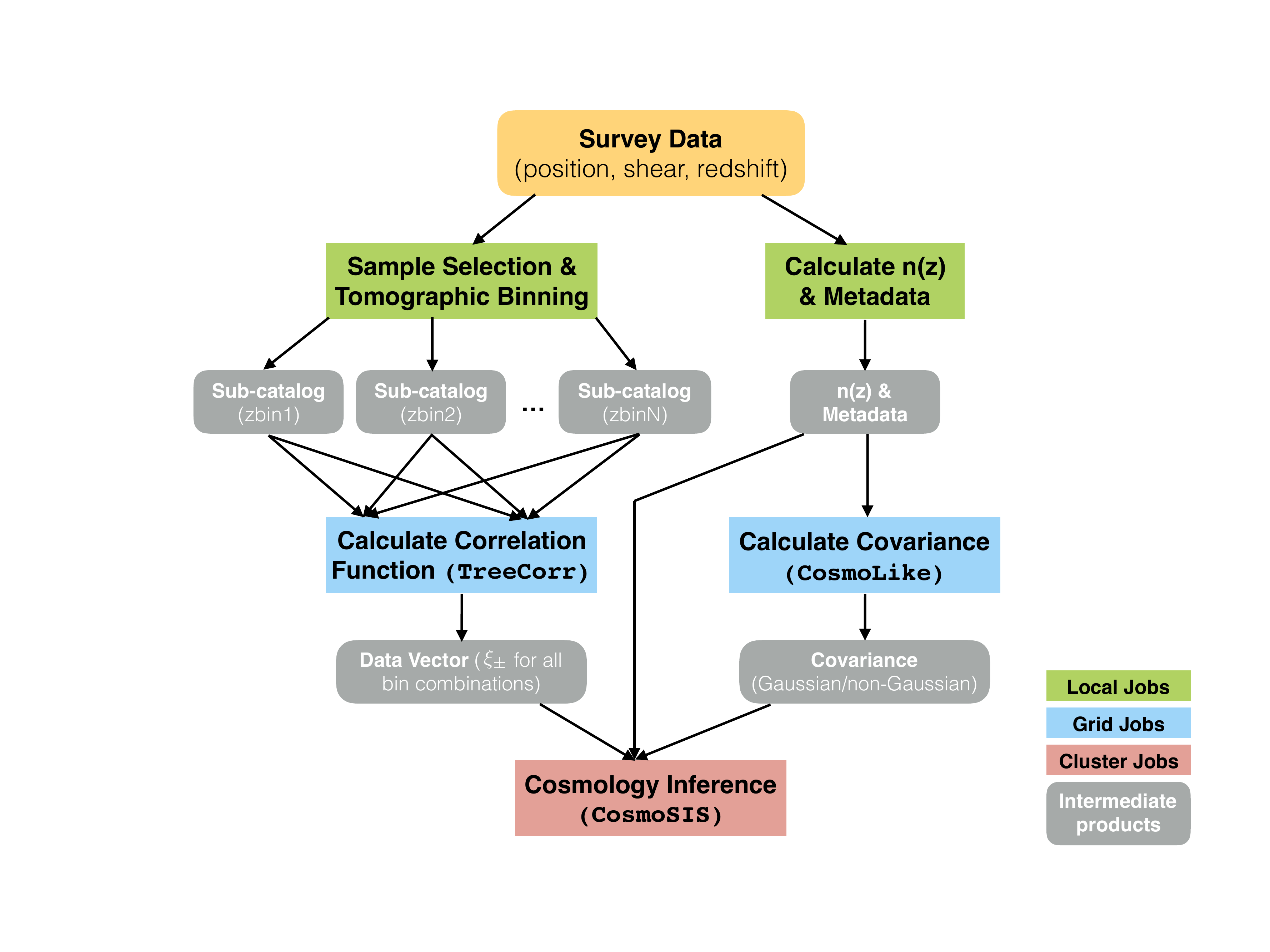}
\caption{Flow chart of steps used in pipeline that goes from survey data to cosmology. The arrow pointing towards the left from the ``$n(z)$ \& Metadata'' box bypasses the covariance calculation -- it refers to the route taken when the survey-provided covariances are used in the inference.}
\label{fig:pipefig}
\end{figure*}

\subsection{KiDS-450: the 450 deg$^2$  Kilo-Degree Survey}
\label{sec:kids}

The KiDS-450 dataset consists of five separate patches covering a total effective area of $\sim 360$ deg$^{2}$. Data was taken using the OmegaCAM CCD Mosaic camera mounted at the Cassegrain focus of the VLT Survey Telescope (VST). There are four SDSS-like filter bands, $u$, $g$, $r$, $i$, and the image depth is approximately 24.3, 25.1, 24.9, 23.8 in each band, respectively (5$\sigma$ limit in 2$\arcsec$ aperture). The median seeing is 0.66$\arcsec$ in $r$, and no $r$-band images have seeing greater than 0.96$\arcsec$.

The cosmology analysis from cosmic shear using KiDS-450 data was presented in \cite{Hildebrandt2017}. The cosmological inference pipeline was largely based on that used in CFHTLenS \citep{Joudaki2017}, while several updates were made to the measurement pipeline. First, the shear calibration to the \textsc{LensFit} shear catalog was based on more sophisticated image simulations \citep{FenechConti2017}. Second, a new approach for estimating photo-z and propagating photo-z uncertainties into cosmological inferences was implemented, which we briefly describe below.

The $n(z)$ estimation in KiDS-450 is based on ideas presented in \citet{Lima2008} and implemented in \citet{Bonnett2016}. This approach is referred to in \citet{Hildebrandt2017} as the ``weighted direct calibration (DIR)'' method. The $n(z)$ is taken directly from the redshift distribution of a spectroscopic sample with appropriate re-weighting in the color-magnitude space to correct for the incompleteness and selection effects in both the shear catalog and the spectroscopic sample. Since the $n(z)$'s are derived from a small number of spectroscopic galaxies, they appear more noisy than the other surveys in \Fref{fig:nofz}, where more traditional photo-z methods (stacked redshift probability distribution functions, or PDFs) are used. 


\section{Pipeline}
\label{sec:pipeline}

A directed acyclic graph (DAG) representing the modular pipeline developed for this analysis is shown in \Fref{fig:pipefig}. The pipeline is implemented using the \textsc{Pegasus} \citep{Deelman2015} workflow management system. The individual components in the DAG are explained in more detail in \Sref{sec:comparison_framework}, but we outline below the basic structure of the pipeline. Starting from the top, catalogs from each survey are fed into the first two branches of the pipeline which are run in parallel. The first branch (the left half of the DAG) starts with performing sample selection and tomographic binning by sorting catalog data into $N_t$ redshift bins and applying appropriate quality cuts, producing one intermediate catalog file per bin. Next, $N_{c}=N_{t}(N_{t}+1)/2$ jobs are launched in parallel to calculate the two-point shear correlation functions using the \textsc{TreeCorr}\footnote{\url{https://github.com/rmjarvis/TreeCorr}} code. The output of all the parallel jobs are collected to form the data vector for the analysis. The second branch (the right half of the DAG) starts with estimating the full redshift distribution $n(z)$ by summing the redshift PDFs for each individual galaxy\footnote{We note that in the later analyses for the individual surveys, some of them do not provide redshift PDFs per galaxy. Instead, they provide the full $n(z)$ for each redshift bin. In those cases (DLS and KiDS) we directly use the survey-provided $n(z)$.}. This approach of stacking the redshift PDFs for cosmological inference is not mathematically correct, but is consistent with the implementation of the four surveys under study. The $n(z)$, together with other metadata from each survey (the effective number densities for each tomographic bin, the total shape noise, the survey area) are fed into the calculation of the analytic covariance corresponding to the data vector using the code \textsc{Cosmolike} \citep{Krause2016}. A total of $N_{c}(2N_{c}+1)$ \textsc{Cosmolike} jobs are launched to calculate each submatrix of the full covariance matrix in parallel. The results for all submatrices are then combined to form the full covariance.

Finally, the outputs from the two branches -- the data vector and the covariance matrix -- are fed into \textsc{CosmoSIS} \citep{Zuntz2014} for inference of the cosmological model. The last step also involves choosing the appropriate theory models, priors and scale cuts within \textsc{CosmoSIS}.

This pipeline is written in a modular and generic fashion that strings together the three main codes that are used: \textsc{TreeCorr}, \textsc{Cosmolike} and \textsc{CosmoSIS}, so that it is easy to substitute different input catalogs, covariances and theory models. Building on this pipeline, it is easy to incorporate other cosmological probes, though that is beyond the scope of this paper. We note also that \textsc{WLPipe} serves as a test ground for experimenting on different pipeline architecture for future DESC cosmology analyses. For example, we have tested \textsc{WLPipe} using other workflow engines such as \textsc{Parsl}\footnote{\url{https://github.com/Parsl/parsl}} \citep{Babuji2018}. A similar pipeline was previously constructed for the recent DES Year 1 weak lensing and large-scale structure analyses \citep{DES2017}, the cosmic shear part \citep{Troxel2017} of which was made available to this project. The DES pipeline, however, did not employ any formal workflow management engine. The two pipelines have since been validated against one another to ensure they produce consistent results.

All plots of the cosmological constraints from \textsc{CosmoSIS} chains are plotted using the software package \textsc{ChainConsumer}\footnote{\url{https://samreay.github.io/ChainConsumer/}} with setting \texttt{kde=1.5}.


\section{Comparison Framework}
\label{sec:comparison_framework}

The focus of this paper is to compare the cosmic shear analyses of the four precursor surveys in multiple aspects, both within the same dataset and across the four datasets. We describe below the different elements that we consider in this work. We note that as our goal was to investigate and compare the various existing (published) datasets, there was no attempt of blinding throughout the analysis. 

\subsection{Two-Point Correlation Functions}
\label{sec:twopoint}
\newcommand\scott[1]{\ \textcolor{red}{SD: #1}}
An important intermediate output of our pipelines is a set of two-point correlation functions for different redshift bins: $\xi^{ij}_\pm(\theta)$ [see \Eref{eq:xipm}]. These together form the data vector for the cosmological parameter fitting. Except for DLS, whose shape catalogs are not in the public domain yet, we can compare the two-point functions output from \textsc{WLPipe} with those obtained by the different survey collaborations. For this work, we use the code \textsc{TreeCorr} to measure the two-point shear correlation function. \textsc{TreeCorr} is a fast tree-based method that allows one to estimate a variety of two- and three-point correlation functions. To measure the two-point shear correlation function, we calculate 
\begin{align}
&\xi^{ij}_{\pm}(\bar{\theta}_{\alpha}) = \notag \\ 
&\frac{\sum_{ab}W_{a}W_{b}\left[e^{t}_{a}(\vec{\theta}_{a}) e^{t}_{b}(\vec{\theta}_{b}) \pm e^{\times}_{a}(\vec{\theta}_{a}) e^{\times}_{b}(\vec{\theta}_{b}) \right]\,\Theta_{\alpha}(|\vec{\theta}_{a}-\vec{\theta}_{b}|)}{\sum_{ab}W_{a}W_{b}S_{a}S_{b}\,\Theta_{\alpha}(|\vec{\theta}_{a}-\vec{\theta}_{b}|)},
\end{align}
where $e^{t}_{a}$ is the tangential component of the ellipticity of galaxy $a$ with respect to the vector $(\vec{\theta}_{a}-\vec{\theta}_{b})$, and $e^{\times}_{a}$ is the cross component; $\bar{\theta}_{\alpha}$ is the mean angular separation between all galaxy pairs in bin $\alpha$; $W$ is the weight associated with each galaxy; and $S$ is an (algorithm-dependent) calibration factor defined by each of the different shear catalogs. The last factor, $\Theta_{\alpha}(|\vec{\theta}_{a}-\vec{\theta}_{b}|)$, is 1 when $|\vec{\theta}_{a}-\vec{\theta}_{b}|$ is inside angular bin $\alpha$ and 0 elsewhere. When using \textsc{TreeCorr}, we set the parameter {\tt binslop}=0, which means there are no approximations in calculating the angular separation between two galaxies. We note also that DES-SV uses \textsc{TreeCorr} to calculate its two-point correlation functions, while  for DLS, CFHTLenS and KiDS-450, the measurements are obtained via \textsc{athena}\footnote{\url{http://www.cosmostat.org/software/athena/}}.

One final subtle point to note is that $\bar{\theta}$ is the weighted mean of the logarithmic angular separation between all pairs of galaxies in a given angular bin\footnote{The shear response should also be included in the weights, but since the shear response is approximately homogeneous across the survey, we do not incorporate it into this calculation.}, or 
\begin{equation}
\bar{\theta}_{\alpha} = \exp \left[ \frac{\sum W_{a} W_{b} \ln |\vec{\theta}_{a}-\vec{\theta}_{b}| \, \Theta_{\alpha}(|\vec{\theta}_{a}-\vec{\theta}_{b}|)}{\sum W_{a} W_{b}\, \Theta_{\alpha}(|\vec{\theta}_{a}-\vec{\theta}_{b}|)} \right].
\label{eq:mean_theta}
\end{equation}
The choice of $\bar{\theta}$ is important because it will be the positions at which the model is evaluated and compared to the $\xi^{ij}_{\pm}(\theta)$ measurements during the parameter inference process. For DLS, CFHTLenS and KiDS-450, this was not taken into account and the geometric mean of the logarithmic angular bins were used\footnote{For CFHTLenS, $\bar{\theta}_{\alpha}$ as described in \Eref{eq:mean_theta} was used in the early study of \citet{Heymans2013}. In the later analysis of \citet{Joudaki2017}, considered in this work, \textsc{athena} had changed this to the geometric mean of the logarithmic angular bins.}. This will result in a small shift in the parameter inference as we discuss later in \Sref{sec:default} and \Aref{sec:theta_bins}. This effect has also been pointed out previously in \citet{Joudaki2018} and \citet{Troxel2018}.

\subsection{Covariance Matrices}
\label{sec:cov}

The covariance matrix is an essential element in the pipeline. The full covariance matrix receives contributions from two terms \citep{Cooray2001,Sato2009,Takada2013}: the Gaussian covariance and the non-Gaussian covariance. The non-Gaussian covariance includes the super-sample covariance \citep{Takada2013}, which describes the uncertainty induced by large-scale density modes outside the survey window.
In this work we use two sets of covariance matrices for each analysis: First, we use the covariance matrices used in the four papers \citep{Jee2016,Abbott2015,Joudaki2017,Hildebrandt2017}, which were provided by the collaborations. Next, we use a theoretical Gaussian covariance matrix produced by the \textsc{Cosmolike} \citep{Krause2016} code. We note that the Gaussian covariance may not be sufficient, especially for DLS, given the smaller area and lower shape noise in this dataset. For further details of the covariance calculation, see \citet{Krause2017}. The \textsc{CosmoLike} covariance calculation requires the following information from each survey:
\begin{itemize}
\item $n(z)$: estimate of redshift distribution for each tomographic bin (see \Fref{fig:nofz})
\item $n_{\rm eff}$: the effective number of source galaxies used in each bin as defined in \citet{Heymans2012} (see \Tref{tab:survey_param})
\item $\sigma_{e}$: standard deviation of the galaxy shape (or, shape noise) for the whole catalog (see \Tref{tab:survey_param})
\item $A_{\rm sky}$: area of footprint (see \Tref{tab:survey_param}).
\end{itemize}
Recently, \citet{Krause2017} and \citet{Troxel2018} also pointed out the importance of accounting for the geometry of the footprint, not just its area $A_{\rm sky}$, by using the survey window function when calculating the analytic covariance. Briefly, one can estimate the effect of the survey geometry by actually counting the number of source galaxy pairs as a function of separation or via an analytic integration of the survey mask. One then uses this information to calculate the shape noise contribution to the covariance instead of the simple geometric calculation based only on the area and mean source number density. We have incorporated this correction to our analytic covariances. Note that this correction does not include the survey geometry correction to the cosmic variance piece of the covariance, which may be important for surveys with low shape noise, such as DLS. The cosmological parameters used to generate all \textsc{CosmoLike} Gaussian covariances in this work are: $\Omega_{\rm m}=0.286$, $\Omega_{\rm b}=0.05$, $\sigma_{8}=0.82$, $h=0.7$, $n_{s}=0.96$.

We first use the survey-provided covariance to check whether we can reproduce the results from the papers. Next we compare the cosmological constraints derived using the survey-provided and the theoretical Gaussian covariance. The four surveys have different approaches to estimate the covariance: for DLS, DES-SV and CFHTLenS, the covariance was estimated via simulations (which are also different between the three cases). For KiDS-450, both simulation and analytic covariances were used and shown to be broadly consistent \citep[though can cause a 1$\sigma$ shift in the $S_{8}$ constraints, see][]{Hildebrandt2017}. The final results were based on the analytic covariance.

For covariances estimated via simulations, we need to apply the Hartlap correction factor $H$ \citep{Hartlap2007} when inverting the covariance to approximately correct for the bias in the inverse covariance estimate coming from the noise in the simulated covariance $\boldsymbol{C}_{\rm sim}$. That is,
\begin{equation}
\boldsymbol{C}^{-1} = H \boldsymbol{C}^{-1}_{\rm sim},
\label{eq:hartlap}
\end{equation}
where
\begin{equation}
H \equiv \frac{N-p-2}{N-1}.
\end{equation}
$N$ is the number of independent simulations and $p$ is the length of the data vector. We note that this gives an unbiased but still noisy estimate of the inverse covariance. In all our analyses, $N$ is large enough so that the noise associated with the resulting inverse covariance is reasonable \citep{Sellentin2016}.

\begin{table*}
\centering
\begin{threeparttable}
\caption{Free parameters in the cosmology inference used in \Sref{sec:default}, i.e. matching certain cases of the published results as closely as possible. The brackets indicate flat priors with [min, max] and the parentheses indicate Gaussian priors with (mean, standard deviation). We note that for CFHTLenS we choose to use the ``fiducial'' setting in \citet{Joudaki2017} as the \textit{Baseline}, which does not consider any systematic effects. In later analyses when we unify the analysis choices across surveys, the shear calibration bias, photo-z bias and IA amplitude will be allowed to vary.}
\label{tab:params} 
\begin{tabular}{lllll}
\hline\hline
&DLS & CFHTLenS & DES-SV & KiDS-450 \\
\hline
\multirow{5}{*}{Cosmology} &  $\Omega_{\rm m}$: [0.01, 1.0] &$\Omega_{\rm c}h^2$: [0.001, 0.99]  &$\Omega_{\rm m}$: [0.05, 0.9]  &$\Omega_{\rm c}h^2$: [0.01, 0.99]  \\
&$\Omega_{\rm b}$: [0.03, 0.06] &$\Omega_{\rm b}h^2$: [0.013, 0.033]  &$\Omega_{\rm b}$: [0.02, 0.07]  &$\Omega_{\rm b}h^2$: [0.019, 0.026]  \\
&$\sigma_{8}$: [0.1, 1.2] &$\ln (10^{10} A_{s})$: [2.3, 5.0]  &$\sigma_{8}$: [0.2, 1.6]  &$\ln (10^{10} A_{s})$: [1.7, 5.0]  \\
&$h$: [0.6, 0.8] &$h$: [0.61, 0.81]$^{*}$  &$h$: [0.3, 1]  &$h$: [0.64, 0.82]$^{*}$ \\
&$n_s$: [0.92, 1.02] &$n_s$: [0.7, 1.3]  &$n_s$: [0.7, 1.3]  &$n_s$: [0.7, 1.3]  \\
\hline
Intrinsic Alignment &  $A_{IA}$: 0.0 & $A_{IA}$: 0.0  & $A_{IA}$: [-5,5]  & $A_{IA}$: [-6,6]  \\
\hline
\multirow{2}{*}{Photo-z bias} & $b_{z}$: [-0.03, 0.03]$^{\dagger}$ & 0 & $b_{z,1}$: (0, 0.05) & $b_{z,1}$: (0, 0.036) \\
                                               & && $b_{z,2}$: (0, 0.05) & $b_{z,2}$: (0, 0.015)  \\
                                               &  && $b_{z,3}$: (0, 0.05) & $b_{z,3}$: (0, 0.01)  \\
                                               &  &&& $b_{z,4}$: (0, 0.006) \\
\hline
\multirow{2}{*}{Shear calibration bias} & $m$: [-0.03, 0.03]$^{\dagger}$ & 0 &  $m_{1}$: (0, 0.05)& $m_{1}$: (0, 0.01)\\
                                                             &  & & $m_{2}$: (0, 0.05) & $m_{2}$: (0, 0.01) \\
                                                             &  & & $m_{3}$: (0, 0.05) & $m_{3}$: (0, 0.01) \\
                                                             &  & & & $m_{4}$: (0, 0.01) \\
\hline\hline
\end{tabular}
\begin{tablenotes}
\item[*] Priors were placed in an intermediate stage.
\item[$\dagger$] Bins are assumed to be 100\% correlated.
\end{tablenotes}
\end{threeparttable}
\end{table*}

\subsection{Cosmological/Nuisance Parameters}
\label{sub:parameters}

In all four survey analyses, of order ten cosmological parameters and parameters modeling systematic effects are varied. The parameters each survey chooses to vary are slightly different and the corresponding priors are also different. In \Tref{tab:params} we summarize the free cosmological/nuisance parameters and priors for the four analyses under the $\Lambda$CDM framework. We note that for CFHTLenS we have chosen to study the ``fiducial'' setting in \citet{Joudaki2017}, which does not consider any systematic effects. In later analyses when we unify the analysis choices across surveys, the shear calibration bias, photo-z bias and IA amplitude will be allowed to vary. We also note that in \Tref{tab:params} there are two classes of parametrization of the free cosmological parameters. For DLS and DES-SV, [$\Omega_{\rm m}$, $\Omega_{\rm b}$, $h$, $\sigma_{8}$, $n_{s}$] was used, whereas for CFHTLenS and KiDS-450, [$\Omega_{\rm c}h^2$, $\Omega_{\rm b}h^2$, $h$, $\ln (10^{10} A_{s})$, $n_{s}$] was used. Here, $\Omega_{\rm b}$ is the baryon density today, $h$ is the unitless Hubble constant ($H_{0} = 100 h$ km/s/Mpc), $\sigma_{8}$ is the amplitude of the (linear) power spectrum on the scale of 8 $h^{-1}$Mpc, $\Omega_{\rm c}$ is the cold dark matter density today, $A_{s}$ is the amplitude of the matter power spectrum, and $n_{s}$ is the spectral index. Since the priors on the varied parameters are taken to be flat, choosing $\Omega_b h^2$ for example instead of $\Omega_{\rm b}$ translates to choosing a differently shaped prior on the $\Omega_{\rm b}-h$ parameter space. Furthermore, for CFHTLenS and KiDS-450, the $h$ prior is an indirect one that depends on $\theta_{\rm MC}$, defined as 100 times the ratio of the sound horizon to the angular diameter distance, and is imposed at an intermediate stage. The effective prior on $h$ is therefore not flat. As discussed in \citet{Hildebrandt2017} and \citet{Troxel2018}, however, this only has a small effect on the tails of the parameter constraints. 

The three classes of nuisance parameters considered here  are defined as follows.
\begin{itemize}
\item \textbf{Intrinsic Alignment:} Most current lensing surveys use the nonlinear alignment model (NLA) proposed by \citet{Hirata2004,Bridle2007,Joachimi2011}. The model assumes that the IA power spectra $P_{\rm II}$ and $P_{\rm GI}$ scale with the nonlinear power spectrum $P_{\delta}$ and can be redshift and luminosity-dependent:
\begin{equation}
P_{\rm II}(k,z) = F^{2} P_{\delta}(k,z); \; P_{\rm GI}(k,z) = F P_{\delta}(k,z),
\end{equation}
where
\begin{eqnarray}
F(z, \bar{L}) &=& -A_{\rm IA}C_{1}\rho_{\rm crit} \frac{\Omega_{\rm m}}{D_{+}(z)}\nonumber\\
&&\times\left( \frac{1+z}{1+z_{0}}\right)^{\eta} \left( \frac{\bar{L}}{L_{0}} \right)^{\beta}.
\end{eqnarray}
Here $A_{\rm IA}$ is a free parameter that dictates the amplitude of the effect, $C_{1}=5\times10^{-14}h^{-2}M_{\odot}^{-1}$Mpc$^{3}$ is a constant, $\rho_{\rm crit}$ is the critical density at redshift zero, and $D_{+}(z)$ is the linear growth factor that is normalized to 1 today. The power laws $\eta$ and $\beta$ determine the redshift and luminosity evolution of the IA effect with $z_{0}$ and $L_{0}$ chosen as the anchoring redshift and luminosity. $\bar{L}$ is the mean luminosity of the sample. In the four surveys considered in this work, DLS varied $A_{\rm IA}$, $\eta$ and $\beta$, the ``MID'' case of CFHTLenS varied $A_{\rm IA}$, $\eta$, while the ``MIN'' case of CFHTLenS, DES-SV and KiDS-450 only varied $A_{\rm IA}$\footnote{The ``MAX'' case of CFHTLenS, DES-SV and KiDS-450 did explore further IA models in \citet{Abbott2015} and \citet{Joudaki2017b}, even though they were not taken as the fiducial case.}.\\

\item \textbf{Photo-z Uncertainty:} The $n(z)$ estimation can be uncertain and one should marginalize over this uncertainty. We parametrize this uncertainty following the approach used in DLS, CFHTLenS and DES-SV \citep[also see][]{Huterer2006}. That is, we assume the true redshift distribution $n(z)$ has the same shape as the measured redshift distribution $n_{\rm obs}(z)$, but has an uncertain shift in the mean of the distribution, $b_{z,i}$, for each redshift bin $i$ so that
\begin{equation}
n_{i}(z) = n_{{\rm obs}, i}(z-b_{z,i}).
\end{equation}
The approach used in KiDS-450 is slightly different, where the variation in the $n(z)$ itself and the correlation between the errors is accounted for directly. This is done by running a large number \citep[750 is used in][]{Hildebrandt2017} of chains for each cosmological inference, where each chain uses a different bootstrap sample of the $n(z)$, and combining all the chains at the very end. As the current \textsc{WLPipe} is not able to implement this operation, we calculate the standard deviation of the mean redshift for each of the 1000 bootstrap $n(z)$'s provided by the collaboration to be [0.036, 0.015, 0.010, 0.006] for each of the redshift bins, and use these values as the priors on the photo-z uncertainty the same way as the other surveys. We find that this approximation gives consistent results to the KiDS-450 approach. The one other subtle point is that in the DLS analysis, the photo-z biases are assumed to be 100\% correlated across redshift bins. \\

\item \textbf{Shear Calibration Uncertainty:} The shear measurements in each catalog can be uncertain due to imperfect calibration \citep{Mandelbaum2015}. A common way of parametrizing this uncertainty is assuming the true shear $\gamma$ scales linearly with the measured shear $\gamma_{\rm obs}$ by a factor $(1+m_{i})$ for each redshift bin $i$, plus an additive term $c_{i}$ \citep{Heymans2006}. That is
\begin{equation}
\gamma_{\rm obs} = \gamma (1+m_{i}) + c_{i}.
\end{equation}

As we will discuss in \Sref{sec:fiducial_kids}, the uncertainty in $m_{i}$ can either be incorporated at the parameter level or directly in the covariance matrix. We choose the former approach but show that the resulting cosmological constraints are identical (see \Fref{fig:kids_cov} in \Aref{sec:kids_cov}). The one other subtle point is that in the DLS analysis, the shear calibration uncertainties are assumed to be 100\% correlated across redshift bins. Finally, all surveys we analyzed assume that any residual additive shear biases, $c_{i}$, are negligible for the scales used.
\end{itemize}
In \Sref{sec:effect_prior}, we compare the cosmological constraints from the four datasets using the same priors on cosmological parameters and IA parameters. To see the effect of varying different combinations of the cosmological parameters discussed above, we run analyses for both the DES-SV priors and the KiDS-450 priors. For the photo-z and shear calibration parameters, however, we do not attempt to match between the surveys, as these are parameters that are characterized using the specific datasets. It would be  incorrect to assume they have identical priors.

One final subtlety on the modeling side concerns the nonlinear matter power spectrum. Amongst the surveys considered here, DLS uses the \citet{Smith2003} \textsc{Halofit} power spectrum, CFHTLenS and KiDS-450 use the \textsc{HMCODE} power spectrum, which is based on \citet{Mead2015}, and DES-SV uses the \citet{Takahashi2012} \textsc{Halofit} power spectrum. The difference in these power spectrum models can result in slightly shifted cosmological constraints, as discussed in \citet{Jee2016,Joudaki2017,Joudaki2017b, MacCrann2015}. In this work we use the \citet{Takahashi2012} \textsc{Halofit} power spectrum.

\subsection{Scale Cuts}
\label{sub:scale}

In the four cosmic shear analyses, choices were made for which scales will be used for the cosmological inference. The choices were often based on considerations of systematic effects and model uncertainties. In general, the minimum scale is determined by model uncertainties such as baryonic physics and the accuracy of the nonlinear power spectrum. The maximum scale cuts are usually related to survey-specific considerations such as the size of the footprint, additive shear bias, and super-sample covariance. For the four surveys considered, different choices of scale cuts were used and listed in \Tref{tab:survey_param}. A few things to point out: For DLS, the same scale cuts were chosen for $\xi_{+}$ and $\xi_{-}$, though a discussion of how the scale cuts would change the cosmological constraints was presented in \citet{Jee2016}. For CFHTLenS and KiDS-450, in addition to the motivations described above, scales with low signal-to-noise were also removed. Also, the use of smaller scales was justified since the effect of baryonic effects were modeled and marginalized over. For DES-SV, the scale cuts are redshift-dependent and the most conservative.

In our final joint analysis we aim for a uniform scale cut across all four datasets to remove the difference in the four analyses coming from this decision. Since the different surveys have different redshift binning strategies, a unified set of scale cuts is not straightforward. We take the approach of choosing a set of scale cuts in physical units and propagating it into the corresponding angular scale cuts for all of the shear correlation functions. This choice is motivated by the fact that the main consideration that goes into the scale cuts is the uncertainties in the model on small scales (nonlinear power spectrum, baryonic effects etc.). The scales on which these effects are important are usually related to the physical size of, for example, dark matter halos. In addition, for cosmic shear measurements, one is not measuring the matter distribution at the redshift of the source galaxies. Instead, it is the matter distribution in the foreground of the source galaxies that we are probing -- in specific, matter at the redshift where the lensing efficiency is high [\Eref{eq:lensing_efficiency}]. As a result, we choose the scale cuts by calculating the corresponding angular scale cut $\theta_{{\rm min}, \pm}$ for some given physical scale $R_{{\rm min},\pm}$ at the redshift of the peak of the lensing kernel $z_{p}$. That is, for $\xi_{\pm}$, we use only angular scales
\begin{equation}
\theta > \theta_{\rm min, \pm} = \frac{R_{{\rm min}, \pm}}{D_{A}(z_{p})},
\label{eq:scale_cut}
\end{equation}
where $D_{A}(z_{p})$ is the angular diameter distance to redshift $z_{p}$.

The physical scale cuts $R_{{\rm min}, \pm}$ chosen in our common analysis are $R_{{\rm min},+}=$1.3 Mpc for $\xi_{+}$ and $R_{{\rm min},-}=$11.4 Mpc for $\xi_{-}$. These choices are equal to the most conservative scale cuts amongst the four datasets and very similar to the DES-SV scale cuts. We note that we use $R_{{\rm min}, \pm}$ to translate the angular scale cuts between different redshift ranges. The reason for a larger $R_{{\rm min}, -}$ is mainly reflecting the difference between the $J_{0}$ and $J_{4}$ Bessel functions in \Eref{eq:xipm}. We also note that for a more rigorous approach of using truly ``linear scale'' cuts, see Sec. 3.5 of \citet{Joudaki2017}.

\subsection{Cosmological Constraints and Comparisons Metrics}
\label{sec:metric}

To obtain cosmological constraints, we vary the full set of cosmological and nuisance parameters $\vec p$ using a Monte Carlo approach where we assume a Gaussian likelihood, which is the prior multiplied by $e^{-\chi^2/2}$, where
\begin{equation}
\chi^2(\vec p) \equiv \sum_{i,j} \left[d_i-t_i(\vec p)\right] C^{-1}{}_{ij} \,\left[d_j-t_j(\vec p)\right].
\end{equation}
$d_i$ ranges over all data points; $t_i(\vec p)$ is the theoretical prediction given the set of parameters; and $C$ the covariance matrix. We use the \textsc{Multinest} Monte Carlo sampler \citep{Feroz2008} implemented in \textsc{CosmoSIS}, which has been shown in \citet{DES2017} and \citet{Krause2017} to agree very well with other sampling methods such as \textsc{emcee} as well as the \textsc{CosmoLike} inference code \citep{Krause2017b}.

This cosmic shear experiments studied in this paper effectively constrains one or at most two cosmological parameters, depending on choices to be discussed below. The parameter that is most tightly constrained is \citep{Jain1997}
\begin{equation}
S_8 \equiv \sigma_8 (\Omega_{\rm m}/0.3)^{\alpha},
\end{equation}
where $\alpha \sim 0.5$ denotes the degeneracy direction in the $\Omega_{\rm m}$-$\sigma_{8}$ plane so that $S_{8}$ gives the most constraining direction of the dataset. The particular value of $\alpha$ depends somewhat on the details of the data and modeling choices. In most existing cosmological analyses, a customary choice is to set $\alpha=0.5$. However, this could lead to slightly misleading results when comparing different datasets, as not all of them would yield the most constraining $S_{8}$ with this choice of $\alpha$. In the following analysis, we will use $\alpha=0.5$ as our fiducial value but discuss in \Sref{sec:S8} the effect of changing $\alpha$.

In the next section, we will focus our comparison discussions surrounding four quantities:
\begin{itemize}
\item {\bf Signal-to-noise (S/N):} This is simply 
\begin{equation}
S/N = \left[\sum_{i,j} d_i C^{-1}{}_{ij} \,d_j \right]^{0.5},
\end{equation}
and it quantifies the statistical significance of the observables.
\item {\bf Goodness of fit ($\chi^2/\nu$, p.t.e.)}: For the best-fit data vector $\hat{D}$, we can calculate the $\chi^2$ per effective number of degree of freedom $\nu$, and the corresponding probability-to-exceed (p.t.e.). It is important to evaluate the goodness of fit for each of the chains in parallel to check for consistency. One disadvantage for using the goodness-of-fit is that the determination of the degree-of-freedom in a high dimensional space is not straightforward. However, for this work the length of the data vector usually dominates over the number of model parameters.  
\item {\bf 1D distance in $S_{8}$ ($\Delta S_{8}$):} We calculate the ratio of the absolute difference between the mean parameter values in the two experiments and the uncertainty in the difference. For two experiments $a$ and $b$, we thus have
\begin{equation}
\Delta S_{8} \equiv \frac{|S_{8}^{a}-S_{8}^{b}|}{\sqrt{\sigma(S_{8}^{a})^2+\sigma(S_{8}^{b})^2}}.
\end{equation}
$\Delta S_{8}$ can roughly be interpreted as an $n$-$\sigma$ difference in $S_{8}$ for the two experiments. This metric inherently assumes Gaussianity in the $S_{8}$ posterior and ignores possible tensions in other parameter projections. It can also overestimate the inferred disagreement when there are strong degeneracies in other parameter dimensions.
\item {\bf Logarithmic Bayes Factor (BF):} Based on \citet{Marshall2006}, we consider the logarithmic ratio of the evidence for the two hypotheses: first that the two experiments are measuring the same cosmological parameters and second that they are measuring different cosmological parameters. That is, we calculate
\begin{equation}
{\rm BF} = \log_{10} \left( \frac{\int d\Omega  P_a(\vec p) P_b(\vec p)}{[\int d\vec p  P_a(\vec p) ][\int d \vec p  P_b(\vec p) ]}\right).
\label{eq:bf}
\end{equation}
Here the posteriors (including the priors) for each experiment $P_{a,b}$ are integrated over all parameters $\vec p$. To properly interpret the BF values, one should evaluate it for cases where the two datasets share the same priors. As a result, we only calculate this at the end of the paper when all analysis choices are unified. We use the criteria ${\rm BF}>-1$ to determine whether two surveys are consistent and can be combined. When ${\rm BF}<-1$, the Jeffrey scale \citep{Jeffreys1961} suggests that there is effectively no evidence that the two datasets can be described by the same model. 

We note, however, that the BF metric is sensitive to the priors on the constrained parameters, and is usually biased towards consistency \citep{Raveri2018}.  
\end{itemize}

\begin{figure*}
\includegraphics[width=1.8\columnwidth]{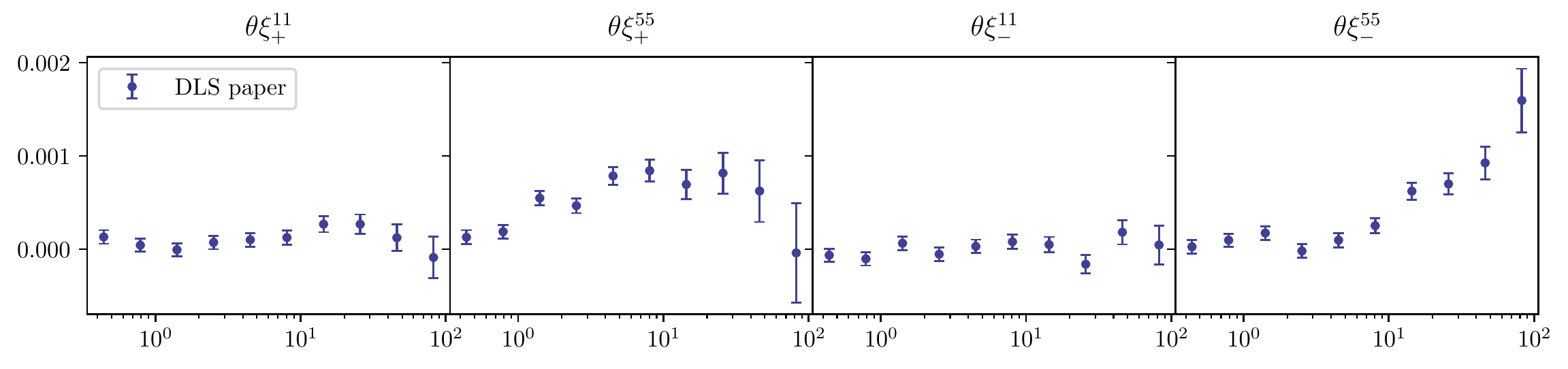}\\
\includegraphics[width=1.85\columnwidth]{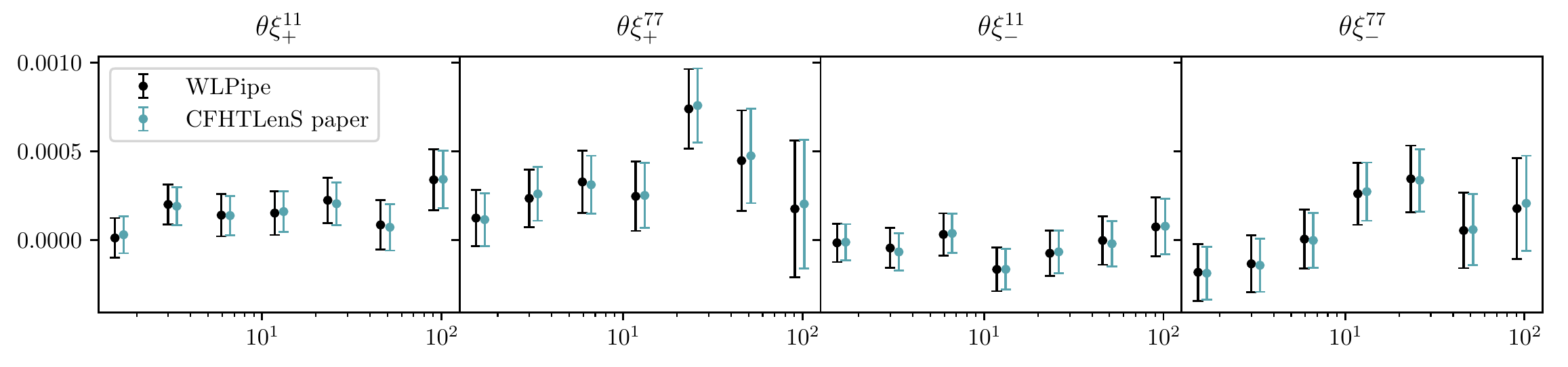}\\
\includegraphics[width=1.9\columnwidth]{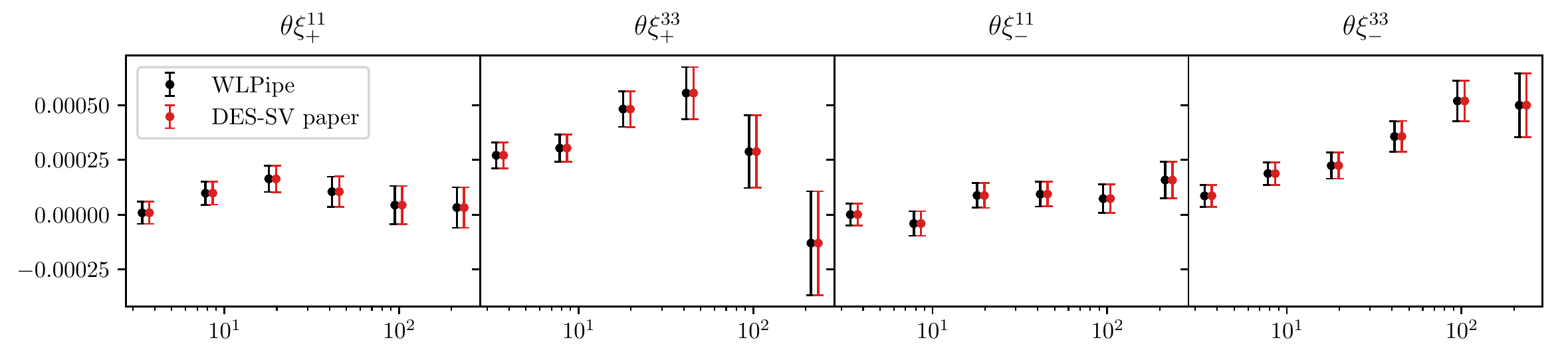}\\
\includegraphics[width=1.85\columnwidth]{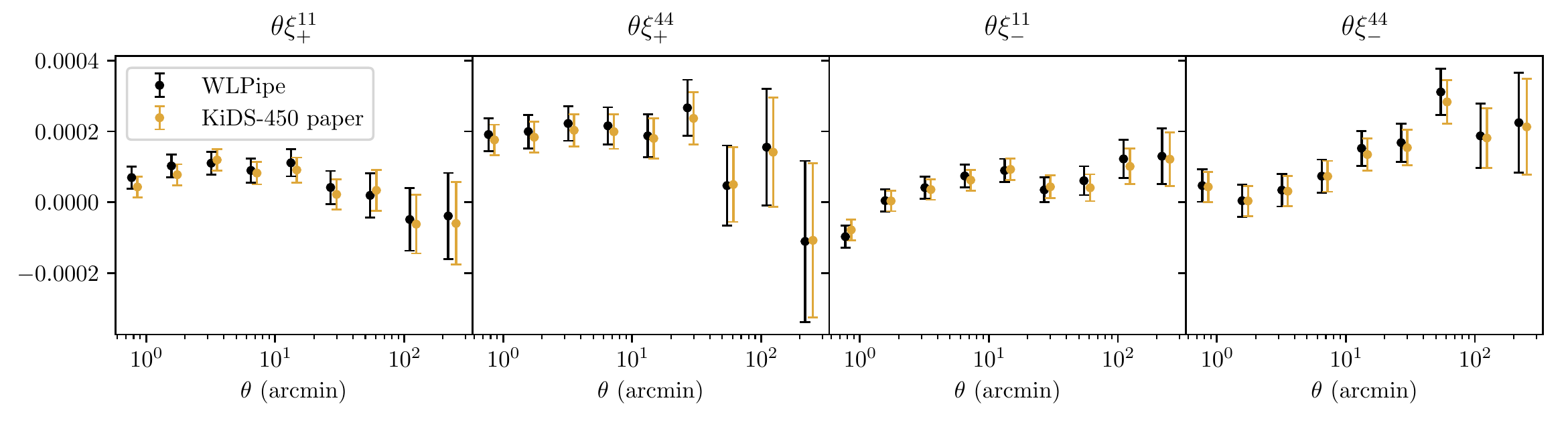}\\
\caption{The 2-point function $\theta \xi_{\pm}(\theta)$ as measured by \textsc{WLPipe} from the catalogs provided by the collaborations compared with the results obtained by the collaborations themselves. For visualization purpose we only show the auto-correlation functions for the lowest and the highest redshift bins, and the colored data points are slightly displaced from the black points. From left to right in each panel is $\xi_{+}$ for the lowest redshift bin, $\xi_{+}$ for the highest redshift bin, $\xi_{-}$ for the lowest redshift bin, and $\xi_{-}$ for the highest redshift bin. From top to bottom are the four surveys: DLS, CFHTLens, DES-SV, and KiDS-450. Since the catalogs from DLS are not public, only the collaboration 2-point functions are shown in the top panel. We also note that the difference in the angular binning discussed in \Sref{sec:default} is not shown in this plot, but explained more clearly in \Aref{sec:theta_bins}.}
\label{fig:xiplus}
\end{figure*}

\begin{figure*}
\includegraphics[width=0.7\columnwidth]{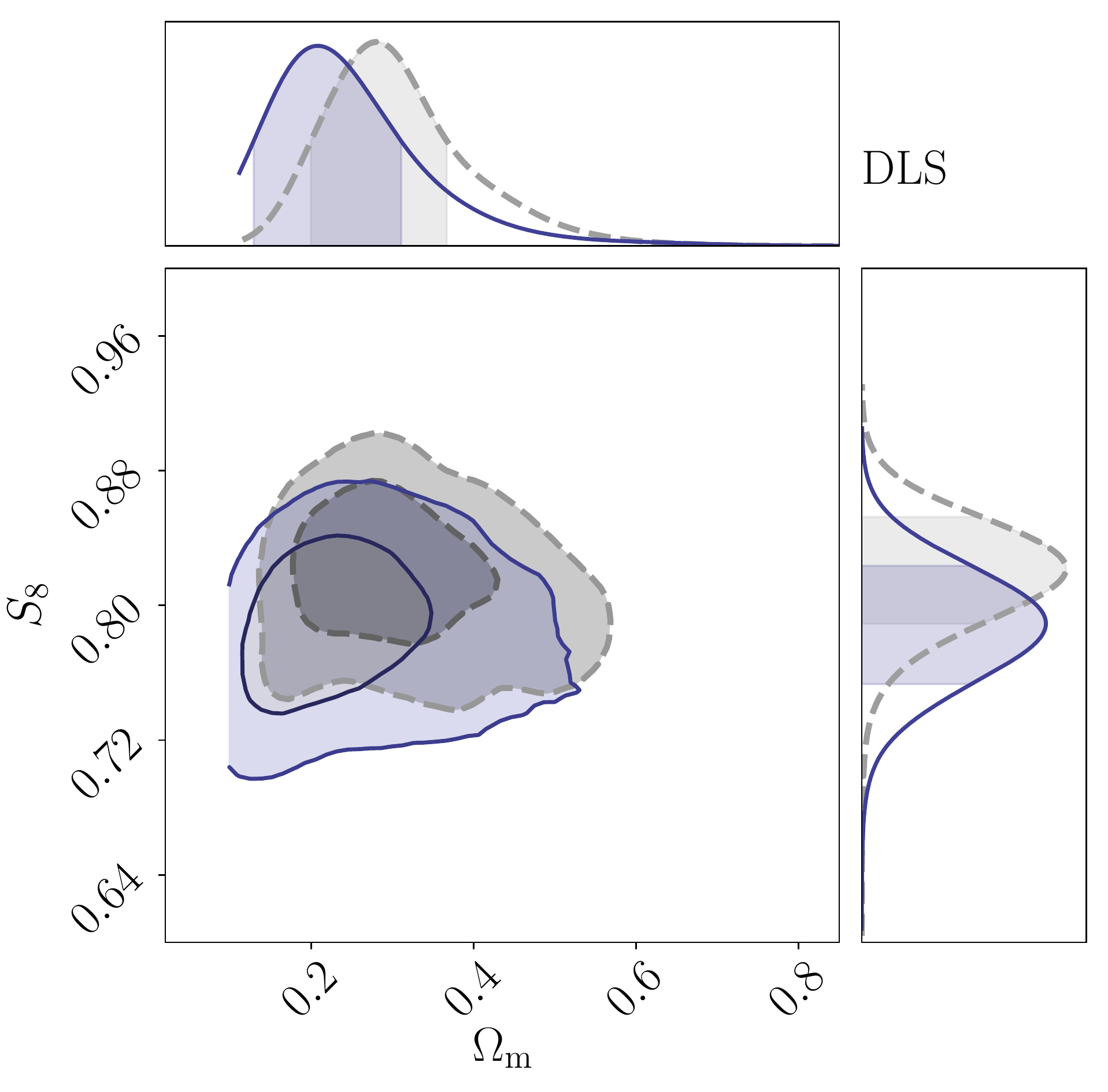}
\includegraphics[width=0.7\columnwidth]{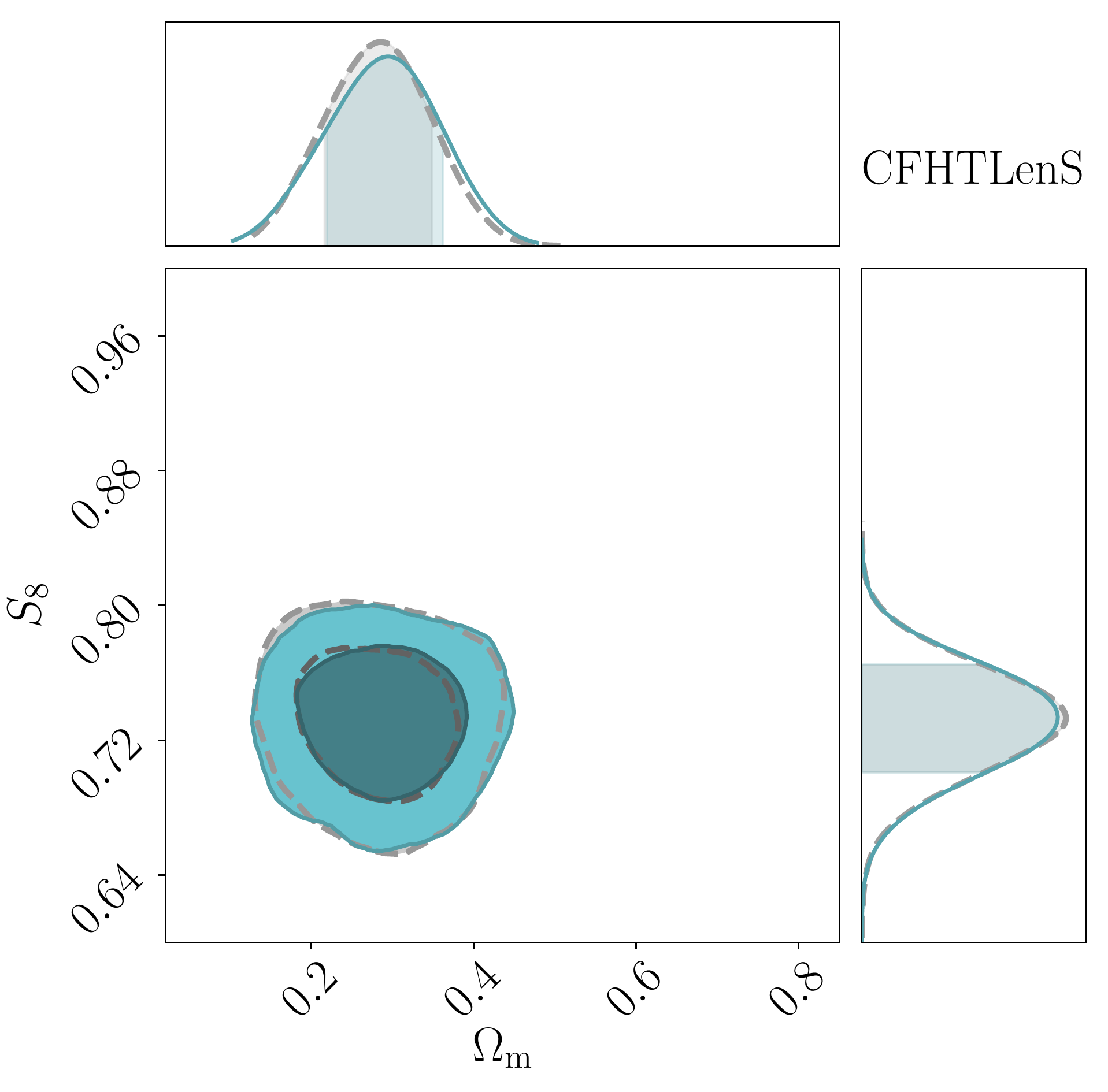}\\
\includegraphics[width=0.7\columnwidth]{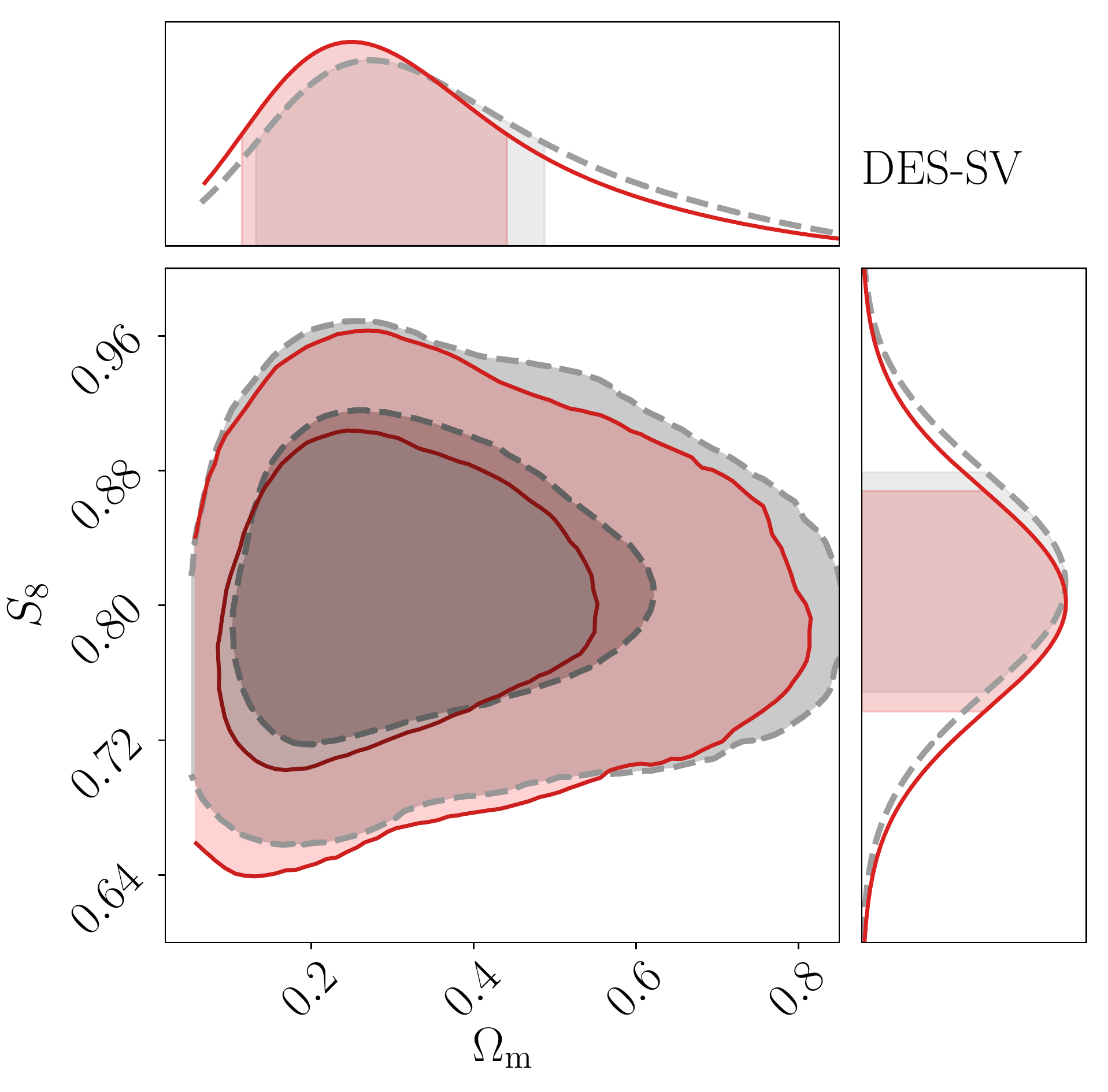}
\includegraphics[width=0.7\columnwidth]{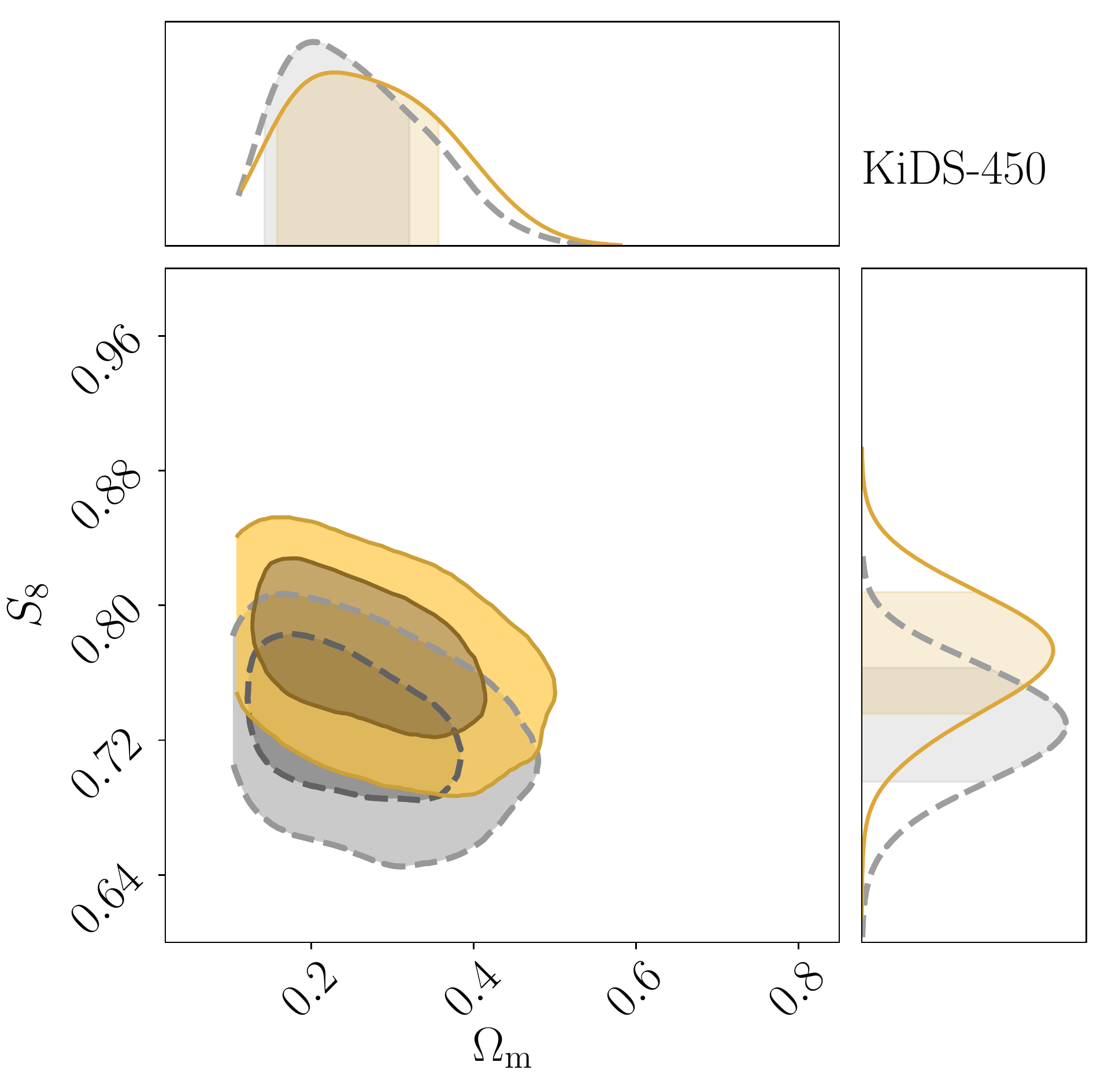}
\caption{Results from reproducing published results with \textsc{WLPipe}. Here we compare the marginalized constraints on $\Omega_{\rm m}$ and $S_8 \equiv \sigma_{8}(\Omega_{\rm m}/0.3)^{0.5}$ obtained from \textsc{WLPipe} (\textit{Baseline}, solid colored contours) to those obtained by the collaborations (\textit{Published Baseline}, dashed grey contours) for four different experiments. One can see clear shifts in the DLS and KiDS-450 contours. \Sref{sec:default} describes all the factors that drive these discrepancies, and any accidental cancellations of effects for the other surveys.}
\label{fig:constraints}
\end{figure*}


\section{Results}
\label{sec:results}

In this section we present the main results of this paper. In \Sref{sec:default} we present results from the \textit{Baseline} case: we set out to reproduce the results from the four published papers and discuss in detail the remaining differences between our reproduction and the published results, which we refer to as the \textit{Published Baseline}. We also calculate several comparison metrics in order to understand the internal (external) consistency within (between) the four datasets. In \Sref{sec:cov_effect}, \Sref{sec:effect_scale} and \Sref{sec:effect_prior}, we investigate individually the effect of changing the covariance estimation, the scale cuts, and the priors on cosmological parameters, and intrinsic alignment treatment. In \Sref{sec:match} we unify the analysis choices and reexamine the comparison metrics. In \Sref{sec:S8} we discuss how the definition of $S_{8}$ may affect the comparison between the surveys.

Throughout, we will also use the term \textit{Nominal Baseline} to refer to the nominal analysis results that each collaboration uses as their most representative cosmological constraints, which for the case of CFHTLenS and KiDS-450 can be slightly different from the \textit{Published Baseline} in terms of the treatment of systematic effects.

\subsection{\textit{Baseline:} Reproducing Literature Results}
\label{sec:default}

The most basic test is the comparison of the literature results with the \textsc{WLPipe}'s measurements using the same catalogs under the same assumptions.

First, we examine the intermediate output of the measured $\xi_{\pm}$ functions. \Fref{fig:xiplus} shows $\xi_+(\theta)$ and $\xi_-(\theta)$ produced by \textsc{WLPipe} using the same binning and angular scales chosen by the collaborations, overlaid on top of results obtained by the collaborations for comparison. We find excellent agreement in all cases for the values of $\xi_{\pm}$\footnote{There is a small discrepancy between the published results and the \textsc{WLPipe} measurements for KiDS-450 due to the fact that we have used the per-object per-patch multiplicative bias correction instead of a constant for each tomographic bin used in \cite{Hildebrandt2017}. We have checked that this does not affect the rest of the analysis.}. Note that for DLS, CFHTLenS and KiDS-450, the angular values for each data point assigned by \textsc{WLPipe} differ from the paper-provided data vectors. This, as we discussed in \Sref{sec:twopoint}, is due to the fact that those paper-provided data vectors used the center of each angular bin instead of the area-weighted center. We show how this propagates into a bias in the cosmological constraints in \Aref{sec:theta_bins}.  

\Fref{fig:constraints} shows the constraints obtained from \textsc{WLPipe} for each experiment compared with those obtained by the collaborations themselves using the same binning, parameters, priors, and covariance matrices used to obtain the published results. In doing this we aim to reproduce the published results. However, \Fref{fig:constraints} shows that there are differences between the published results and the \textsc{WLPipe} results, which we discuss in detail in the following subsections. The \textsc{CosmoSIS} configuration files and data files for these \textit{Baseline} results are publicly available \citep{chang2018_baseline_cosmosis_config}.

\subsubsection{DLS}
\label{sec:fiducial_dls}
From the upper left panel of \Fref{fig:constraints}, we see that the \textit{Published Baseline} constraints from DLS are about 0.7$\sigma$ higher in $S_{8}$ and 0.5$\sigma$ higher in $\Omega_{\rm m}$ than the \textit{Baseline} constraints obtained via \textsc{WLPipe}. Differences in angular binning cannot be an issue here, since we are using the collaboration-computed $\xi_\pm$. Two differences in the analysis explain the offset: First, the nonlinear power spectrum used in the original DLS analysis of \citet{Jee2016} comes from an older version of \textsc{Halofit} \citep{Smith2003}, while in \textsc{CosmoSIS} we use the nonlinear power spectrum of \citet{Takahashi2012}. As shown in \citet{MacCrann2015} and also discussed in Sec. 6.3 of \citet{Jee2016}, switching from the \citet{Smith2003} model to the \citet{Takahashi2012} model causes the inferred $\sigma_{8}$ to be lowered by $\sim 0.02$ at $\Omega_{\rm m}=0.3$. Second, as we have not implemented the particular IA model used in \citet{Jee2016} in \textsc{WLPipe}, we are assuming no IA in the \textsc{WLPipe} case. According to Figure 12 of \citet{Jee2016}, this results in a $\sim 0.02$ lower $\Omega_{\rm m}$ (with approximately the same $S_{8}$). Accounting for these two factors brings the two contours to better agreement -- where the \textsc{WLPipe} reproduction gives a slightly lower $\Omega_{\rm m}$, but almost exactly the same $S_{8}$ compared to the published results.

\subsubsection{CFHTLenS}
\label{sec:fiducial_cfhtlens}
From the upper right panel of \Fref{fig:constraints}, we see that the published constraints from CFHTLenS are consistent with \textsc{WLPipe} in both the $\Omega_{\rm m}$ and $S_{8}$ directions. We note that we have chosen to compare the ``fiducial'' chain in \citet{Joudaki2017}, which does not include IA, baryons, photo-z uncertainties or shear calibration uncertainties. Three factors need to be accounted for here: First, the angular values used in the paper-provided chains (the center of the bin) are different from that in the \textsc{WLPipe} chain [\Eref{eq:mean_theta}]. As we show in \Aref{sec:theta_bins}, using the area-weighted angular values would shift the contours up by about 0.4$\sigma$. Second, whereas \textsc{CosmoSIS} uses the \citet{Takahashi2012} model in \textsc{Halofit}, \citet{Joudaki2017} used the slightly more accurate \textsc{HMCODE} \citep{Mead2016} for the nonlinear power spectrum. As shown in Figure 10 of \citet{Joudaki2017}, the \textsc{HMCODE} version used at that time moves the contour higher in $S_8$ by about 0.4$\sigma$ compared to \textsc{Halofit}. These first two effects cancel, bringing the paper-provided chains and the \textsc{WLPipe} reproduction to perfect agreement. The final difference in our approaches is more subtle. As we noted in \Sref{sub:parameters}, CFHTLenS and KiDS-450 uses \textsc{CosmoMC}, which does not sample $h$ directly. Instead, it samples a wide flat prior in $\theta_{\rm MC}$ (which is connected to $h$) and imposes the $h$ priors after the fact. This means that the real $h$ prior in the paper-provided chains is not exactly flat. This difference has been found to be small  \citep{Hildebrandt2017, Troxel2018}.

In Sec 5.1-5.4, we compare with the ``fiducial'' case in \citet{Joudaki2017} for simplicity. This assumes no systematic uncertainties, which according to Fig. 12 of \citet{Joudaki2017} and \Fref{fig:bars}, is close to the MID case in \citet{Joudaki2017} ($S_{8}$ is lower by 0.1$\sigma$). We do incorporate systematic uncertainties for CFHTLenS in \Sref{sec:match} based on \citet{Kilbinger2017, Choi2016}.

\subsubsection{DES-SV}
\label{sec:fiducial_des}
We expect the \textsc{WLPipe} reproduction of the DES-SV \textit{Published Baseline} results to be perfect up to noise in the sampling, since the analysis pipeline is almost identical in the two analyses \citep[\textsc{WLPipe} uses slightly updated versions of \text{TreeCorr}, \textsc{Cosmolike} and \textsc{CosmoSIS} compared to that used in][]{Abbott2015}. As shown in the lower left panel of \Fref{fig:constraints}, this is indeed the case -- the two contours agree very well.

\subsubsection{KiDS-450}
\label{sec:fiducial_kids}
From the lower right panel of \Fref{fig:constraints}, we see that the \textit{Published Baseline} constraints from KiDS-450 agree with the \textit{Baseline} constraints from \textsc{WLPipe} in the $\Omega_{\rm m}$ direction and are about 0.9$\sigma$ higher in the $S_{8}$ direction. Several factors contribute to this discrepancy at different levels. First, the angular values used in the paper-provided chains (the center of the bin) are different from those in the \textsc{WLPipe} chain [\Eref{eq:mean_theta}]. Changing the bin values shifts the paper-provided chains up by about 0.4$\sigma$ as shown in \Fref{fig:ang}. Second, similar to CFHTLenS, the paper-provided chain uses \textsc{HMCODE} for the nonlinear power spectrum while \textsc{WLPipe} uses \textsc{Halofit}. However, while \citet{Joudaki2017} used the original version of \textsc{HMCODE} \citep{Mead2015b}, a newer version of \textsc{HMCODE} \citep{Mead2016} was used in \citet{Hildebrandt2017}. In this newer version, the fitting parameters were updated to allow for better fits when considering massive neutrino cosmologies, at the expense of slightly worse fits in standard $\Lambda$CDM. This newer version of \textsc{HMCODE} agrees more strongly with \textsc{Halofit}, and the resulting parameter constraints from KiDS-450 when using either prescription are almost identical (when excluding baryonic feedback). Third, similar to CFHTLenS, $\theta_{\rm MC}$ is varied in the analysis while $h$ is a derived parameter. Fourth, the covariance used in \citet{Hildebrandt2017} is designed to include the marginalization over multiplicative bias $m$, but was not implemented correctly \citep[see also \Fref{fig:kids_cov} and][]{Troxel2018}. This moves the $S_{8}$ constraints up by about $0.5\sigma$. Finally, as described in \Sref{sec:kids}, the photo-z uncertainties are incorporated differently in \textsc{WLPipe} compared to \citet{Hildebrandt2017}. We have checked, however, that this does not generate any noticeable effect in the cosmological constraints. 

\begin{figure*}
\includegraphics[width=0.95\columnwidth]{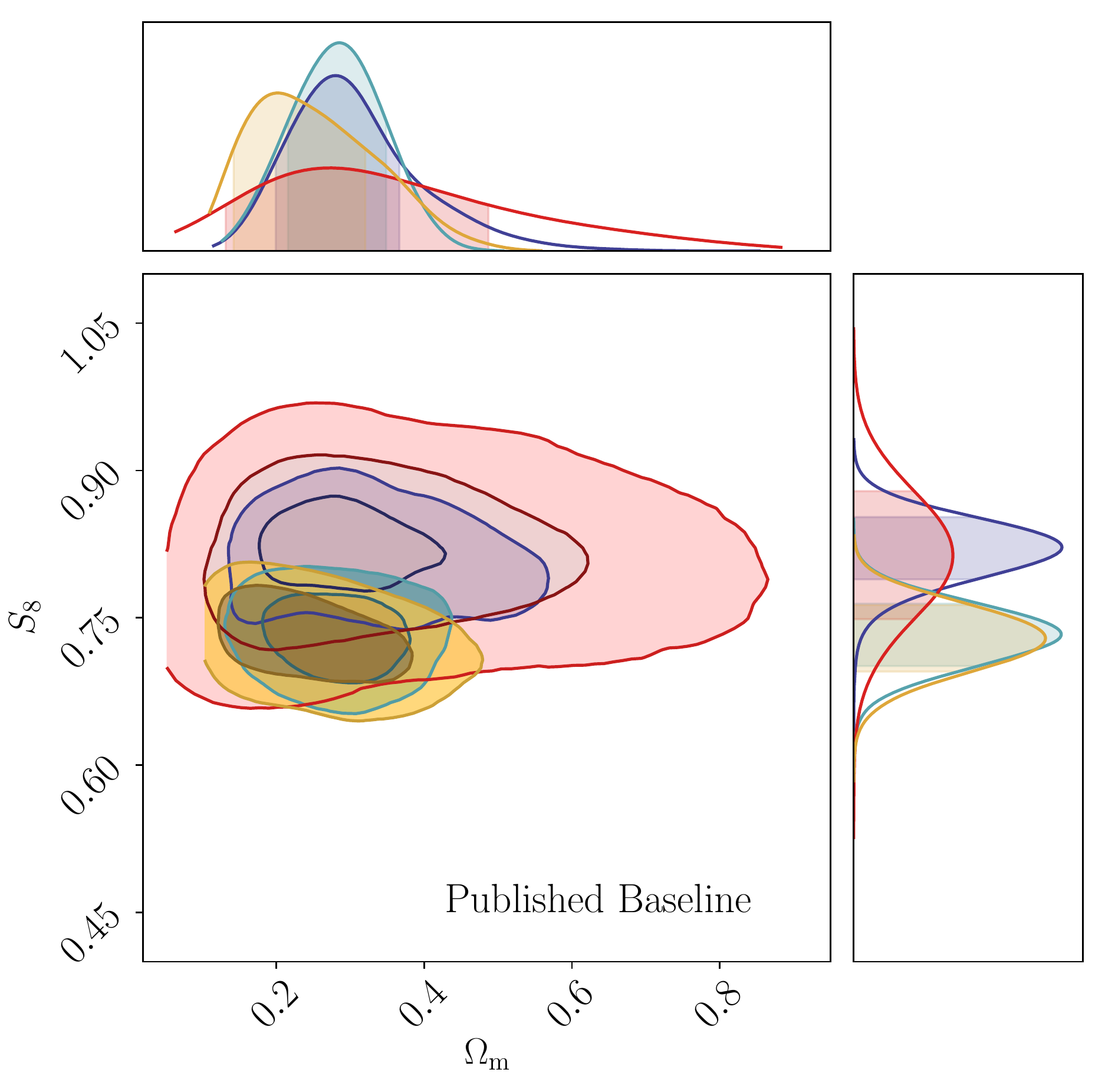}
\includegraphics[width=0.95\columnwidth]{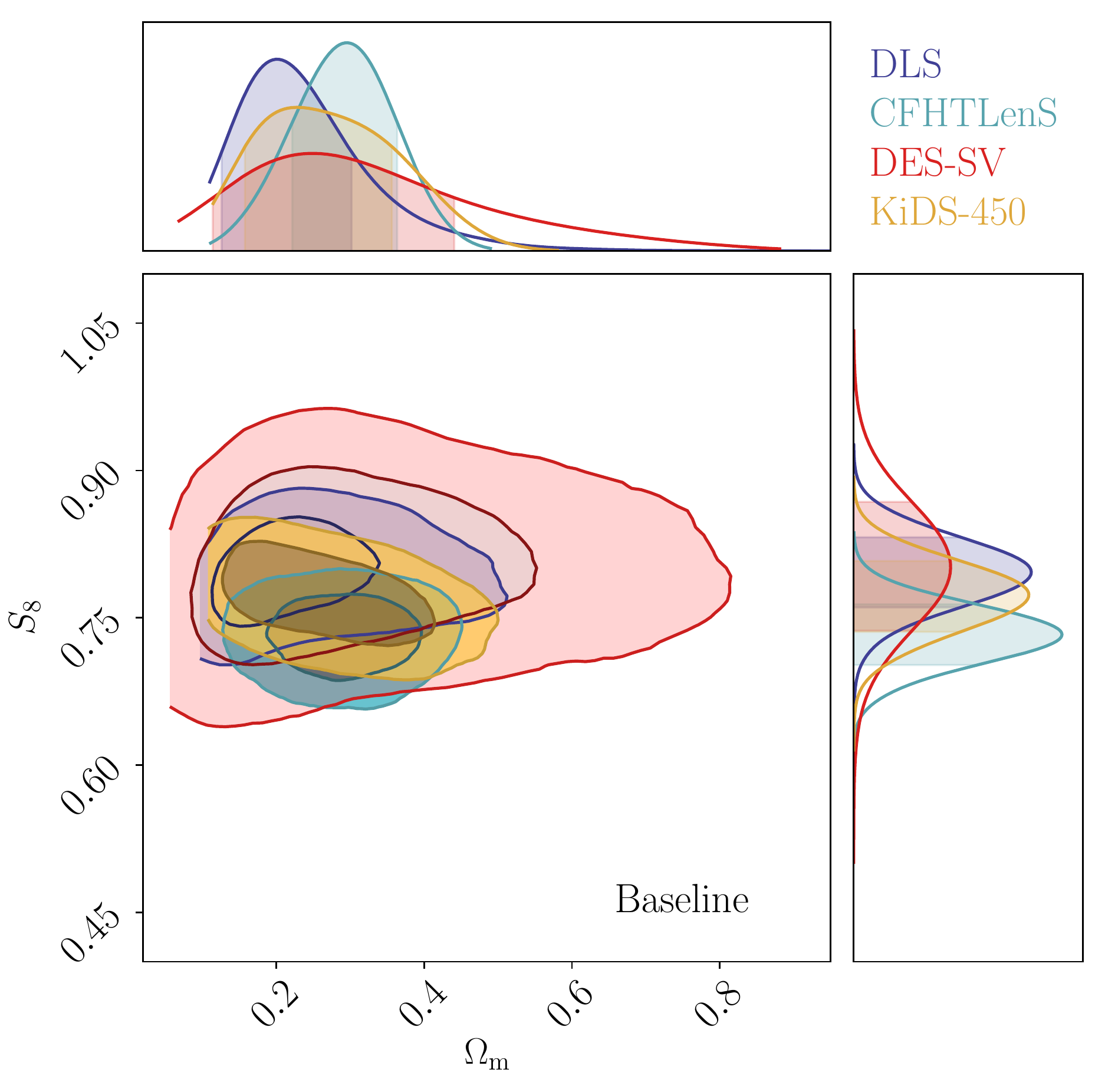}
\caption{Here we compare the constraints of the four surveys from the published results and the \textsc{WLPipe} reanalysis. We show the marginalized constraints on $\Omega_{\rm m}$ and $S_8 \equiv \sigma_{8}(\Omega_{\rm m}/0.3)^{0.5}$ from the paper-provided chains (the \textit{Published Baseline} case, left panel) and from \textsc{WLPipe} in the \textit{Baseline} case. Note that compared to the \textit{Published Nominal} results, here the KiDS-450 contours do not include baryonic effects, while the CFHTLenS contours do not include any systematic uncertainties. }
\label{fig:comp4}
\end{figure*}

In summary, we are able to reproduce the \citet{Hildebrandt2017} results in both $\Omega_{\rm m}$ and $S_{8}$ when considering these factors. We note here that the fiducial analysis of \citet{Hildebrandt2017} includes modeling of the baryonic effects on small scales whereas we do not here. As a result we compare with their DIR chain, which as shown in Fig. 8 of \citet{Hildebrandt2017} and \Fref{fig:bars}, gives a $S_{8}$ value $0.3\sigma$ lower than the \textit{Nominal Baseline} case. Later when we unify the analysis choices, since we make much more conservative scale cuts, we do not expect the effect of baryons to be important. 

\subsubsection{Comparison of all four surveys}
\label{sec:fiducial_all}

The right panel of \Fref{fig:comp4} shows the \textit{Baseline} results from the four experiments using \textsc{WLPipe} in one plot, i.e., we overlay the colored contours in \Fref{fig:constraints} together. We note that here we have used the different analysis choices based on each of the collaborations, therefore the four contours cannot be compared on an equal footing. In this picture, we find good agreement between the four surveys in the $\Omega_{\rm m}-S_{8}$ plane, with CFHTLenS slightly lower than the other three surveys. DES-SV has the largest contour (weakest constraining power), whereas the other three surveys have contours of similar sizes. The degeneracy directions of the four surveys are somewhat different, as expected from the different redshift ranges they probe. For comparison, we also show in the left panel of  \Fref{fig:comp4} the \textit{Published Baseline} results from the corresponding survey-provided chains, or the four grey dashed contours in \Fref{fig:constraints}. The main difference from \Fref{fig:constraints} is (1) the shifting of the KiDS-450 contours in the $S_{8}$ direction, which comes from the change in the angular bin values and the covariance, as we discussed in \Sref{sec:fiducial_kids} above, and (2) the DLS contours shifted to lower $S_{8}$ due to the change in the nonlinear power spectrum and the IA model, as we discussed in \Sref{sec:fiducial_dls} above. This can also be seen more clearly comparing the \textit{Published Baseline} and \textit{Baseline} cases in \Fref{fig:bars}. 

\begin{table}
\begin{center}
\caption{Comparison metrics for all pairs of surveys in the \textit{Baseline} analysis case: \textsc{WLPipe} chains that are designed to match the published analyses, or the \textit{Published Baseline} case. For the $S_{8}$ values, we list the mean and the 16\% and 84\% confidence intervals. We note that here we have used the different analysis choices based on each of the collaborations, so these metrics are not on equal footing. Later in \Tref{tab:metric_match} we show similar metrics that \textit{can} be compared directly.} 
\begin{tabular}{lcccc}
\hline\hline
& (1) DLS & (2) CFHTLenS & (3) DES-SV & KiDS-450 \\
\hline
$S_{8}$ &  $0.80_{-0.032}^{+0.032}$ &  $0.73_{-0.028}^{+0.028}$ &  $0.80_{-0.058}^{+0.059}$ & $0.77_{-0.034}^{+0.033}$ \\
S/N   &  21.5 & 22.7  & 10.6  & 16.3 \\
$\chi^2/\nu$  & 334.8/235 & 417.6/275 & 26.9/30 & 122.4/124\\
p.t.e. & 2.0$\times 10^{-5}$ & 6.0$\times 10^{-8}$ & 0.63 & 0.52 \\
$\Delta S_{8}$-(1) & -- & 1.5 & 0.10 & 0.56 \\
$\Delta S_{8}$-(2) & -- & -- & 1.1  & 0.87  \\
$\Delta S_{8}$-(3) & -- & -- & -- & 0.48 \\
\hline\hline
\label{tab:metric_wlpipe}
\end{tabular}
\end{center}
\end{table}

\begin{table}
\begin{center}
\caption{Comparison metrics for all pairs of surveys in the \textit{Published Baseline} analysis case: constraints from the individual collaborations that we choose as baseline to reproduce. For the $S_{8}$ values, we list the mean and the 16\% and 84\% confidence intervals. For CFHTLenS and KiDS-450, these are different from the \textit{Published Nominal} analysis case: constraints from the individual collaborations that can be viewed as the representative results.} 
\begin{tabular}{lcccc}
\hline\hline
& (1) DLS & (2) CFHTLenS & (3) DES-SV & (4) KiDS-450 \\
\hline
$S_{8}$ &  $0.82_{-0.030}^{+0.030}$   & $0.73_{-0.030}^{+0.030}$ & $0.81_{-0.058}^{+0.059}$ & $0.73_{-0.032}^{+0.033}$ \\
$\Delta S_{8}$-(1) & -- & 2.1 & 0.076 & 2.1 \\
$\Delta S_{8}$-(2) & -- & -- & 1.2  & 0.087  \\
$\Delta S_{8}$-(3) & -- & -- & -- & 1.3 \\
\hline\hline
\label{tab:metric_paper}
\end{tabular}
\end{center}
\end{table}

We list the comparison metrics (as described in \Sref{sec:metric}) for all the surveys as well as combinations of survey pairs for the chains in the \textit{Baseline} case in \Tref{tab:metric_wlpipe}. First, looking at the S/N, we notice that in the data configuration used in the individual surveys, the raw statistical power of the measurement is similar for DLS and CFHTLenS, while DES-SV is about half the S/N and KiDS-450 is in between. One interesting observation is that DLS achieves the high S/N even with a significantly smaller area -- this highlights the power of having high-redshift data. A slightly worrying point is that the goodness-of-fits for DLS and CFHTLenS are quite low. For the pair-wise $\Delta S_{8}$, we find trends reflecting what is seen from the figures -- all four surveys are broadly consistent with \Tref{tab:metric_wlpipe} showing some low-level discrepancies (1.5$\sigma$) in $S_{8}$ between CFHTLenS and DLS. 

For the \textit{Published Baseline} chains, we list the $S_{8}$ constraints and $\Delta S_{8}$ values in \Tref{tab:metric_paper}. We do not list the goodness-of-fit here since they are not all available in the papers, and are not directly comparable with the values in \Tref{tab:metric_wlpipe}. We just quote two numbers that available: in \citet{Joudaki2017}, the reduced $\chi^{2}$ for the fiducial CFHTLenS analysis best-fit is 1.5, whereas in \citet{Hildebrandt2017}, the reduced $\chi^{2}$ for the fiducial KiDS-450 analysis best-fit is 1.3. In \citet{Troxel2018}, it was shown that the reduced $\chi^2$ for the fiducial KiDS-450 improves to 1.0 when accounting for the survey geometry. 

\begin{figure*}
\includegraphics[width=0.7\columnwidth]{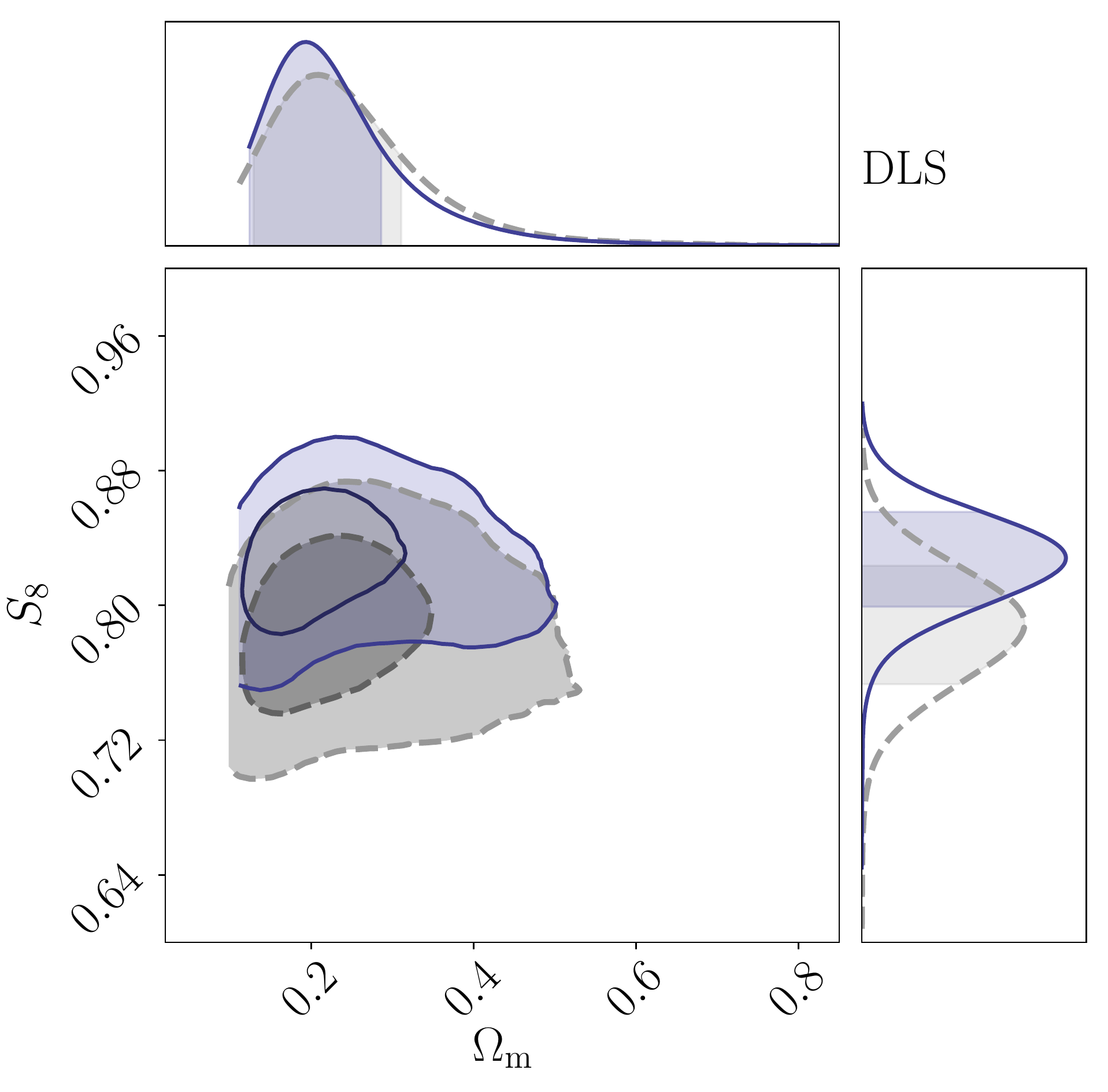}
\includegraphics[width=0.7\columnwidth]{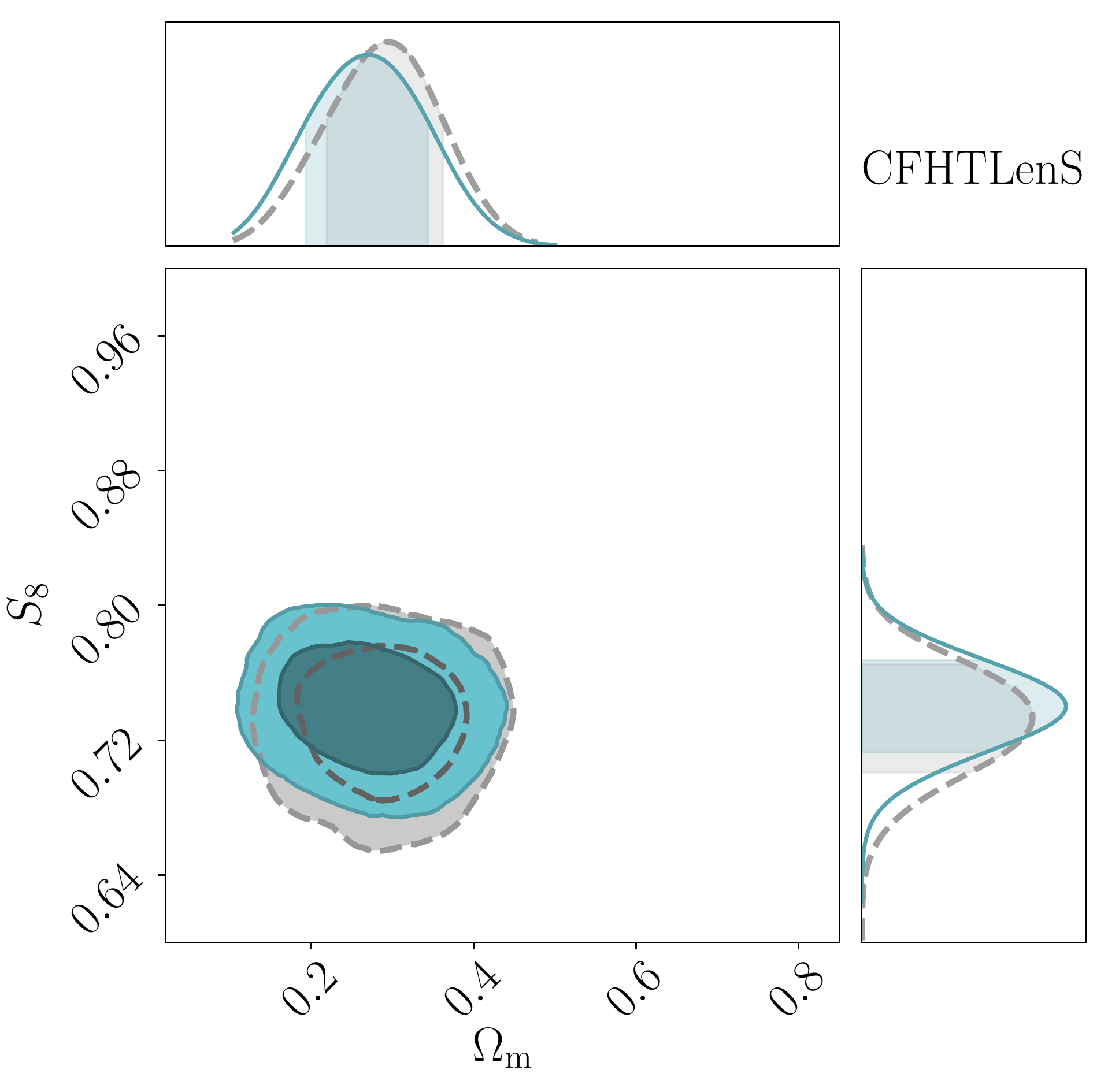}\\
\includegraphics[width=0.7\columnwidth]{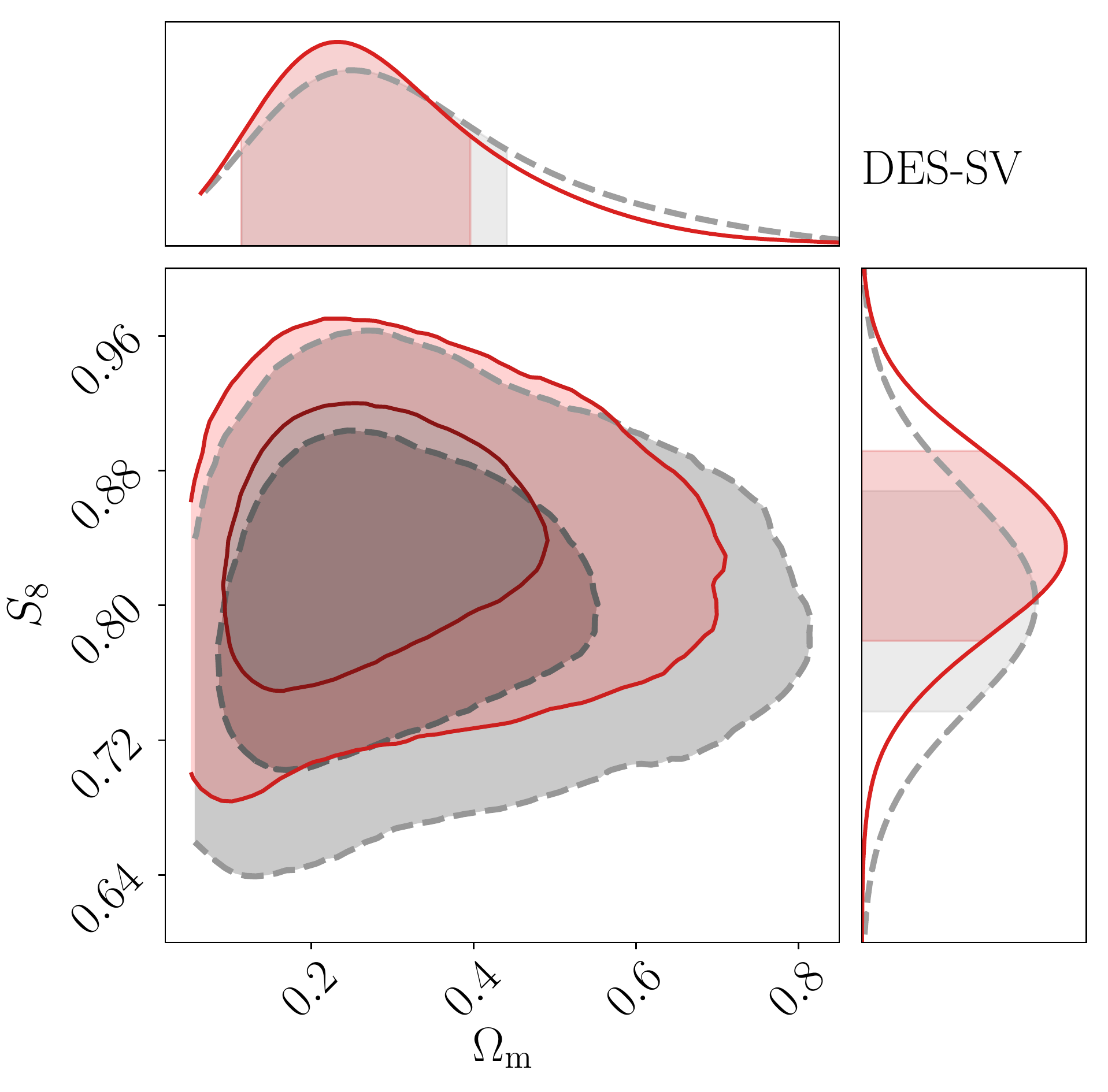}
\includegraphics[width=0.7\columnwidth]{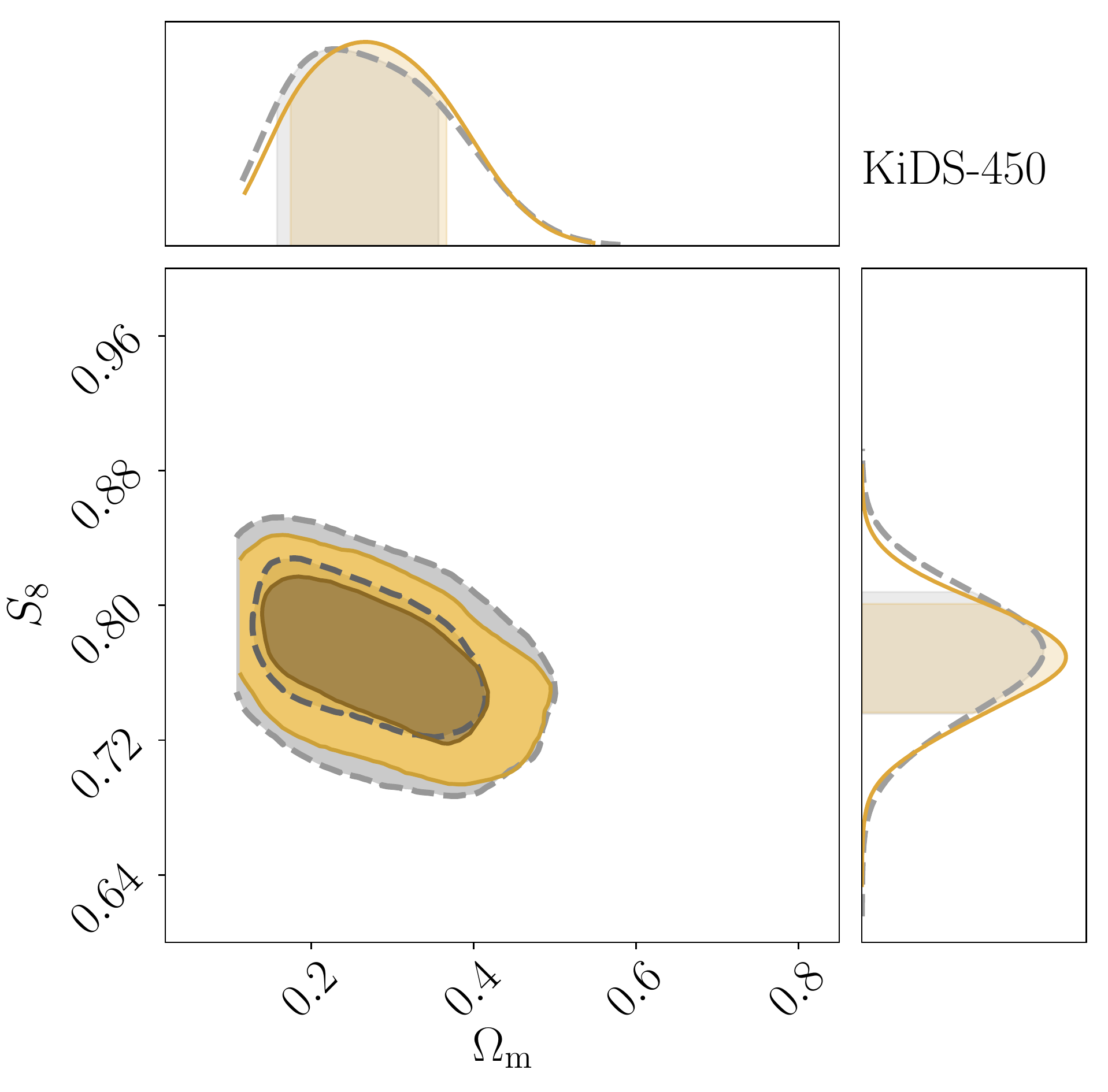}
\caption{Effect of the covariance estimation -- marginalized constraints on $\Omega_{\rm m}$ and $S_8 \equiv \sigma_{8}(\Omega_{\rm m}/0.3)^{0.5}$ obtained from \textsc{WLPipe} using the survey-provided covariance (grey dashed) for four different experiments and the \textsc{CosmoLike} Gaussian analytic covariance (colored solid). We can see a shift in the contours for DLS, CFHTLenS and DES-SV.}
\label{fig:cov}
\end{figure*}

\begin{table}
\begin{center}
\caption{$S_{8}$ constraints, S/N and goodness of fit when we change one analysis choice at a time in the analysis pipeline from the \textit{Baseline} case (see \Tref{tab:metric_wlpipe}). For the $S_{8}$ values, we list the mean and the 16\% and 84\% confidence intervals. The sections of this table correspond to discussions in \Sref{sec:cov_effect}, \Sref{sec:effect_scale} and \Sref{sec:effect_prior}.}
\begin{tabular}{lcccc}
\hline\hline
& (1) DLS & (2) CFHTLenS & (3) DES-SV & (4) KiDS-450 \\
\hline
& \multicolumn{4}{c}{Gaussian covariance matrix (\Sref{sec:cov_effect})}\\
$S_{8}$ &  $0.84_{-0.030}^{+0.030}$ & $0.74_{-0.025}^{+0.024}$ & $0.83_{-0.050}^{+0.052}$ & $0.77_{-0.030}^{+0.030}$ \\
S/N   & 26.0  & 22.2  & 12.7  & 20.4 \\
$\chi^2/\nu$  & 412.5/235 & 344.3/275 & 34.6/30 & 133.0/124 \\
p.t.e. & 7.0$\times 10^{-12}$ &0.0028 & 0.26 & 0.27 \\
\hline
& \multicolumn{4}{c}{Conservative scale cuts (\Sref{sec:effect_scale})}\\
$S_{8}$ &  $0.93_{-0.050}^{+0.050}$ & $0.73_{-0.050}^{+0.052}$ & $0.80_{-0.069}^{+0.068}$ & $0.75_{-0.055}^{+0.055}$ \\
S/N   &15.4 & 16.6 & 10.0 & 10.5 \\
$\chi^2/\nu$  & 112.1/89 & 228.3/132 & 28.4/25 & 62.8/56 \\
p.t.e. & 0.050 & 4.0$\times 10^{-7}$ & 0.29 & 0.24 \\
\hline
& \multicolumn{4}{c}{DES-SV priors (\Sref{sec:effect_prior})}\\
$S_{8}$ & $0.85_{-0.042}^{+0.042}$ & $0.66_{-0.052}^{+0.052}$ & $0.80_{-0.058}^{+0.059}$ & $0.76_{-0.038}^{+0.038}$ \\
$\chi^2/\nu$  & 319.5/235 & 412.2/275& 26.9/30 & 121.5/124\\
p.t.e. & 2.0$\times 10^{-4}$ & 1.6$\times 10^{-7}$ & 0.63 & 0.55 \\
\hline
& \multicolumn{4}{c}{KiDS-450 priors (\Sref{sec:effect_prior})}\\
$S_{8}$ & $0.82_{-0.033}^{+0.033}$  & $0.68_{-0.039}^{+0.039}$ & $0.81_{-0.059}^{+0.059}$ & $0.77_{-0.033}^{+0.033}$ \\
$\chi^2/\nu$  & 323.6/235 & 412.5/275 & 27.0/30 & 122.2/124\\
p.t.e. & 1.1$\times 10^{-4}$ & 1.5$\times 10^{-7}$ & 0.63 & 0.53\\
\hline\hline
\label{tab:metric_tests}
\end{tabular}
\end{center}
\end{table}

\begin{figure}
\includegraphics[width=0.95\columnwidth]{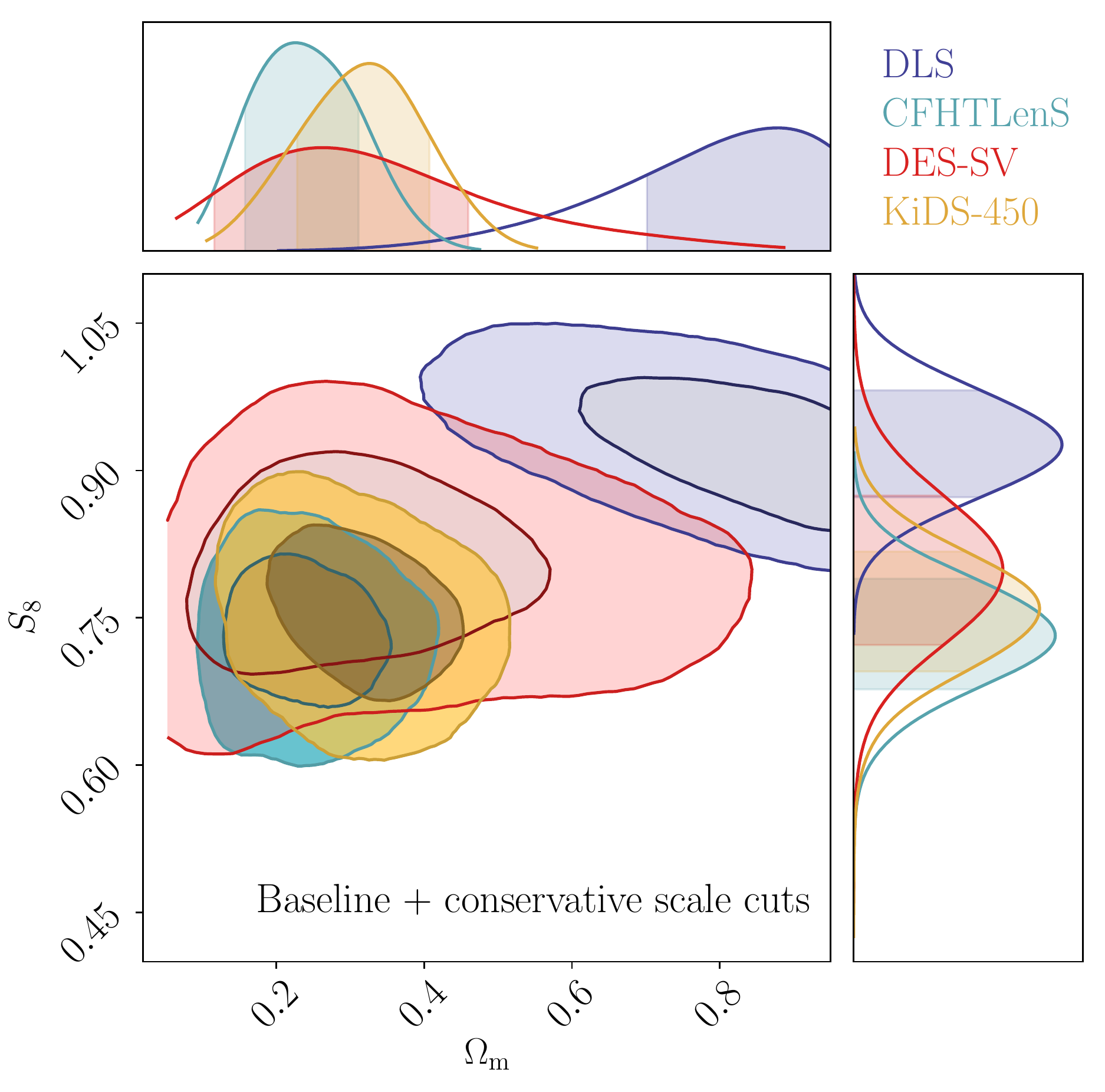}
\caption{Effect of scale cuts compared to \textit{Baseline} (right panel of \Fref{fig:comp4}) -- we show the marginalized constraints for $\Omega_{\rm m}$ and $S_{8}\equiv \sigma_{8} (\Omega_{\rm m}/0.3)^{0.5}$ when small scale data are removed from the fit (requiring $R_{\rm min, +}>$1.3 Mpc and $R_{\rm min, -}>$11.4 Mpc). DLS shifts to large $\Omega_{\rm m}$ and $S_{8}$ values, while the other surveys show enlarged contours compared to \textit{Baseline}. We note that the four contours should not be compared directly here, as the analysis choices are not unified.}
\label{fig:angularscale}
\end{figure}

\subsection{Effect of the Covariance Matrix}
\label{sec:cov_effect}

Now we investigate the effect of the covariance matrix estimation. As discussed in \Sref{sec:cov}, the four surveys have different approaches to covariance estimation. We eliminate these differences by generating a Gaussian analytical \textsc{CosmoLike} covariance matrix for each survey. 

\Fref{fig:cov} shows the changes in the contours in the four experiments when analytic covariance matrices are used in place of those provided by the collaborations. The corresponding comparison metrics are listed in \Tref{tab:metric_tests}. We notice a shifts of the contours in the $S_{8}$ constraints for some of the surveys. Overall, the Gaussian analytic covariance leads to slightly tighter constraints compared to covariance matrices estimated from simulations. This could be partially due to the fact that we have not accounted for the non-Gaussian piece of the analytic covariance.

For DLS, we see a significant shift in the mean of the constraints towards higher $S_{8}$ values; DES-SV and CFHTLenS also show some shifts in $S_{8}$, but less significant. We note that, since the data vector is noisy, we do not expect the contours to agree exactly. However, we believe the shift for DLS is more than what is expected from statistical fluctuation. The DLS field is much smaller and contains a lower level of shape noise compared to the other surveys. In addition, one of the fields contains a galaxy cluster. These factors mean that the covariance is challenging to model and the simple Gaussian covariance used here may not be a good approximation for the dataset. It is possible that neither the survey-provided covariance nor the Gaussian \textsc{CosmoLike} covariance from \textsc{WLPipe} captures these complications. We also note that for the three cases where simulation covariance is used, DES-SV has the smallest Hartlap factor ($H_{\rm DLS}=$0.88, $H_{\rm CFHTLenS}=0.86$, $H_{\rm DES-SV}=$0.7). This means that the inverse of the simulation covariance in DES-SV is expected to be noisier (but unbiased) compared to the other two simulation covariances \citep{Dodelson2013}.

Finally, it is also worth noting that since the survey-provided covariance from KiDS-450 is also an analytic covariance matrix, the agreement between the dashed and the solid contours in the bottom right of \Fref{fig:cov} is a good check on the analytic calculation for the covariance. We have checked that the slightly smaller contours from \textsc{WLPipe} is partially reflecting the difference between the Gaussian and non-Gaussian covariance.

\begin{figure}
\includegraphics[width=0.95\columnwidth]{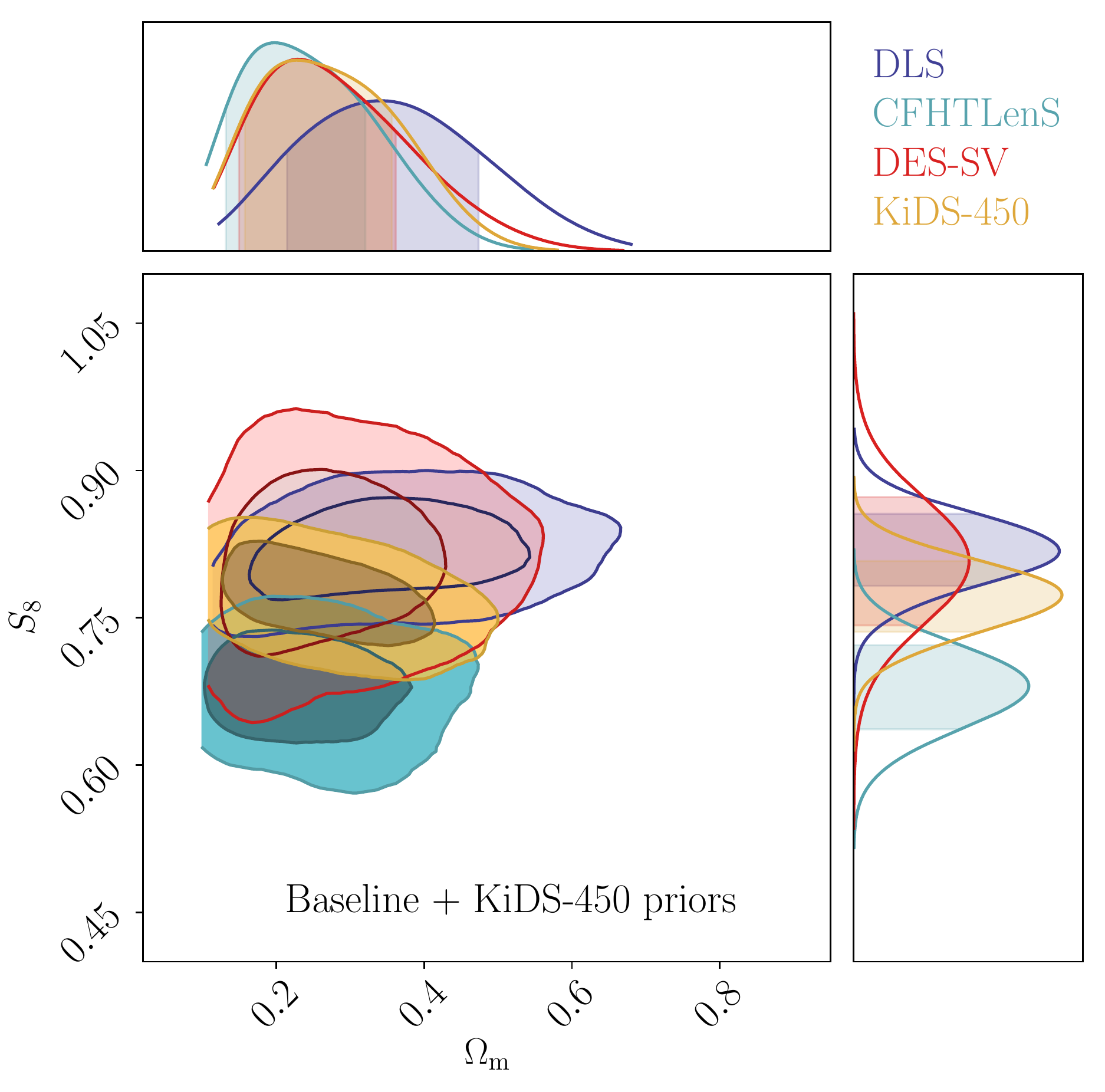}
\includegraphics[width=0.95\columnwidth]{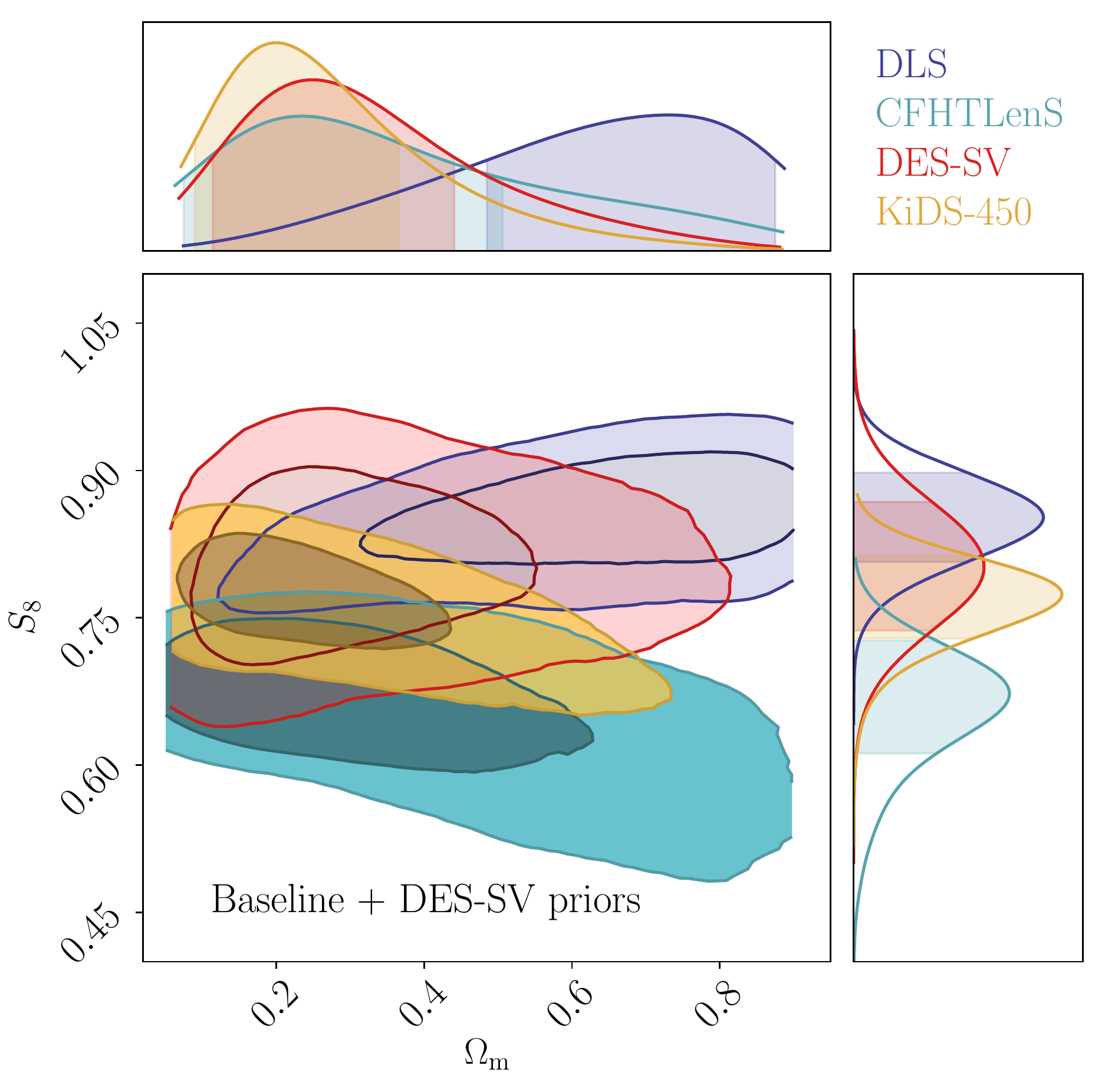}
\caption{Impact of cosmological priors and IA treatments compared to \textit{Baseline} (right panel of \Fref{fig:comp4}) -- we show the marginalized constraints for $\Omega_{\rm m}$ and $S_{8}\equiv \sigma_{8} (\Omega_{\rm m}/0.3)^{0.5}$ when unifying the priors on the cosmological and IA parameters. We first unify to the KiDS-450 priors (top), then to the DES-SV priors (bottom). The constraints in the $\Omega_{\rm m}$ direction is heavily affected by the priors, while in the $S_{8}$ direction, there is a larger effect for surveys with a strong degeneracy in the $\Omega_{\rm m}-S_{8}$ plane. We note that the four contours should not be compared directly here, as the analysis choices are not unified.}
\label{fig:prior_effect}
\end{figure}

\begin{figure}
\includegraphics[width=0.95\columnwidth]{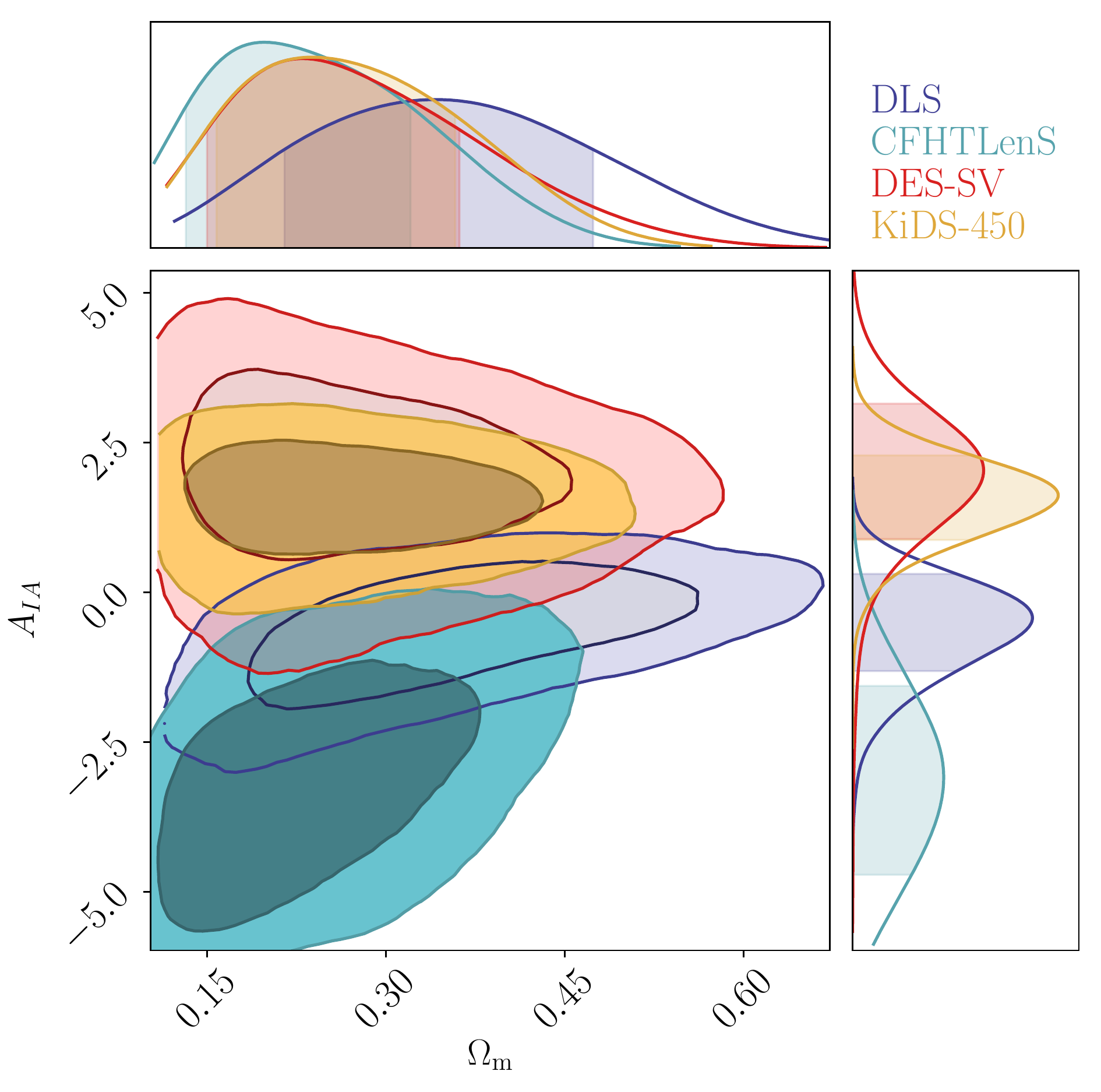}
\caption{Constraints on IA amplitude -- we show the marginalized constraints for $\Omega_{\rm m}$ and $A_{IA}$ when unifying the priors to the KiDS-450 priors (keeping all other analysis choices as in the \textit{Baseline} case). We find a degeneracy between $A_{IA}$ and $\Omega_{\rm m}$ for DLS and CFHTLenS. We also find that both these surveys prefer negative IA amplitudes in this set-up.}
\label{fig:Om_IA}
\end{figure}

\subsection{Effect of Scale Cuts}
\label{sec:effect_scale}

In this section we investigate the effect of scale cuts. Following \Sref{sub:scale}, we choose to match all scale cuts to the most conservative scale cuts in the four datasets ($R_{\rm min, +}>$1.3 Mpc and $R_{\rm min, -}>11.4$ Mpc, see \Eref{eq:scale_cut}). The results are shown in \Fref{fig:angularscale}, with the corresponding metrics listed in \Tref{tab:metric_tests}. The exact cuts used in each bin are tabulated in \Aref{sec:scale_table}, \Tref{tab:scales}. In all these tests, everything else in the analysis stays the same as the \textit{Baseline} case in \Sref{sec:default}.

In \Fref{fig:angularscale}, the first thing that draws the eye is the DLS contours, which shift to very large  $\Omega_{\rm m}$ values, as well as a higher $S_{8}$. All the other surveys appear consistent with the original case in \Fref{fig:comp4}, but with looser constraints due to the fact that we have removed information. 

We note that the goodness-of-fit for DLS improved significantly when applying the conservative scale cuts compared to the \textit{Baseline} case. After a more careful look at the DLS measurements, it appears that the small-scale data points for $\xi_{-}$ is the source of the contour shift -- those data points prefer a lower amplitude compared to the rest of the data points. Therefore, when applying the conservative scale cuts, the model amplitude increases (so does $S_{8}$), and the goodness-of-fit improves. This could also be a hint that the small-scale covariance is underestimated, as already discussed in \Sref{sec:cov_effect}, that the characteristics of the DLS data makes it difficult to model the covariance. We note that some of these issues were discussed in \citet{Jee2012} and \citet{Jee2016}, and a similar trend in $S_{8}$ was seen in Fig.13 of \citet{Jee2016}. Here we caution that since the DLS contours are far from $\Omega_{\rm m}=0.3$ and clipped by the priors ($\Omega_{\rm m}<1$), the $S_{8}$ values quoted are not very meaningful.

\subsection{Impact of Cosmological Priors and IA Treatment}
\label{sec:effect_prior}

Next, we consider the impact of different cosmological priors and IA treatments. To address this, we impose identical priors on all surveys, first using those from DES-SV and then from KiDS-450 (see \Tref{tab:params}) since they roughly represent the two approaches of handling the parameters: DES-SV has priors that are relatively conservative, and in the parametrization of  [$\Omega_{\rm m}$, $\Omega_{\rm b}$, $h$, $\sigma_{8}$, $n_{s}$], whereas KiDS-450 has more restrictive priors and uses the parametrization [$\Omega_{\rm c}h^2$, $\Omega_{\rm b}h^2$, $h$, $\ln (10^{10}A_{s})$, $n_{s}$]. We moreover allow for intrinsic alignments in the case of CFHTLenS and DLS. For all surveys, we consider either the IA amplitude prior $-5 < A_{\rm IA} < 5$ used by DES-SV or the prior $-6 < A_{\rm IA} < 6$ used by KiDS-450. Note that aside from these changes to the cosmological priors and IA treatments, we keep all other analysis choices the same as in the \textit{Baseline} case of \Sref{sec:default}.

The two panels of \Fref{fig:prior_effect} show the effect of unifying the cosmological priors and IA treatments from that chosen as \textit{Baseline}, with the corresponding metrics listed in \Tref{tab:metric_tests}. Looking at CFHTLenS, DES-SV and KiDS-450, it is apparent that the constraints in the $\Omega_{\rm m}$ direction are largely dominated by cosmological priors. Specifically, the prior on $h$, which is wider for DES-SV compared to KiDS-450, leads to large changes in the $\Omega_{\rm m}$ posterior. The constraints on $S_{8}$, on the other hand, are relatively robust to cosmological priors, consistent with previous findings \citep{Kilbinger2012,Joudaki2017}. This again is showing that cosmic shear measurements for these four datasets are mainly constraining only the amplitude of the power spectrum and not the detailed shape of it. The uncertainty on $S_8$ decreases for CFHTLenS when moving to tighter cosmological priors, however, this is largely due to the fact that the $S_{8}$ definition here is not optimal for the CFHTLenS dataset. We will discuss this point in \Sref{sec:S8}. 

\begin{figure*}
\includegraphics[width=1.9\columnwidth]{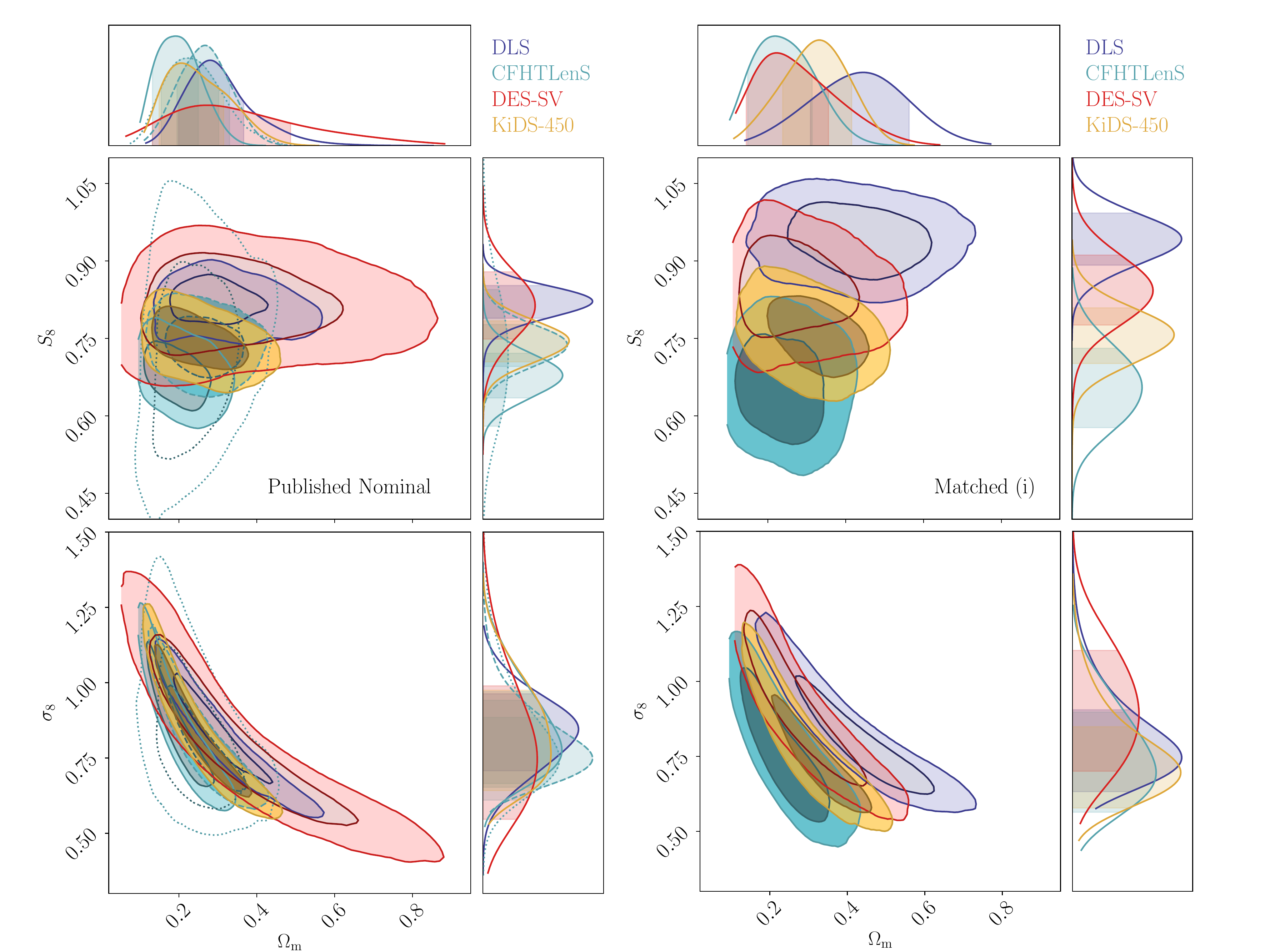}
\caption{Final comparison of the cosmological constraints from the four surveys according to the literature (\textit{Published Nominal}, left) and according to our unified analysis framework (right, \textit{Matched}) -- we show the marginalized constraints for $\Omega_{\rm m}$, $S_{8}\equiv \sigma_{8}(\Omega_{\rm m}/0.3)^{0.5}$ and $\sigma_{8}$ for the four cosmic shear surveys. In the right panel, we use Gaussian analytic covariances, conservative scale cuts and the KiDS-450 priors. We note that for the CFHTLenS \textit{Published Nominal} constraints, we show all three settings MIN (solid), MID (dashed) and MAX (dotted) in \citet{Joudaki2017}.}
\label{fig:final}
\end{figure*}

A few other effects of the cosmological priors and IA treatment are visible in \Fref{fig:prior_effect}. First, for DLS, when imposing the DES-SV priors, $\Omega_{\rm m}$ moves to high values while $S_{8}$ remains roughly the same. When imposing KiDS-450 priors, the $\Omega_{\rm m}$ constraints appear similar to the \textit{Baseline} case. This behavior, together with what is shown in \Sref{sec:effect_scale}, suggests that the DLS constraints on $\Omega_{\rm m}$ are sensitive to the scales used and the priors. For CFHTLenS, the $S_{8}$ constraints move to lower values using both DES-SV and KiDS-450 priors. This comes from the fact that compared to the \textit{Baseline} case, here there is additional freedom in the IA amplitude. We examine the IA amplitude when using the KiDS-450 priors, as shown in \Fref{fig:Om_IA}, and find that the CFHTLenS favors a negative IA amplitude at the $2\sigma$ level. This, based on previous work in measurements of IA, suggests that we may be fitting to some systematic effects that appear to behave like IA \citep{Kilbinger2017,Choi2016,vanUitert2018}. This is consistent with Fig. 8 and Fig. 9 of \citet{Joudaki2017}, where they show that this negative IA shifts the $S_{8}$ constraints to lower values. There is also a similar (but less severe) trend in the DLS data.

\subsection{Common Covariances, Angular Scale Cuts, Cosmological Priors, and IA Treatments}
\label{sec:match}

After investigating the individual effects in \Sref{sec:cov_effect}, \Sref{sec:effect_scale} and \Sref{sec:effect_prior}, we now combine all of them and perform a uniform analysis on all four surveys. We study two cases, both using \textsc{CosmoLike} Gaussian covariances, conservative scale cuts, the same IA treatments, and we use two sets of priors: 
\begin{enumerate}
\item[(i)] KiDS-450 priors and 
\item[(ii)] DES-SV priors. 
\end{enumerate}

As discussed in \Sref{sec:fiducial_cfhtlens}, in this subsection we incorporate the photo-z and shear calibration bias uncertainties for CFHTLenS. As summarized in Sec. 4.3 of \citet{Kilbinger2017}, a number of improvements to CFHTLenS have been identified since the public release of the catalogues in 2013. Of importance to this study is the analysis by \citet{Choi2016} who showed that significant biases existed in the reported photo-z distributions, the result from \citet{Kuijken2015} that the CFHTLenS shear calibration corrections were in general underestimated and the finding by \citet{FenechConti2017} that the previously unexplored area of galaxy selection bias results in a few percent overestimation of the shear calibration correction. The conclusion of all these works was that any future analyses of CFHTLenS should include conservative systematic error terms to account for these effects. In this section, we therefore marginalize over an uncertainty in the mean redshift of each bin with zero-mean top-hat prior of full-width $0.2$, and an uncertainty in the shear calibration correction zero-mean top-hat prior of full-width $0.1$.

\begin{figure}
\includegraphics[width=0.95\columnwidth]{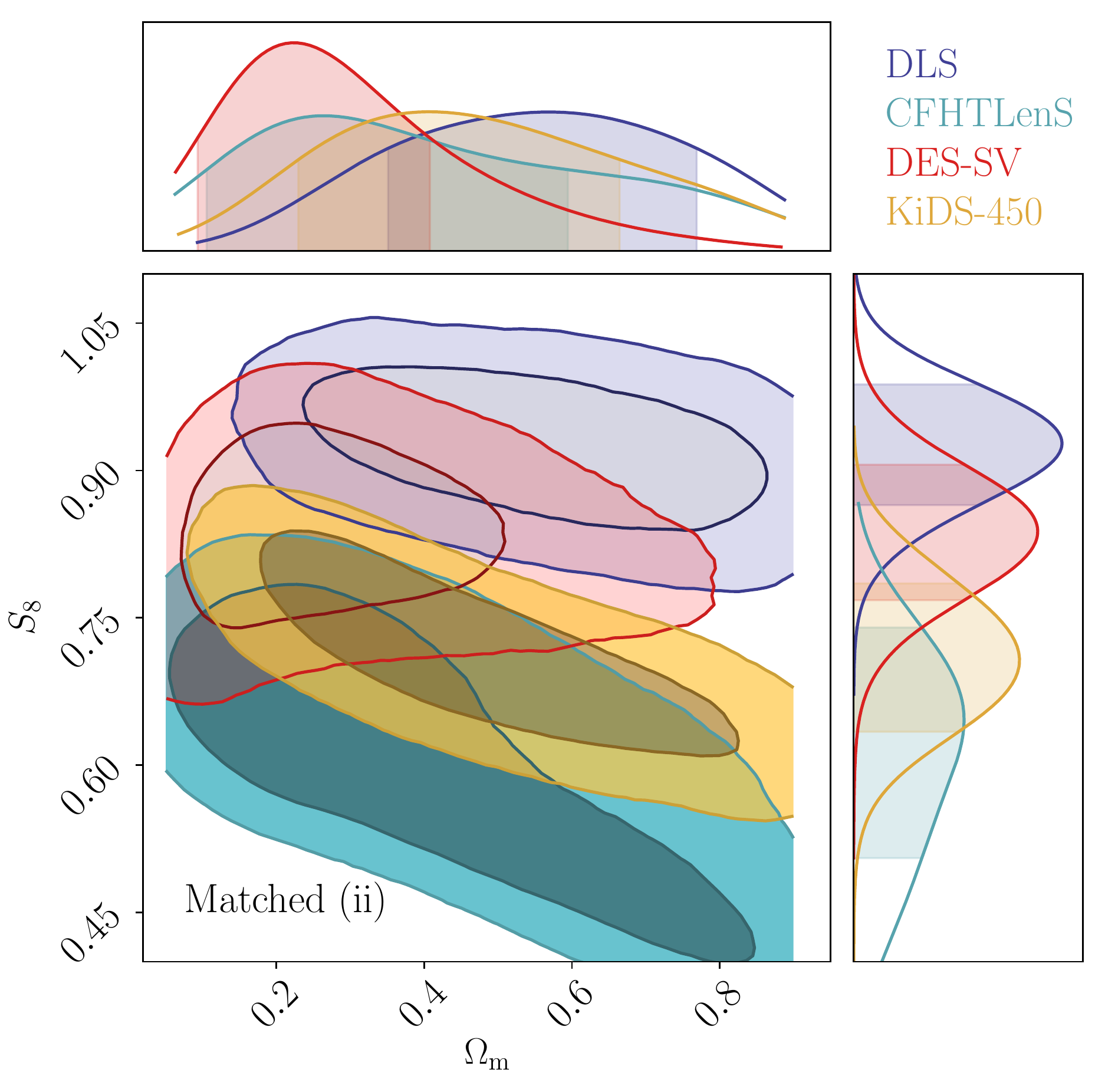}
\caption{Same as the upper right panel of \Fref{fig:final}, but now using DES-SV priors.}
\label{fig:final2}
\end{figure}

In \Fref{fig:final} we show the comparison between the \textit{Published Nominal} contours and case (i) listed above. The \textit{Published Nominal} contours present the view one would have on the four cosmic shear surveys after reading the individual papers \citep{Jee2016,Joudaki2017,Abbott2015,Hildebrandt2017}, while the \textit{Matched} contours present what the cosmological constraints are when analysed through a unified analysis framework. One can also compare with \Fref{fig:comp4} to understand the nature of the different changes in the contours. The contours for (ii) are shown in \Fref{fig:final2}. We choose to focus on (i) here as it is not as affected by the $S_{8}$ definition (see \Sref{sec:S8}) as (ii). The four surveys in the right panel of \Fref{fig:final} can now be compared on equal footing. The comparison metrics are shown in \Tref{tab:metric_match}.  

In general, we observe the same effects seen individually in \Sref{sec:cov_effect}, \Sref{sec:effect_scale} and \Sref{sec:effect_prior}. But when put together, the discrepancies between the different surveys coming from the different effects accumulate and become larger. Looking at \Fref{fig:final} and the $\Delta S_{8}$ statistics in \Tref{tab:metric_match}, we find that essentially none of the surveys have $S_{8}$ constraints that agree within 1$\sigma$ and the extreme cases differ more than 3$\sigma$. If we look at the change in the $S_{8}$ constraints for the individual surveys from the left panel to the right panel, it is clear that the main effect is that DLS moves to larger $S_{8}$. CFHTLenS is consistent with the MIN and MAX case but not the MID case, which is expected given that the choice of IA models we use is the same as the MIN case, and that the MAX case has little constraining power. DES-SV and KiDS-450 stay roughly the same. 

\begin{table}
\begin{center}
\caption{Comparison metrics corresponding to the right panel of \Fref{fig:final}. That is, all analysis choices matched: Gaussian \textsc{CosmoLike} covariance matrix, conservative scale cuts, same IA treatments, and KiDS-450 cosmological priors. For the $S_{8}$ values, we list the mean and the 16\% and 84\% confidence intervals. }
\begin{tabular}{lcccc}
\hline\hline
& (1) DLS & (2) CFHTLenS & (3) DES-SV & (4) KiDS-450 \\
\hline
$S_{8}$   &  $0.94_{-0.045}^{+0.046}$ & $0.66_{-0.070}^{+0.071}$ & $0.84_{-0.061}^{+0.062}$ & $0.76_{-0.049}^{+0.048}$ \\
S/N                & 17.4 & 15.1 & 11.6 & 12.1 \\
$\chi^2/\nu$  & 137.8/89  & 176.3/132 & 32.7/26 & 71.5/56 \\
p.t.e.              & 7.0$\times 10^{-4}$  & 0.0060 & 0.17 & 0.079 \\
$\Delta S_{8}$-(1) & -- & 3.4 & 1.3 & 2.9 \\
$\Delta S_{8}$-(2) & -- & -- & 2.0 & 1.1  \\
$\Delta S_{8}$-(3) & -- & -- & -- &  1.2 \\
BF-(1) & -- & -1.1 & 1.6 & -0.50 \\
BF-(2) & -- & -- & 0.70  & 1.3 \\
BF-(3) & -- & -- & -- & 1.1 \\
\hline\hline
\label{tab:metric_match}
\end{tabular}
\end{center}
\end{table}

 \begin{figure*}
\includegraphics[width=1.99\columnwidth]{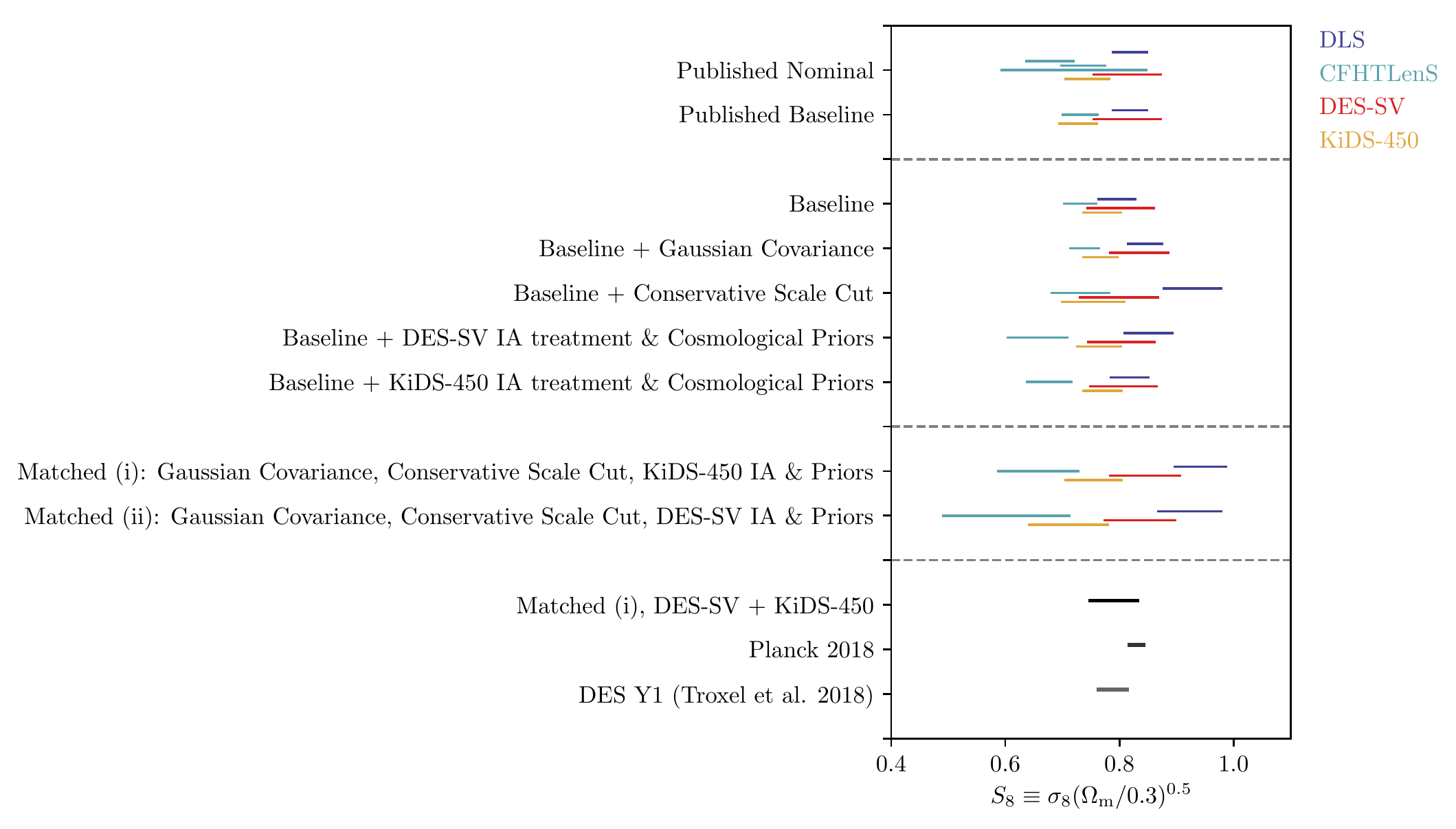} \\
\vspace{0.1in}
\begin{tabular}{|l|l|}
\hline
\textit{Published Nominal} & constraints from the individual collaborations that can be viewed as the representative results  \\
\textit{Published Baseline} & constraints from the individual collaborations that we choose as our baseline to reproduce  \\
\textit{Baseline} & constraints from \textsc{WLPipe} that matches all analysis choices in \textit{Published baseline} \\
\textit{Baseline} + X & same as \textit{Baseline} but changing one analysis choice \textit{X} \\
\textit{Matched} & constraints from \textsc{WLPipe} with all analysis choice matched \\
\hline
\end{tabular}
\caption{Summary of $S_{8}=\sigma_{8}(\Omega_{\rm m}/0.3)^{0.5}$ constraints from the four cosmic shear surveys studied in this work. Each bar shows the 16\% and 84\% confidence interval of the $S_{8}$ constraints. The plot is divided by dashed horizontal lines into four sections. The first section shows chains provided by the respective collaborations. The second section shows constraints derived from \textsc{WLPipe}, but are not fully matched in all analysis steps, ergo not directly comparable. The third section shows constraints derived from \textsc{WLPipe} and have analysis steps matched, so can be compared directly. The last section shows the constraints from combining the two consistent surveys DES-SV and KiDS-450, the \textit{Planck} satellite \citep{Planck2018} and DES Y1 cosmic shear \citep{Troxel2017}. The naming conventions are explained in the table below the figure. We note that for the CFHTLenS \textit{Published Nominal} constraints, we show all three settings MIN (top), MID (middle) and MAX (bottom) in \citet{Joudaki2017}.} 
\label{fig:bars}
\end{figure*}

Next we turn to the other statistics in \Tref{tab:metric_match}. We note that the signal-to-noise for the four datasets change slightly, but the relative power stays roughly the same, with DLS being the highest and DES-SV being the lowest. We note that the goodness-of-fit for DLS and CFHTLenS improved from the \textit{Baseline} case but is still quite low. The largest $\Delta S_{8}$ is about 3.4$\sigma$ between DLS and CFHTLenS, which is also apparent from \Fref{fig:final}. Next we look at the BF statistic [\Eref{eq:bf}] between pairs of surveys. Here when evaluating the numerator in \Eref{eq:bf} for BF, we only require the cosmological parameters to be shared amongst the two experiments being compared and keep the IA amplitude, shear calibration parameter and photo-z uncertainty separate. We find that the message from the BF statistics is similar to that captured by the $\Delta S_{8}$ metric in this case, though the message of consistency/inconsistency is somewhat weaker -- the only BF value that fails the requirement of combining is DLS and CFHTLenS.  

We now combine the two surveys, DES-SV and KiDS-450, under this matched setting. These two datasets are consistent under the same model assumption according the BF metric, and have reasonable goodness-of-fit values. The combined constraint of DES-SV and KiDS-450 is $S_{8}=0.79^{+0.042}_{-0.041}$. Compared to the cosmic shear results from the first year DES data \citep{Troxel2017}, $S_8=0.79^{+0.024}_{-0.026}$, we find excellent agreement. Compared to the state-of-the-art CMB constraints \citep{Planck2018}, $S_{8}=0.83_{-0.013}^{+0.013}$, we find reasonably consistent results with roughly $1\sigma$ lower $S_{8}$. These results are in good agreement with that found in \citet{Troxel2018}. 

\subsection{A side note on the $S_8$ definition}
\label{sec:S8}
As discussed briefly in \Sref{sec:metric}, $S_{8}$ is defined as $\sigma_{8}(\Omega_{\rm m}/0.3)^\alpha$, where $\alpha$ is designed to remove the degeneracy between $\sigma_{8}$ and $\Omega_{\rm m}$. That is, if $\alpha$ is chosen optimally, it characterizes the direction orthogonal to the $\Omega_{\rm m}-\sigma_{8}$ contours. For datasets of different redshift distribution, the optimal $\alpha$ is different. 

Throughout our analysis, we have fixed $\alpha$ to be 0.5, which may not be optimal for all datasets. This implies that for datasets where $\alpha$ is further from 0.5, the projected uncertainties on $S_{8}=\sigma_{8}(\Omega_{\rm m}/0.3)^{0.5}$ are going to be slightly larger than if the optimal $\alpha$ were used, and that when comparing the different surveys they will tend towards being consistent. This can be seen clearly in\Fref{fig:final2}, where the contours for CFHTLenS and KiDS-450 are tilted leading to larger uncertainties in the $S_{8}$ direction. The effect is much reduced when a tighter prior is imposed as in the right-hand panels of \Fref{fig:final}. Roughly, we find that with the DES-SV priors (corresponding to \Fref{fig:final2}), the optimal $\alpha$ values are 0.56 (DLS), 0.71 (CFHTLenS), 0.51 (DES-SV) and 0.67 (KiDS-450). With the KiDS-450 priors (corresponding to the right panels of \Fref{fig:final}), the optimal $\alpha$ values are 0.52 (DLS), 0.52 (CFHTLenS), 0.52 (DES-SV) and 0.58 (KiDS-450). That is, we expect the discrepancies between the surveys in the single parameter that quantifies the amplitude to be sensitive to the priors and likely larger if an optimal $\alpha$ is used. On the other hand, the BF metric is insensitive to the choice of $\alpha$ so is a more robust measure of consistency.

\section{Summary and Discussion}
\label{sec:conclusion}

In this paper we use a generic cosmic shear pipeline, \textsc{WLPipe}, that takes in galaxy shear catalogs, calculates the two-point shear-shear correlation function via the software package \textsc{TreeCorr} and the associated covariance matrix via the software package \textsc{CosmoLike}, and finally carries out cosmological parameter inference via the software package \textsc{CosmoSIS}. The \textsc{WLPipe} framework is constructed using the \textsc{Pegasus} workflow engine, which takes care of data and code transfer between different computing resources seamlessly. This pipeline also serves as a prototype pipeline for future analysis pipelines in DESC. 

We apply this pipeline to four existing cosmic shear surveys: the Deep Lens Survey (DLS), the Canada-France-Hawaii Telescope Lensing Survey (CFHTLenS), the Science Verification data from the Dark Energy Survey (DES-SV), and the 450 deg$^{2}$ release of the Kilo-Degree Survey (KiDS-450). The goal is to first reproduce the literature results, investigate the effect of different analysis choices adopted in each of the surveys, and finally unifying these different analysis choices in order to perform an apples-to-apples comparison of the survey results. In \Fref{fig:bars} we summarize the constraints on $S_{8} \equiv \sigma_{8} (\Omega_{\rm m}/0.3)^{0.5}$ from all the cases studied in this work. We summarize our main findings below: 
\begin{itemize}
\item We are able to reproduce a specific set of the published results from the four collaborations when following the same analysis choices to well within the uncertainties. In this \textit{Baseline} case, the four surveys appear to be broadly consistent in terms of their constraints on $S_{8}$: $S_{8}=0.80_{-0.032}^{+0.032}$ (DLS), $0.73_{-0.028}^{+0.028}$ (CFHTLenS), $0.80_{-0.058}^{+0.059}$ (DES-SV), and $0.77_{-0.034}^{+0.033}$ (KiDS-450). However, we note that not all the model fits are good descriptions of the data -- for DLS and CFHTLenS, the p-values for the fits are low, while for KiDS-450, the p-value is acceptable, but only after incorporating recent improvements for the covariance.
\item In reproducing the published results, we investigate several issues in the published results: the angular bin values used in the data vector, the incorporation of nuisance parameters in the covariance, and the nonlinear power spectrum model and others. We find these details can shift the cosmological constraints by $\sim 0.5 \sigma$. Most of these issues are known, but analyzing all four experiments systematically in this work gives a big picture view of how the four analyses agree and differ. 
\item Effect of the covariance matrix: constraints based on simulation-based covariances can be shifted from analytic covariances due to noise. In addition, the DLS covariance may not be well approximated by a Gaussian covariance due to the complexity of the data, the small area and the low shape noise. 
\item Effect of scale cuts: sensitivity of the cosmological constraints to scale cuts could indicate internal inconsistency of datasets or further issues with the covariance. It could also point to potential failures in the models at small scales (e.g. IA, nonlinear matter power spectrum, baryonic physics).
\item Effect of priors: for parameters that are not constrained (e.g. $\Omega_{\rm m}$), the priors have an effect on the constraints, but for parameters that are constrained (e.g. $S_{8}$), the effect of priors is smaller, but not negligible. A wide prior on the IA amplitude can absorb other sources of systematic issues, which could explain the slightly negative IA amplitude constrained by CFHTLenS.
\item When unifying all analysis choices discussed above, the four surveys give the following constraints: we find $S_{8}\equiv \sigma_{8}(\Omega_{\rm m}/0.3)^{0.5}$ to be $0.94_{-0.045}^{+0.046}$ (DLS), $0.66_{-0.070}^{+0.071}$ (CFHTLenS), $0.84_{-0.061}^{+0.062}$ (DES-SV) and $0.76_{-0.049}^{+0.048}$ (KiDS-450). Specifically, DLS moves to higher $S_{8}$ while CFHTLenS moves to lower $S_{8}$ compared to the \textit{Baseline}. The change in the DLS constraints is primarily due to the scale cuts and covariance, while the change in CFHTLenS is due to the change in the IA treatment, and could be an indication of residual issues in the photo-z estimation. The goodness-of-fit values for DLS and CFHTLenS improved but is still low. 
\item We calculate the $\Delta S_{8}$ statistics and the Bayesian evidence ratio (BF) between each of the two surveys (when analysis choices are unified). The $S_8$ constraints from the two most discrepant cases (DLS and CFHTLenS) differ by 3.4$\sigma$. The $S_{8}$ constraints for DES-SV and KiDS-450 in the final matched analysis appear consistent with the \textit{Baseline} analysis as well as with each other. They also seem to be robust to the various analysis choices tested. Together with the more reasonable goodness-of-fit values and IA constraints, this is an encouraging indication for the field given that DES-SV and KiDS-450 are the most recent work amongst the four surveys.
\item Based on all the above information, we decide to combine the DES-SV and KiDS-450 datasets (based on the goodness-of-fit, IA constraints and consistency). The combined constraint is $S_{8}=0.79^{+0.042}_{-0.041}$, which is in agreement with both the cosmic shear constraints from the first year of DES data in \citet{Troxel2017}, and the CMB constraints from \citet{Planck2018}.
\end{itemize}

Cosmic shear measurements hold great promise in terms of the constraining power in cosmology. In order to fully exploit this power in upcoming and future cosmic shear surveys (DES, KiDS, HSC, Euclid, LSST, WFIRST), it is important to learn from the experiences accumulated over the past years in the community across the different collaborations and datasets. We have demonstrated that a number of analysis choices can result in significant changes in the cosmological constraints and should therefore be treated with care for the future analyses. 

\input{main.bbl}

\appendix

\section{Effect of Angular Values}
\label{sec:theta_bins}

\begin{figure}
\includegraphics[width=0.85\columnwidth]{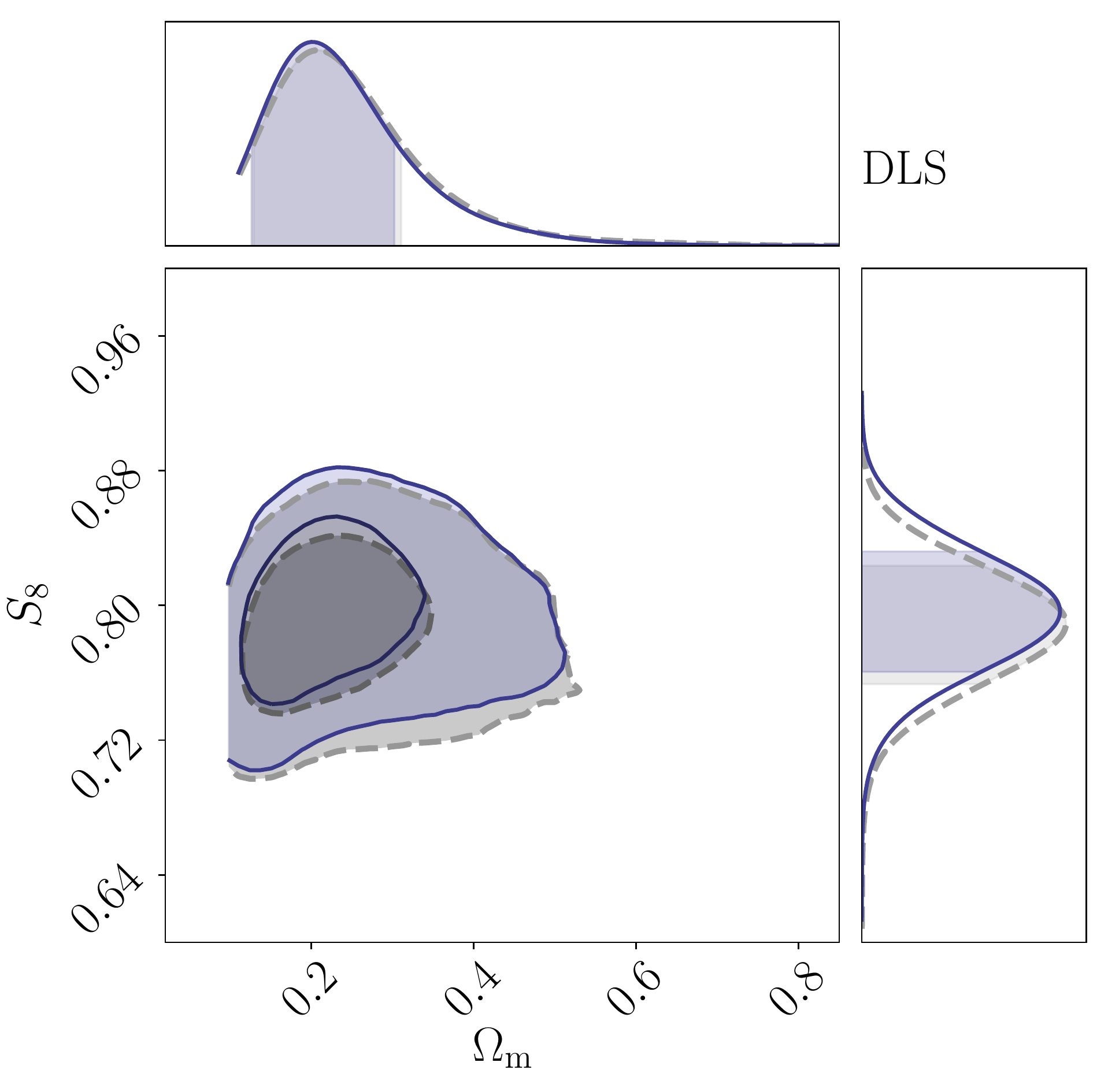}
\includegraphics[width=0.85\columnwidth]{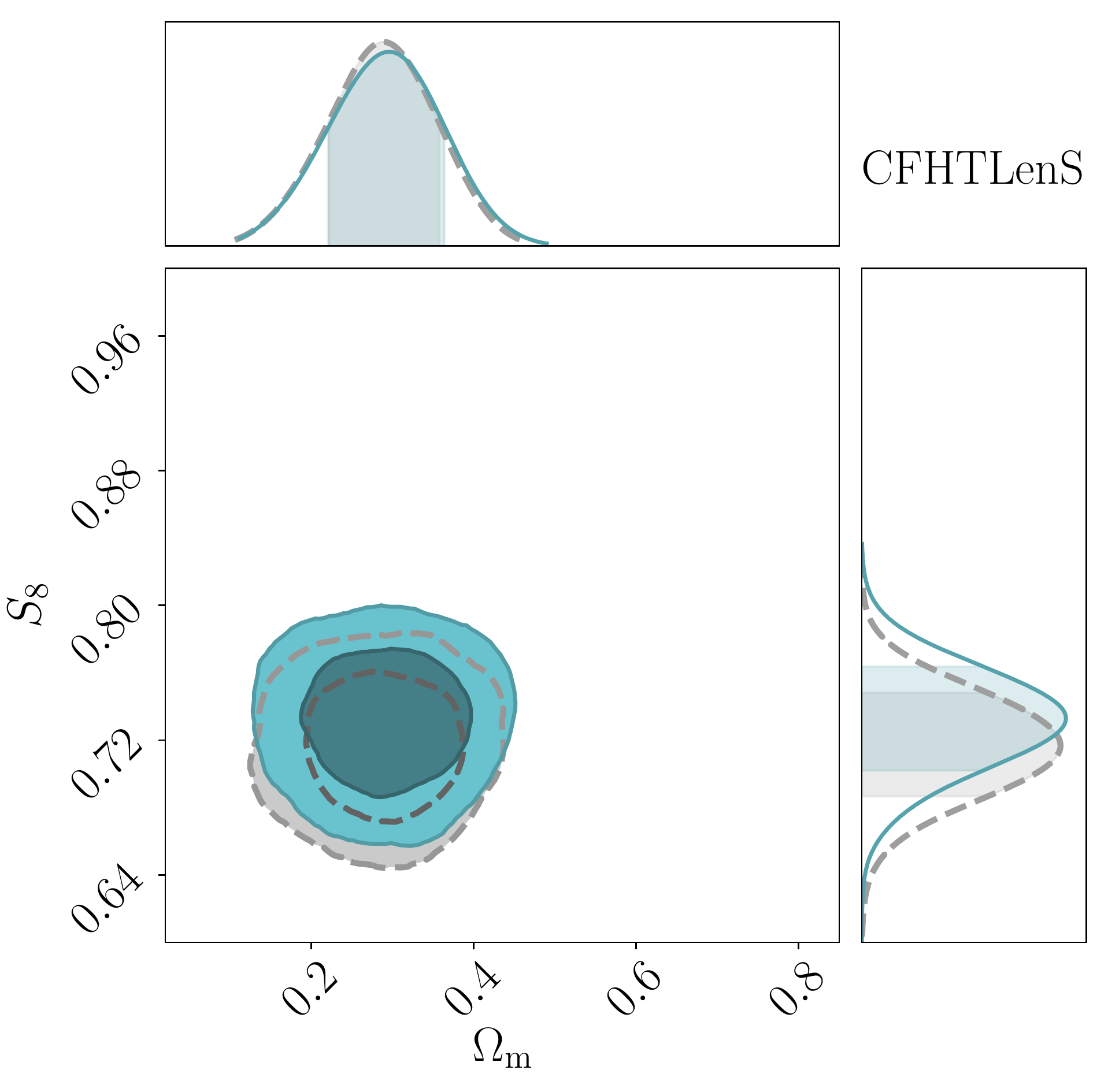}
\includegraphics[width=0.85\columnwidth]{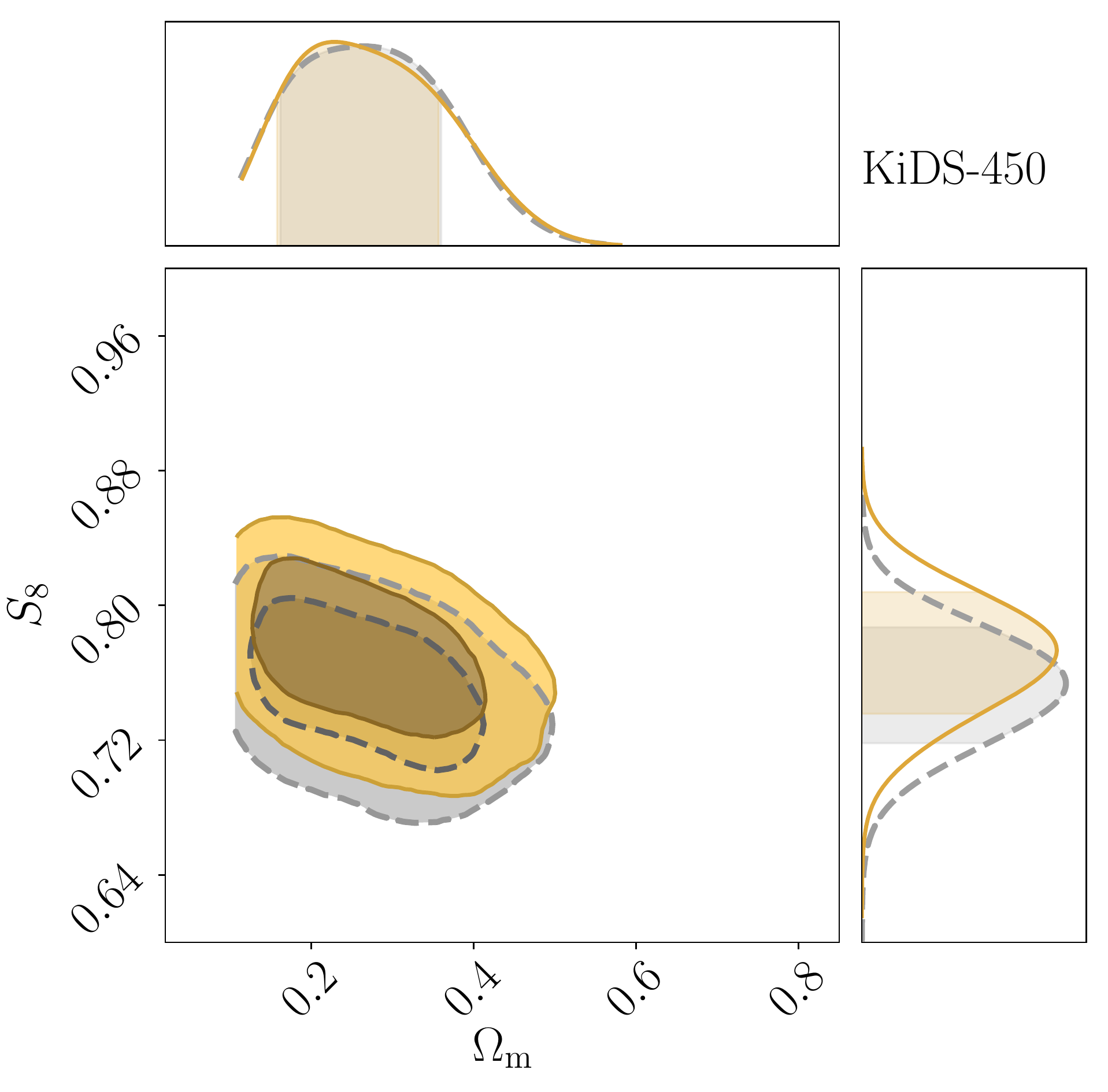}
\caption{This plot illustrates the change in the cosmological constraints when we change the angular values used in the data vector: the grey dashed contour shows the constraints using the bin centers, while the colored contour shows the constraints using the weighted centers.}
\label{fig:ang}
\end{figure}

As discussed in \Sref{sec:twopoint}, out of the four surveys, three of them did not use the weighted angular values in the data vector [\Eref{eq:mean_theta}]. Correcting for this, the DLS angular values changed from [0.40, 0.72, 1.28, 2.29, 4.10, 7.32, 13.09, 23.40, 41.84, 74.79] to [ 0.43, 0.77, 1.37, 2.45, 4.38, 7.84, 14.01, 25.05, 44.78, 80.04] arcmin\footnote{Since we do not have the catalogs, this was estimated using Eq. 9 of \citet{Krause2017}.}, the CFHTLenS angular values changed from  [1.41, 2.79, 5.53, 11.0, 21.7, 43.0, 85.2] to [1.51, 3.00, 5.93, 11.77, 23.25, 45.77, 90.62] arcmin, and the KiDS-450 angular values changed from [0.71, 1.45, 2.96, 6.02, 12.25, 24.93, 50.75, 103.30, 210.27] to [0.77, 1.57, 3.19, 6.50, 13.25, 26.88, 54.50, 110.45, 219.36] arcmin. In \Fref{fig:ang} we show how this change in the angular bin values affect the cosmological constraints. In general, using the weighted bin centers give slightly higher $S_{8}$ values than using the bin centers.

\section{The KiDS-450 Covariance}
\label{sec:kids_cov}

\begin{figure}
\includegraphics[width=0.95\columnwidth]{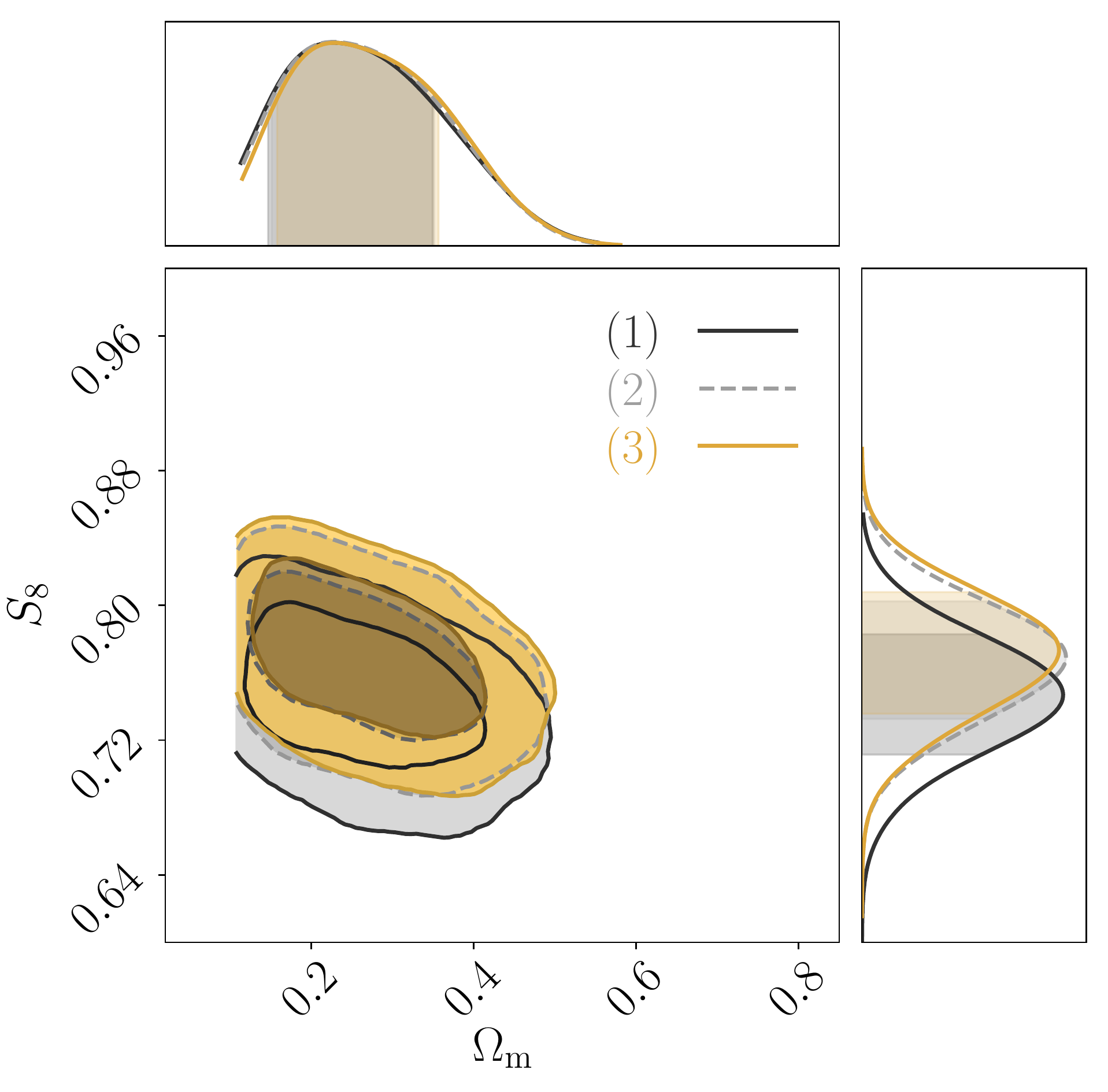}
\caption{KiDS-450 constraints when using different approaches to marginalize $\sigma_{m}$.}
\label{fig:kids_cov}
\end{figure}

In \Fref{fig:kids_cov} we show the effect of different approaches in marginalizing $\sigma_{m}$ for the \textit{Baseline} KiDS-450 case. For (1), we use the survey-provided covariance that includes $\sigma_{m}$. For (2) we use the survey-provided covariance without $\sigma_{m}$, and include $\sigma_{m}$ using Eq. 12 of \citet{Hildebrandt2017}. For (3), we use the survey-provided covariance without $\sigma_{m}$, and marginalize over $\sigma_{m}$ at the parameter level. We find that (2) and (3) are in very good agreement, demonstrating that the two approaches are effectively equivalent, while (1) is shifted. The reason for the shift is due to the fact that (1) uses the noisy data vector in Eq. 12 of \citet{Hildebrandt2017}, instead of a theoretical data vector. This is also explained in \citet{Troxel2018}. 

\section{Scale Cut Values}
\label{sec:scale_table}
In \Tref{tab:scales} we list the scale cuts used in \Sref{sec:effect_scale}. We also show the redshift $z_{p}$ used to calculate these scale cuts, as well as the number of data points remaining after making the scale cuts. 

\begin{table}
\begin{center}
\caption{\textit{Top:} The redshift of the peak of the lensing efficiency for each redshift bin in each survey. $z_{p}$ is used to calculate the scale cuts in the bottom table. \textit{Bottom:} Scale cuts applied in each survey. Each row is for a combination of bins. The bins are ordered as (bin1, bin1), (bin1, bin2) ... (bin1, bin$N$), (bin2, bin2) ... (bin2, bin$N$) ... (bin$N$, bin$N$), where $N$ is the maximum number of bins and bin$N$ is the highest redshift bin. The maximum scale cuts are fixed to the survey-specified scale cuts. ``--'' indicates no data points are used after the scale cut. The last row lists the remaining number of data points after the scale cut.}
\label{tab:scales}
\begin{tabular}{llcccc}
\hline\hline
&& DLS & CFHTLenS & DES-SV & KiDS-450  \\
\hline
$z_{p}$ & bin1 & 0.13 & 0.13 & 0.19 & 0.13 \\
& bin2 & 0.22 & 0.18 & 0.29 & 0.23  \\
& bin3 & 0.31 & 0.23 & 0.42 & 0.28 \\
& bin4 & 0.42 & 0.28 & & 0.38  \\
& bin5 & 0.57 & 0.38 & &  \\
& bin6 & & 0.43 & &  \\
& bin7 & & 0.48 & &  \\
\hline\hline
\vspace{0.1in}
\end{tabular}

\begin{tabular}{lcccc}
\hline\hline
& \multicolumn{4}{c}{Conservative}  \\
\hline
& DLS & CFHTLenS & DES-SV & KiDS-450 \\
\hline
$\theta_{\rm min, \pm}$ (arcmin)&8.7/78.6&8.7/78.6&6.2/56.5& 9.0/81.3 \\ 
&8.7/78.6&8.7/78.6&6.2/56.5&9.0/81.3  \\ 
&8.7/78.6&8.7/78.6&6.2/56.5&9.0/81.3  \\ 
&8.7/78.6&8.7/78.6&4.6/41.6&9.0/81.3  \\ 
&8.7/78.6&8.7/78.6&4.6/41.6&5.6/50.4  \\ 
&5.7/51.3&8.7/78.6&3.6/33.0&5.6/50.4  \\ 
&5.7/51.3&8.7/78.6&&5.6/50.4  \\ 
&5.7/51.3&6.6/60.0&&4.8/43.5  \\ 
&5.7/51.3&6.6/60.0&&4.8/43.5  \\ 
&4.4/40.0&6.6/60.0&&3.9/35.4  \\ 
&4.4/40.0&6.6/60.0&&  \\
&4.4/40.0&6.6/60.0&&  \\ 
&3.6/33.0&6.6/60.0&&  \\ 
&3.6/33.0&5.5/49.6&&  \\ 
&3.1/28.0&5.5/49.6&&  \\ 
&&5.5/49.6&& \\ 
&&5.5/49.6&& \\ 
&&5.5/49.6&& \\
&&4.8/43.0&& \\ 
&&4.8/43.0&& \\ 
&&4.8/43.0&& \\ 
&&4.8/43.0&& \\ 
&&3.9/35.1&& \\ 
&&3.9/35.1&& \\
&&3.9/35.1&& \\ 
&&3.6/32.6&& \\ 
&&3.6/32.6&& \\ 
&&3.4/30.6&& \\ 
\hline 
\# of data points& 94 & 137 & 25 & 52 \\
\hline\hline
\end{tabular}
\end{center}
\end{table}

\section*{Acknowledgement}
We thank Hendrik Hildebrandt and Martin Kilbinger for helping with the KiDS-450 reanalysis. We thank Sebastian Bocquet, Francois Lanusse and Javier Sanchez for help in the early stages of this project. We thank Karan Vahi and Mats Rynge for support in using the Pegasus software. This paper has undergone internal review in the LSST Dark Energy Science Collaboration, with Bhuvnesh Jain, Niall MacCrann, and Marco Raveri as internal reviewers.

CC was supported in part by the Kavli Institute for Cosmological Physics at the University of Chicago through grant NSF PHY-1125897 and an endowment from Kavli Foundation and its founder
Fred Kavli. 
CH acknowledges support from the European Research Council under grant number 647112. 
Support for MS was provided by the University of California Riverside Office of Research and Economic Development through the FIELDS NASA-MIRO program. A portion of this research was carried out at the Jet Propulsion Laboratory, California Institute of Technology, under a contract with the National Aeronautics and Space Administration.
M.J.J. acknowledges support for the current research from the National Research Foundation of Korea under the program 2017R1A2B2004644 and 2017R1A4A1015178. 
RM is supported by the Department of Energy Cosmic Frontier program, grant DE-SC0010118.
SJ acknowledges support from the Beecroft Trust and ERC 693024. 
AIM is advised by David W. Hogg and was supported by National Science Foundation grant AST-1517237.

The DESC acknowledges ongoing support from the Institut National de Physique Nucl\'eaire et de Physique des Particules in France; the Science \& Technology Facilities Council in the United Kingdom; and the Department of Energy, the National Science Foundation, and the LSST Corporation in the United States.  DESC uses resources of the IN2P3 Computing Center (CC-IN2P3--Lyon/Villeurbanne - France) funded by the Centre National de la Recherche Scientifique; the National Energy Research Scientific Computing Center, a DOE Office of Science User Facility supported by the Office of Science of the U.S.\ Department of Energy under Contract No.\ DE-AC02-05CH11231; STFC DiRAC HPC Facilities, funded by UK BIS National E-infrastructure capital grants; and the UK particle physics grid, supported by the GridPP Collaboration.  This work was performed in part under DOE Contract DE-AC02-76SF00515.

Below we acknowledge the data sources:

\textbf{CFHTLenS:} This work is based on observations obtained with MegaPrime/MegaCam, a joint project of CFHT and CEA/IRFU, at the Canada-France-Hawaii Telescope (CFHT) which is operated by the National Research Council (NRC) of Canada, the Institut National des Sciences de l'Univers of the Centre National de la Recherche Scientifique (CNRS) of France, and the University of Hawaii. This research used the facilities of the Canadian Astronomy Data Centre operated by the National Research Council of Canada with the support of the Canadian Space Agency. CFHTLenS data processing was made possible thanks to significant computing support from the NSERC Research Tools and Instruments grant program.

\textbf{DES-SV:} This project used public archival data from the Dark Energy Survey (DES). Funding for the DES Projects has been provided by the U.S. Department of Energy, the U.S. National Science Foundation, the Ministry of Science and Education of Spain, the Science and Technology FacilitiesCouncil of the United Kingdom, the Higher Education Funding Council for England, the National Center for Supercomputing Applications at the University of Illinois at Urbana-Champaign, the Kavli Institute of Cosmological Physics at the University of Chicago, the Center for Cosmology and Astro-Particle Physics at the Ohio State University, the Mitchell Institute for Fundamental Physics and Astronomy at Texas A\&M University, Financiadora de Estudos e Projetos, Funda{\c c}{\~a}o Carlos Chagas Filho de Amparo {\`a} Pesquisa do Estado do Rio de Janeiro, Conselho Nacional de Desenvolvimento Cient{\'i}fico e Tecnol{\'o}gico and the Minist{\'e}rio da Ci{\^e}ncia, Tecnologia e Inova{\c c}{\~a}o, the Deutsche Forschungsgemeinschaft, and the Collaborating Institutions in the Dark Energy Survey.
The Collaborating Institutions are Argonne National Laboratory, the University of California at Santa Cruz, the University of Cambridge, Centro de Investigaciones Energ{\'e}ticas, Medioambientales y Tecnol{\'o}gicas-Madrid, the University of Chicago, University College London, the DES-Brazil Consortium, the University of Edinburgh, the Eidgen{\"o}ssische Technische Hochschule (ETH) Z{\"u}rich,  Fermi National Accelerator Laboratory, the University of Illinois at Urbana-Champaign, the Institut de Ci{\`e}ncies de l'Espai (IEEC/CSIC), the Institut de F{\'i}sica d'Altes Energies, Lawrence Berkeley National Laboratory, the Ludwig-Maximilians Universit{\"a}t M{\"u}nchen and the associated Excellence Cluster Universe, the University of Michigan, the National Optical Astronomy Observatory, the University of Nottingham, The Ohio State University, the OzDES Membership Consortium, the University of Pennsylvania, the University of Portsmouth, SLAC National Accelerator Laboratory, Stanford University, the University of Sussex, and Texas A\&M University.
Based in part on observations at Cerro Tololo Inter-American Observatory, National Optical Astronomy Observatory, which is operated by the Association of Universities for Research in Astronomy (AURA) under a cooperative agreement with the National Science Foundation.

\textbf{KiDS-450:} This work is based on data products from observations made with ESO Telescopes at the La Silla Paranal Observatory under programme IDs 177.A-3016, 177.A-3017 and 177.A-3018. We use cosmic shear measurements from the Kilo-Degree Survey (Kuijken et al. 2015, Hildebrandt \& Viola et al. 2017, Fenech Conti et al. 2016), hereafter referred to as KiDS. The KiDS data are processed by THELI (Erben et al. 2013) and Astro-WISE (Begeman et al. 2013, de Jong et al 2015). Shears are measured using lensfit (Miller et al. 2013), and photometric redshifts are obtained from PSF-matched photometry and calibrated using external overlapping spectroscopic surveys (see Hildebrandt et al. 2016).


The contributions from the primary authors are listed below. C.C. led the main analysis and writing of this paper. M.W. wrote the main software package WLPipe which integrates the python scripts using the Pegasus workflow engine. S.D. helped with the pipeline development, covariance assessment, comparison metrics, pipeline testing, and editing. 

\end{document}